\documentclass[12pt]{article}
\usepackage{amssymb,amsmath}
\usepackage{latexsym}
\usepackage{graphicx}

\renewcommand{\theequation}{\arabic{section}.\arabic{equation}}

\def\IR{\mathbb{R}}

\def\cF{{\cal F}}

\def\ds{\displaystyle}

\newcommand{\be}{\begin{equation}}
\newcommand{\ee}{\end{equation}}
\newcommand{\bea}{\begin{eqnarray}}
\newcommand{\eea}{\end{eqnarray}}

\def\ov{\overline}

\begin{document}

\begin{titlepage}

\begin{flushright}
IPhT-T08/211
\end{flushright}

\bigskip
\bigskip
\centerline{\Large \bf Supertubes in Bubbling Backgrounds:}
\medskip
 \centerline{\Large \bf Born-Infeld Meets Supergravity}
\bigskip
\bigskip
\centerline{{\bf Iosif Bena$^1$, Nikolay Bobev$^2$, Cl\'{e}ment
Ruef$^{\, 1}$ and Nicholas P. Warner$^2$}}
\bigskip
\centerline{$^1$ Institut de Physique Th\'eorique, }
\centerline{CEA/Saclay, CNRS-URA 2306,}
 \centerline{Orme des Merisiers, F-91191 Gif sur Yvette, France}
\bigskip
\centerline{$^2$ Department of Physics and Astronomy}
\centerline{University of Southern California} \centerline{Los
Angeles, CA 90089, USA}
\bigskip
\centerline{{\rm iosif.bena@cea.fr, bobev@usc.edu,
clement.ruef@cea.fr, warner@usc.edu} }
\bigskip \bigskip

\begin{abstract}

We discuss two ways in which one can study two-charge supertubes
as components of generic three-charge, three-dipole charge
supergravity solutions. The first is using the Born-Infeld action
of the supertubes, and the second is via the complete supergravity
solution.  Even though the Born-Infeld description is only a probe
approximation, we find that it gives exactly the same essential
physics as the complete supergravity solution. Since supertubes
can depend on arbitrary functions, our analysis strengthens the
evidence for the existence of three-charge black-hole microstate
geometries that depend on an infinite set of parameters, and sets
the stage for the computation of the entropy of these backgrounds.
We examine numerous other aspects of supertubes  in  three-charge, three-dipole charge supergravity backgrounds, including chronology protection during mergers, the contribution of supertubes to the charges and angular momenta, and the enhancement of their entropy. In particular, we find that entropy enhancement affects supertube fluctuations both along the internal and the spacetime directions, and we prove that the charges that give the enhanced entropy can be much larger than the asymptotic charges of the solution.
We also re-examine the embedding of five-dimensional black rings in Taub-NUT, and show that in different coordinate patches a ring can correspond to different four-dimensional black holes. Last, but not least, we show that all the three-charge black hole microstate geometries constructed so far can be embedded in $AdS_3 \times S^3$, and hence can be related to states of the D1-D5 CFT.

\end{abstract}

\topmargin=-0.8in \oddsidemargin=-0.4in \textheight=9.0in
\textwidth=6.8in

\end{titlepage}

\topmargin=-0.4in \oddsidemargin=-0.2in \textheight=8.8in
\textwidth=6.8in


{\tableofcontents}

\section{Introduction}

The physics of two-charge supertubes is an essential ingredient in
understanding the microstates of the D1-D5 system.  Indeed,
supergravity solutions for two charge supertubes with D1 and D5
charges and KKM dipole charge are smooth in six dimensions and can
have arbitrary shape. Hence, they have an infinite dimensional
classical moduli space, which, upon quantization, gives the entropy
one expects from counting at weak-coupling: $S = 2 \pi \sqrt{2
N_1N_5}$
\cite{Mateos:2001qs,Palmer:2004gu,Lunin:2001fv,Lunin:2001jy,Lunin:2002iz,Rychkov:2005ji,Kanitscheider:2007wq}.

While this entropy is considerable, it is nowhere near the entropy
of a black hole with three charges:  $S =  2 \pi \sqrt{N_1 N_5
N_P}$  \cite{stromvafa}. Hence,  if one's goal is to prove that in
the regime of parameters where the classical black hole exists one
can find a very large number of string/supergravity configurations
that realize enough microstates to account for the
Bekenstein-Hawking entropy of this black hole
\cite{Mathur:2005zp,Bena:2007kg,Skenderis:2008qn,Balasubramanian},
the entropy coming from two-charge supertubes  does not appear to
be large enough.

However, it has recently been found that the humble two-charge
supertube has more to it than meets the eye:  In a scaling
supergravity background with large magnetic dipole fluxes it can
undergo {\it entropy enhancement} \cite{enhance}. That is, if one
uses the Born-Infeld action to compute the entropy of a probe
two-charge supertube placed in a background with three charges and
three dipole charges, one finds that such a supertube can have an
entropy that is {\it much} larger than that of the same supertube
in empty space. The magnetic dipole-dipole interactions between
the supertube and the background can greatly increase the capacity
of the supertube to store entropy. Hence, the interaction with the supergravity background
can enhance (or decrease) the entropy coming from the fluctuating
shape of a supertube.

As yet,  the fully back-reacted solution corresponding to a
supertube of arbitrary shape has not been constructed and so the
entropy enhancement calculation has only been done in a probe
approximation.  Nevertheless, in the absence of the fully
back-reacted  solutions, one can still pose a very sharp question,
whose answer can tilt the balance one way or another in the quest
to understand whether the black hole is a thermodynamic
description of a very large number of horizonless microstates:
``{\it Do two-charge supertubes that are solutions of the
Born-Infeld equations of motion correspond  to smooth solutions of
supergravity once the back-reaction is included?}''

If the answer to this question is {\it yes}, then all the
supertube microstates that were counted in \cite{enhance} give
smooth microstate solutions of supergravity, valid in the same
regime of parameters where the classical black hole exists. Since
the Born-Infeld counting might give a macroscopic
(black-hole-like) entropy, this would imply that the same entropy
could come from smooth supergravity solutions. Our goal in this
paper is to show that the Born-Infeld description of a supertube
does indeed capture all the essential physics of the complete
supergravity solution and argue that the corresponding
supergravity solution will be smooth in the D1-D5 duality frame.

First, we establish that when one has {\it both} a Born-Infeld and
a supergravity description of supertubes in a three-charge,
three-dipole-charge background, the two descriptions agree to the
last detail. As we will see, this agreement can be rather subtle.
For example, a supertube that is merging with a black ring appears
to merge at an angle that depends on its charges but when this
merger is described in supergravity, the merger appears to be
angle-independent.  The resolution of this rests upon the correct
identification of constituent charges and the fact that such
charges can depend upon ``large'' gauge transformations.

Another important fact we establish is that the solutions of the
Born-Infeld action are always such that the corresponding
solutions of supergravity are smooth in the duality frame where
the supertube has D1 and D5 charges. Indeed, upon carefully
relating the Born-Infeld and the supergravity charges, we will
find that the equations that insure that a supertube is a solution
of the Born-Infeld action are identical to the equations that
insure that the corresponding supergravity solution is smooth.

One could take the position that our analysis here only implies
the smoothness of round supertubes, which have both Born-Infeld
and supergravity descriptions. It is possible that the wiggly
supertubes (which, upon entropy enhancement, might give a
black-hole-like entropy) could give rise to singular solutions
when brought to the supergravity regime. While such a possibility
cannot be fully excluded before the construction of the fully
back-reacted wiggly supertubes, we have some rather strong reasons
to believe it is highly unlikely.  Indeed,  if one investigates
the conditions for smoothness of the supergravity solution and
compares them to the Born-Infeld conditions, one finds that both
the supergravity conditions and the Born-Infeld conditions are
{\it local}. Hence, since any curve can be locally approximated as
flat, our analysis indicates that no local properties of wiggly
supertubes (like the absence of regions of high curvature) will differ
from the local properties of round or flat supertubes. Thus one
has a very reasonable expectation that supertubes of arbitrary
shape will source smooth supergravity solutions.

In particular, if one considers supertubes of arbitrary shape in
flat space, the solutions of the Born-Infeld action always give
smooth supergravity solutions \cite{Lunin:2001jy,Lunin:2002iz}. If
one now considers a three-charge, three-dipole charge solution
containing supertubes whose wiggling scale is much smaller than
the variation scale of the gauge fields of the background, one can
perform a gauge transformation that locally removes the gauge
fields and transforms a portion of this supertube into a portion
with many wiggles of a supertube in flat space. Since the latter
supertube is smooth, and since gauge transformations do not affect
the smoothness of solutions, this implies that the original wiggly
supertube is also giving a smooth solution.

Obviously the foregoing conclusion is restricted to the domain of
validity of supergravity. If a supertube of arbitrary shape is
very choppy, the local curvature will be roughly proportional to
the inverse of the scale of the choppiness, and hence if the
choppiness is Planck-sized then the curvature of the solution will
also be Planck-sized.   Such solutions are thus outside the domain
of validity of supergravity.    The main conclusion of our
analysis is that supertubes whose wiggles are {\it not}
Planck-sized will give smooth, low-curvature supergravity
solutions.

Our analysis does not establish whether the typical microstates of
a certain black hole will have high curvature or will be well
described in supergravity. However, it does establish that if the
wiggles of the Born-Infeld supertubes that gave the typical
microstates are not Planck-sized, the corresponding supergravity
solutions will not be either.

The second aim of this paper is to clarify several issues related
to embedding of black rings in Taub-NUT, and to the relation
between the electric charges of the ring and those of the
corresponding four-dimensional black holes. We show that when
embedding a black ring solution in Taub-NUT one needs to use at
least two coordinate patches. From the perspective of one patch,
the electric charges are the ones found in \cite{Bena:2005ni}, and
the ring ``angular'' momentum along the Taub-NUT fiber
(corresponding to the D0 charge in four dimensions) is given by
the difference of the two five-dimensional angular momenta. The
entropy is given by the $E_{7(7)}$ quartic invariant of these
charges \cite{Bena:2004tk}, as common for four-dimensional BPS black holes \cite{KalloshKol}.

From the perspective of the other patch, the charges and the
Kaluza-Klein angular momentum of the corresponding
four-dimensional black hole are shifted, to certain values that
have no obvious five-dimensional interpretation\footnote{The
asymptotic five-dimensional electric charge is the average between
the four-dimensional electric charges in the two patches.}. The
entropy of the black ring is again given by the $E_{7(7)}$ quartic
invariant, but now as a function of the shifted charges. The two
four-dimensional black holes corresponding to the black ring are
related by a gauge transformation, which shifts the Dirac string
in the gauge potentials from one side of the ring to
another\footnote{Note that we can also perform a gauge
transformation that shifts the four-dimensional electric charges
to the asymptotic five-dimensional charges of the black ring
\cite{Hanaki:2007mb}. The corresponding
four-dimensional solution has two Dirac strings in the gauge
potentials}.

A third result in this paper is to verify chronology protection
when supertubes and black rings are merged. While  chronology
protection is expected to be valid for this merger, the way it
works is subtle.   We compute the merger condition between a
supertube and a black ring, and find that this condition depends
on the position on the $S^2$ of the black ring where the supertube
merges. We also find that neither very large nor very small
supertubes can merge with the ring, for obvious reasons. If one
varies the charge of the supertubes we find that mergers happen
when the charge lies in a certain interval:  At one extreme the
supertube barely merges on the exterior of the ring while at the
other it barely merges on the interior of the ring.

We also discuss a subtlety in identifying the constituent charges
carried  into the black ring by a merging supertube. We find that
when the  $S^1$ of the supertube curves around the $S^2$ of the black ring horizon, the
charge brought in by a given supertube {\it must} depend on the
$S^2$ azimuthal angle at which the supertube merges with the ring.
Otherwise chronology is not protected.  It would be most
interesting to see how this comes about  in the full supergravity
merger solution.

The fourth aim of this paper is to present in detail, and to
extend, the entropy enhancement calculation of \cite{enhance}. Our analysis establishes that supertube entropy enhancement can come from supertube oscillation modes in {\it both} the internal space of the solution ($T^4$ in our calculations) and from oscillations of supertubes in the transverse spacetime directions. We analyze entropy enhancement in 
black-ring backgrounds, in which the detailed computation is more straightforward than in generic solutions with a Gibbons-Hawking base.
We find that, despite the presence of different (large) factors in the mode expansions, the fluctuations in the plane transverse to the ring give a
contribution to the entropy that is identical to that coming from
the fluctuations along the compactification torus.    

If, as we expect, the enhanced entropy coming from these fluctuations will be black-hole-like, and therefore the fluctuating supertubes will give the typical microstates of the corresponding black hole, our analysis establishes that these microstates will have a non-trivial transverse size.  
We believe it important to calculate the amount of entropy enhancement coming from {\it all} the oscillations of the
supertube.  If the other transverse oscillations are more entropic than
the torus ones, this would suggest that five-dimensional supergravity may be enough to capture the typical states of the black hole. On the other hand, 
if the torus and the transverse fluctuations are
equally entropic (as hinted by our partial analysis), the typical
states will probably  have a curvature set by the compactification scale. 
Even if this scale is at the Planck scale, the microstate geometries constructed in supergravity
will give a pretty good approximation of the rough features of the
typical states (like the size, the density profile, the multipole
moments).  Hence the smooth microstate geometries will act as
{\it  representatives} of the typical black hole microstates  \cite{enhance,Warner:2008ma}.

We begin in Section 2 by presenting the general three-charge
three-dipole-charge solutions in various duality frames that will
be used throughout the paper.  In particular, we give these
solutions in the type IIA frame where the three charges correspond
to D0 branes, D4 branes and F1 strings (the D0-D4-F1 frame), and
in the type IIB duality frame where the three charges correspond
to D1 branes, D5 branes and momentum (the D1-D5-P frame).  We also
obtain in these frames (for the first time to our knowledge) the
exact form of the RR potentials when the base of the solution is a
Gibbons-Hawking metric.

In Section 3 we explore the regularity of the supergravity
solutions corresponding to two-charge supertubes with D1 and D5
charges placed in three-charge three-dipole charge solutions. 
We find two {\it local} conditions that insure
the absence of singularities near the supertube profile.

In Section 4 we study probe two-charge supertubes in general
three-charge solutions: black holes, black rings, and bubbling
solutions with a Gibbons-Hawking base.   We present
a detailed analysis of two-charge and three-charge supertube
probes in the background of a supersymmetric three-charge black
ring. We also relate the supergravity and Born-Infeld charges of
supertubes, and show that the supergravity smoothness conditions
derived in Section 3 agree with the ones derived from the
Born-Infeld action.  In Section 5 we study mergers of the
supertube with the black ring and discuss chronology protection
and black hole thermodynamics during these mergers.

Section 6 contains an in-depth derivation of the entropy coming
from oscillations of supertubes, illustrating the entropy
enhancement mechanism presented in \cite{enhance} for black rings,
and general solutions with a Gibbons-Hawking base. Section 7 is
devoted to conclusions.

In Appendix A we give the details of the T-duality transformations
of three-charge three-dipole charge solutions in various duality
frames.  We also show how to compute the RR potentials
corresponding to these solutions in various duality frames.  In
Appendix B we take a decoupling limit for general three-charge
three-dipole charge solutions in D1-D5-P frame, which leads to an asymptotically
$AdS_3\times S^3\times T^4$ geometry. This establishes that all the black hole and black ring microstate solutions constructed so far are dual to states of the D1-D5 CFT, and serves as a starting point for analyzing these microstates using holographic anatomy in the context of the $AdS_3/CFT_2$
correspondence \cite{Skenderis:2006ah}.  In Appendix C we compute
the angular momentum of a supertube in several three-charge
backgrounds and in Appendix D we give the units and conventions
used throughout our calculations.

\section{Review of three-charge solutions}
\label{3ChgSummary}

\subsection{Three-charge solutions in the M2-M2-M2 (M-theory) frame}
\label{MtheoryBPS}

Three-charge solutions with four supercharges are most simply
written in the M-theory duality frame in which the three charges
are treated most symmetrically and correspond to three types of M2
branes wrapping three $T^2$'s inside $T^6$ \cite{Bena:2004de}. The
metric is:
\begin{multline}
ds_{11}^2  = - \left( Z_1 Z_2  Z_3 \right)^{-{2 \over 3}} (dt+k)^2
+ \left( Z_1 Z_2 Z_3\right)^{1 \over 3} \, ds_4^2  \\+ \left(Z_2
Z_3 Z_1^{-2}  \right)^{1 \over 3} (dx_5^2+dx_6^2) + \left( Z_1 Z_3
Z_2^{-2} \right)^{1 \over 3} (dx_7^2+dx_8^2)  + \left(Z_1 Z_2
Z_3^{-2} \right)^{1 \over 3} (dx_9^2+dx_{10}^2) \,,
\label{11Dmetric}
\end{multline}
where $ds_4^2$ is a four-dimensional hyper-K\"{a}hler metric
\cite{Bena:2004de,Gutowski:2004yv,Gauntlett:2002nw}\footnote{This
metric can have regions of signature  $+4$ and signature  $-4$
\cite{gm-ambipolar,Bena:2005va,Berglund:2005vb,oleg,gms}, and for this reason
we usually refer to it as ambipolar.}. The solution has a
non-trivial three-form potential, sourced both by the M2 branes
(electrically) and by the M5 dipole branes (magnetically):
\begin{equation}
\mathcal{A} = A^{(1)}\wedge dx_5 \wedge dx_6 + A^{(2)}\wedge dx_7
\wedge dx_8 + A^{(3)}\wedge dx_9 \wedge dx_{10}
\label{11Dthreeform}.
\end{equation}
The magnetic contributions can be separated from the electric ones
by defining the ``magnetic field strengths:"
\begin{equation}
\Theta^{(I)} \equiv dA^{(I)} ~+~
d\left(\ds\frac{(dt+k)}{Z_I}\right)\,, \qquad I=1,2,3.
\end{equation}
Finding supergravity solutions for this system then boils down to
solving the following system of BPS equations\footnote{These
equations also give supersymmetric solutions when the $T^6$ is
replaced by a Calabi-Yau three-fold, and $C_{IJK}$ is replaced by
the triple intersection numbers of this three-fold.}:
\begin{equation}
\begin{array}{l}
\Theta^{(I)} = \star_4 \Theta^{(I)}\,, \\\\
\nabla^2 Z_{I} = \ds\frac{1}{2}C_{IJK} \star_{4} (
\Theta^{(J)}\wedge
\Theta^{(K)})\,, \\\\
dk + \star_4 dk = Z_{I} \Theta^{I} \,.
\end{array} \label{BPSequations}
\end{equation}
In these equations, $\star_4$ is the Hodge dual in the
four-dimensional base space, $ds^2_4$, and
$C_{IJK}=|\epsilon_{IJK}|$.   If the four-dimensional base manifold
has a triholomorphic $U(1)$ isometry then the metric on the base can
be put in a Gibbons-Hawking (GH) form
\cite{GibbonsZT,Gibbons:1987sp}:
%
%
\begin{equation}
ds_4^2 ~=~ V^{-1} \, \big( d\psi + A)^2  ~+~ V\,  d \vec y \cdot d
\vec y \,, \label{GHmetric}
\end{equation}
where $V$ is a harmonic function on the  $\mathbb{R}^3$ spanned by
$(y_1,y_2,y_3)$ and $\vec{\nabla}\times\vec{A} = \vec{\nabla}V$. For
such metrics, the BPS equations (\ref{BPSequations}) can be solved
explicitly \cite{Bena:2005ni,Gauntlett:2004qy}. The most general
solution can be written in terms of eight harmonic functions $(V,
K^I, L_I, M)$ on the $\mathbb{R}^3$ base of the GH
space\footnote{For M-theory compactifications on a generic
Calabi-Yau three-fold the number of harmonic functions will be
$2h^{1,1}+2$.  See \cite{Cheng:2006yq} for a discussion of such
solutions.}. It is convenient to introduce the vielbeins:
\begin{equation}
\hat e^1~=~ V^{-{1\over 2}}\, (d\psi ~+~ A) \,, \qquad \hat
e^{a+1} ~=~ V^{1\over 2}\, dy^a \,, \quad a=1,2,3 \,,
\label{GHvierbeins}
\end{equation}
then one has
\begin{equation}
\Theta^{(I)} ~=~ - \sum_{a=1}^3 \, \left(\partial_a \left(
V^{-1}\, K^I \right)\right) \, \left( \hat{e}^1  \wedge
\hat{e}^{a+1} ~+~ \ds\frac{1}{2}\, \epsilon_{abc}\,\hat{e}^{b+1}
\wedge \hat{e}^{c+1} \right) \,. \label{GHdipoleforms}
\end{equation}
The three gauge fields, $A^{(I)}$, can be written as
\begin{equation}
A^{(I)} =   B^{(I)}   - \ds\frac{1}{Z^{I}}(dt + k)~,
\label{GHMaxwellpot}
\end{equation}
where
\begin{equation}
 B^{(I)}=V^{-1} K^{I} \, (d \psi + A) ~+~ \vec{\xi}^{(I)}\cdot d\vec{y} \,,
 \qquad \vec \nabla \times \vec \xi^{(I)}  ~\equiv~ - \vec \nabla K^I \,.
\label{vecpotdefns}
\end{equation}
The functions $Z_I$ and the angular momentum one-form $k$ are
given by
\begin{equation}
Z_I =  \frac{C_{IJK}}{2}\ds\frac{K^{J}K^{K}}{V} + L_I  \,, \qquad
k = \mu (d\psi+ A) + \vec{\omega}\cdot d\vec{y} \,,
\label{GHparts}
\end{equation}
where
\begin{equation}
\mu ~=~ \frac{C_{IJK}}{6}\ds\frac{K^{I}K^{J}K^{K}}{V^2} ~+~
\ds\frac{K^{I}L_{I}}{2V} ~+~  M
\end{equation}
and $\vec{\omega}$ satisfies the equation
\begin{equation}
\vec{\nabla}\times \vec{\omega} = V\vec{\nabla}M - M\vec{\nabla}V
+\ds\frac{1}{2} \left( K^{I}\vec{\nabla}L_{I} -
L_{I}\vec{\nabla}K^{I}\right). \label{GHomegaeqn}
\end{equation}

This solution can describe five-dimensional black holes, circular
black rings and supertubes, as well as smooth ``bubbling
solutions'' and an arbitrary superposition of these objects. Upon
compactifying to four dimensions, all these reduce to BPS
multi-center black-hole configurations
\cite{BenaKrausKKM,Balasubramanian:2006gi} of the type first
considered in \cite{denef1}.

The harmonic functions are usually chosen to be sourced by simple
poles:
\begin{equation}
\begin{array}{l}
V ~=~ \epsilon_0 ~+~ \ds\sum_{j=1}^{N} \ds\frac{q_j}{r_{j}}~,
\qquad\qquad K^{I} ~=~ \kappa_0^{I} ~+~ \ds\sum_{j=1}^{N}
\ds\frac{k_j^{I}}{r_{j}}~, \\\\
L_{I} ~=~ l_0^{I} ~+~ \ds\sum_{j=1}^{N} \ds\frac{l_j^{I}}{r_{j}}~,
\qquad\qquad M ~=~ m_0 ~+~ \ds\sum_{j=1}^{N} \ds\frac{m_j}{r_{j}}
\,,
\end{array}\label{GHharmonicfn}
\end{equation}
where $r_{j} = |\vec{y}-\vec{y}_j|$ and $N$ is the number of
centers.  We think of the residues of the poles of these functions
as defining the GH charges of the corresponding solution. As was
discussed in \cite{Bena:2008wt}, gauge transformations and
spectral flow can reshuffle these charges, but this produces
physically equivalent solutions.

A necessary (but not sufficient) condition for the solutions to be
free of closed timelike curves (CTC's) is to satisfy the
``integrability equations,'' or ``bubble equations,''
\cite{Bena:2005va,Berglund:2005vb,denef1}:
\begin{equation}
\ds\sum_{j=1, j\neq i}^{N}~ \ds\frac{\langle\hat Q_i|\hat
Q_j\rangle }{r_{ij}} ~=~ 2 (\varepsilon_0 m_i - m_0 q_i) +
\ds\sum_{I=1}^{3} (k_0^I l_i^I - l_0^Ik_i^I) \label{GHbubbleeqns}
\end{equation}
where $\langle\hat Q_i|\hat Q_j\rangle$ is the symplectic
product\footnote{This product is sometimes called the
Dirac-Schwinger-Zwanziger product.} between the eight-vectors of
charges at the points $i$ and $j$
\begin{equation}
\langle\hat Q_i|\hat Q_j\rangle \equiv 2 (m_j q_i -q_j m_i ) +
\ds\sum_{I=1}^{3} (l_j^Ik_i^I - k_j^I l_i^I)~.
\end{equation}
For smooth solutions with multiple GH centers the parameters of
the solution must also satisfy the additional regularity
constraints:
\begin{equation}
l_{j}^{I} =
-\frac{C_{IJK}}{2}\ds\frac{k_{j}^{J}k_{j}^{K}}{q_{j}}~, \qquad~
m_{j} = \ds\frac{k_{j}^{1}k_{j}^{2}k_{j}^{3}}{2q_{j}^{2}} \,,
\end{equation}
These are required to cancel the singularities in $Z_I$ and $\mu$
and with these choices the integrability equations
(\ref{GHbubbleeqns}) reduce to the bubble equations considered in
\cite{Bena:2005va,Berglund:2005vb}.

One can arrange for the global absence of CTC's by requiring that
there is a well-defined, global time function \cite{Berglund:2005vb}.  This is much more
stringent than the bubble equations (which only eliminate CTC's in
the neighborhood of the GH points) and means that the following
inequality should be satisfied globally \cite{Bena:2005va,Berglund:2005vb}:
\begin{equation}
Z_1Z_2Z_3V ~-~ \mu^2V^2 ~-~|\omega|^2 ~\geq~ 0 \label{GHnoCTCcond}
\,.
\end{equation}
This condition is very hard to check in general and usually has to
be checked numerically for particular solutions.

As we mentioned earlier, in order to study two-charge supertubes
in backgrounds like those presented here, it is useful to dualize
to a frame in which the two-charge supertube action is simple. One
such frame is where the three electric charges correspond to D0
branes, D4 branes and F1 strings and the supertube carries  D0 and
F1 electric charges and D2 dipole charge \cite{Mateos:2001qs}. On
the other hand, in order to study the supergravity solutions
describing supertubes in black-ring or bubbling backgrounds, it is
useful to work in a duality frame in which the supergravity
solution for the supertubes is smooth. In this frame the electric
charges of the background correspond to D1 branes, D5 branes, and
momentum P, and the supertube carries D1 and D5 charges, with KKM
dipole charge.  We therefore dualize the foregoing M-theory
solution to these frames and give all the details of the solutions
explicitly.

\subsection{Three-charge solutions in the D0-D4-F1 duality frame}
\label{D0D4F1BPS}

Here we will present the three-charge solutions in the duality
frame in which  they have electric  charges corresponding to D0
branes, D4 branes, and F1 strings, and dipole charges
corresponding to D6, D2 and NS5 branes. We use the T-duality rules
(given in Appendix A) to transform field-strengths. It should be
emphasized that our results are correct for any three-charge
solution (including those without a tri-holomorphic $U(1)$
\cite{Bena:2007ju}), however, finding the explicit form of the RR
and NS-NS {\it potentials} (which is crucial if we want to
investigate this solution using probe supertubes) is
straightforward only when the solution can be written in
Gibbons-Hawking form.

Label the coordinates by $(x^0, \dots, x^8,z)$\footnote{See
Appendix A for more details about the brane configuration that we
use.}. The electric charges $N_1, N_2$ and $N_3$ of the solution then
correspond to:
\begin{equation}
N_1: \, D0   \qquad  N_2: \, D4 \, (5678) \qquad N_3: \, F1 \, (z)
\label{D0D4F1}
\end{equation}
where the numbers in the parentheses refer to spatial directions
wrapped by the branes and $z\equiv x^{10}$. The  magnetic dipole
moments of the solutions correspond to:
\begin{equation}
n_1: \, D6 \, (y5678z) \qquad  n_2: \, D2 \, (yz) \qquad n_3: \,
NS5 \, (y5678)  \,, \label{d6d2ns5}
\end{equation}
where $y$ denotes the brane profile in the spatial base, $(x^1,
\dots, x^4)$. The metric of the solution is:
\begin{equation}
ds^2_{IIA} = - \ds\frac{1}{Z_3\sqrt{Z_1Z_2}}(dt+k)^2 +
\sqrt{Z_1Z_2}ds^2_4 +\ds\frac{\sqrt{Z_1Z_2}}{Z_3}dz^2 +
\ds\sqrt{\frac{Z_1}{Z_2}}(dx_5^2+dx_6^2+dx_7^2+dx_8^2) \,.
\label{D0D4F1metric}
\end{equation}
The dilaton and the Kalb-Ramond fields are:
\begin{equation}
\Phi = \ds\frac{1}{4} \log \left(
\ds\frac{Z_1^3}{Z_2Z_3^2}\right), \qquad\qquad B = -dt\wedge
dz-A^{(3)}\wedge dz \,. \label{D0D341NSNS}
\end{equation}
The RR field strengths are
\begin{equation}
F^{(2)} = - \mathcal{F}^{(1)} \,, \qquad  \widetilde{F}^{(4)} =
-\left(\ds\frac{Z_2^5}{Z_1^3Z_3^2}\right)^{1/4} \star_5
(\mathcal{F}^{(2)}) \wedge dz   \,, \label{D0D4F1RR}
\end{equation}
where we define  $\mathcal{F}^{(I)}\equiv dA^{(I)}$ and $\star_5$
is the Hodge dual with respect to the five dimensional metric:
\begin{equation}
ds^2_{5} = - \ds\frac{1}{Z_3\sqrt{Z_1Z_2}}(dt+k)^2 +
\sqrt{Z_1Z_2}ds^2_4 \,. \label{5dmetric}
\end{equation}
The foregoing results are valid for any three-charge solution with
an arbitrary hyper-K\"{a}hler base. As we show in Appendix A, when
the base has a Gibbons-Hawking metric one can easily find the RR
3-form potential:
\begin{equation}
C^{(3)} ~=~ \big(\zeta_a ~+~  V^{-1} K^3 \xi^{(1)}_a  \big) \,
\Omega_-^{(a)}  \wedge dz  - \big(Z_3^{-1} ( dt +k) \wedge B^{(1)}
~+~ dt \wedge A^{(3)}  \big) \wedge dz \,, \label{C3IIA}
\end{equation}
where $\xi^{(1)}_a$ and $\zeta_a$ are defined by equations
(\ref{vecpotdefns}) and (\ref{zetadefn}). Thus we have the full
three-charge supergravity solution in the D0-D4-F1 duality frame.
In Section 4 we will perform a probe analysis in this class of
backgrounds using the  DBI action for supertubes  with D0 and F1
electric and D2 dipole charge.

\subsection{Three-charge solutions in the D1-D5-P duality frame}
\label{D1D5PBPS}

One can T-dualize the solution above along $z$ to obtain a
solution with D1, D5 and momentum charges:
\begin{equation}
N_1: \, D1 \,(z)  \qquad  N_2: \, D5 \, (5678z) \qquad N_3: \, P
\, (z) \label{D1D5P}
\end{equation}
and dipole moments corresponding to wrapped D1 branes, D5 branes
and Kaluza Klein Monopoles (kkm):
\begin{equation}
n_1: \, D5 \, (y5678) \qquad  n_2: \, D1 \, (y) \qquad n_3: \, kkm
\, (y5678z) \,. \label{d5d1kkm}
\end{equation}
The metric is
\begin{eqnarray}
ds^2_{IIB}  = - \ds\frac{1}{Z_3\sqrt{Z_1Z_2}}\,(dt+k)^2  &+&
\sqrt{Z_1Z_2}\, ds^2_4 + \ds\frac{Z_3}{\sqrt{Z_1Z_2}}\,(dz+A^{(3)})^2  \\
&+&  \ds\sqrt{\frac{Z_1}{Z_2}}\, (dx_5^2+dx_6^2+dx_7^2+dx_8^2)
\label{D1D5Pmetric}
\end{eqnarray}
and the dilaton and the Kalb-Ramond  field are:
\begin{equation}
\Phi = \ds\frac{1}{2} \log \left( \ds\frac{Z_1}{Z_2}\right),
\qquad\qquad B = 0\,. \label{D1D5PNSNS}
\end{equation}
The only non-zero RR three-form field strength is:
\begin{equation}
F^{(3)} = -\left(\ds\frac{Z_2^5}{Z_1^3Z_3^2}\right)^{1/4} \star_5
(\mathcal{F}^{(2)}) - \mathcal{F}^{(1)}\wedge (dz - A^{(3)}) \,.
\label{D1D5PRR}
\end{equation}
If we specialize  our general result to the supersymmetric black
ring solution in the  D1-D5-P frame then it agrees (up to
conventions) with \cite{Elvang:2004ds}. It is also elementary to
find the  RR two-form potential for a general BPS solution with GH
base in D1-D5-P frame.  This can be done by T-dualizing the IIA
D0-D4-F1 result (\ref{C3IIA}), to obtain:
\begin{eqnarray}
C^{(2)} ~=~ \big(\zeta_a ~+~  V^{-1} K^3 \xi^{(1)}_a  \big) \,
\Omega_-^{(a)}    &-& \big(Z_3^{-1} ( dt +k) \wedge B^{(1)} ~+~ dt
\wedge A^{(3)}  \big) \\
&+& A^{(1)}\wedge(A^{(3)}-dz -dt) ~+~ dt\wedge(A^3-dz) \,,
\end{eqnarray}
where again $\xi^{(1)}_a$ and $\zeta_a$ are defined in equations
(\ref{vecpotdefns}) and (\ref{zetadefn}). This is the full
three-charge supergravity solution in the D1-D5-P duality frame.
As shown in \cite{Lunin:2002iz}, two-charge supertubes in flat
space are regular only in this duality frame, so our general
result can be used to analyze the regularity of two charge
supertubes in a general three-charge solution. This will be the
subject of the next section.

\section{Regularity of supertubes in supergravity}

\subsection{Constraints from supertube regularity}

Consider the D1-D5-P solutions in which one of the centers has
vanishing GH charge, and non-trivial D1 and D5 electric charges.
Generally such a solution is not regular and can have a horizon or
a naked singularity. However, the solution will be regular if one
arranges the charges at this point to be those of a two-charge supertube.

Suppose that at $r_1= 0$  we have a round two-charge supertube
with one dipole charge.  We take the latter to be $k^3_1$ and so
we have $k^1_1 = k^2_1 =0$ and $l^3_1 =0$.  This means that in the
neighborhood of a two-charge supertube at  $r_1 =0$, we must have:
\begin{equation}
Z_I ~\sim~ {\cal O}(r_1^{-1}) \,, \quad I =1,2 \,; \qquad \qquad
V, Z_3 ~\sim~ {\rm finite} \,.   \label{neartube}
\end{equation}
The six-dimensional metric in IIB frame can be re-written as:
\begin{equation}
ds_6^2 ~=~ - \ds\frac{1}{Z_3\sqrt{Z_1Z_2}}\, (dt+k)^2  +
\sqrt{Z_1Z_2}\,ds^2_4 + \ds\frac{Z_3}{\sqrt{Z_1Z_2}}\,
(dz+A^{(3)})^2 \,. \label{newsixmetric}
\end{equation}

To check regularity along the supertube one must examine potential
singularities along the $\psi$-fiber by collecting all the $(d\psi
+  A)^2$ terms in (\ref{newsixmetric}):
\begin{equation}
(Z_1 Z_2)^{-\frac{1}{2}} \, V^{-2} \left[ Z_3\, (K^3)^2 ~-~ 2 \mu
V K^3 ~+~ Z_1 Z_2 V \right] \, (d\psi +  A)^2 \,.
\label{fibercoeff}
\end{equation}
For regularity as $r_1 \to 0$, one must have:
\begin{equation}
\lim_{r_1 \to 0}\,  r_1^2 \left[ Z_3\, (K^3)^2 ~-~ 2 \mu V K^3 ~+~
Z_1 Z_2 V \right] ~=~ 0 \,. \label{regconda}
\end{equation}

Next there is a potential problem with CTC's coming from Dirac
strings in $\omega$. For $\omega$ to have a Dirac string
originating  at $r_1 =0$, the source terms in the equation for
$\vec{\omega}$ must have a piece that behaves as a constant multiple
of $ \vec \nabla \frac{1}{r_1}$. To examine this, it is easier to
use (\ref{GHomegaeqn}) and recall that $Z_3$, $K^1, K^2$ and $V$
are finite as $r_1 \to 0$.  Thus the only sources of ``dangerous
terms'' are $V \vec \nabla  \mu$ and $Z_3 \vec \nabla K^3$. Since
$V$ and $Z_3$ are finite at $r_1 =0$, there will be no Dirac
strings starting at $r_1 =0$ if and only if:
\begin{equation}
\lim_{r_1 \to 0}\,  r_1  \left[ V  \mu ~-~Z_3\,  K^3 \right] ~=~ 0
.\label{regcondb}
\end{equation}%

The two conditions, (\ref{regconda}) and (\ref{regcondb}),
guarantee that the supertube smoothly caps off the spatial
geometry and are the generalization to three-charge three-dipole
backgrounds of the conditions for smooth cap-off in
\cite{Lunin:2002iz}.

One can massage these conditions using (\ref{regcondb}) to
eliminate all the explicit $K^3$ terms in (\ref{regconda}).  The
condition (\ref{regconda}) may then be written as
\begin{equation}
\lim_{r_1 \to 0}\,  r_1^2\, {\cal Q}  ~=~ 0  \,. \label{regcondc}
\end{equation}
where ${\cal Q}$ is the $E_{7}$ invariant that determines the
four-dimensional horizon area \cite{Bena:2004tk, Bena:2005ni}:
\begin{eqnarray} \mathcal{Q}  &\equiv&  Z_1 Z_2 Z_3 V ~-~ \mu^2 \, V^2 \\
&=& -   M^2\,V^2   - \ds\frac{1}{3}\,M\,C_{IJK}{K^I}\,{K^J}\,{K^k}
- M\,V\,{K^I}\,{L_I} - \ds\frac{1}{4}(K^I L_I)^2 \nonumber \\
&&+ \ds\frac{1}{6} V C^{IJK}L_I L_J L_K +\ds\frac{1}{4}
C^{IJK}C_{IMN}L_J L_K K^M K^N \,. \label{cQdefn}
\end{eqnarray}
We will therefore refer to (\ref{regcondc}) as the quartic
constraint. Note that the right-hand side of (\ref{GHomegaeqn}) is
the quadratic $E_7$ invariant, and so we may view (\ref{regcondb})
as the ``quadratic constraint.''  It is, however, convenient to
rewrite this constraint by eliminating $\mu$ from (\ref{regconda})
using (\ref{regcondb}). One then obtains:
\begin{equation}
\lim_{r_1 \to 0}\,  r_1^2 \left[ V Z_1 Z_2  ~-~ Z_3 \, (K^3)^2
\right] ~=~ 0  \,.\label{regcondd}
\end{equation}
We will use (\ref{regcondb}) and (\ref{regcondd}) as the
independent constraints because they are simplest to apply.

In flat space the supertube solution has $V= \ds\frac{1}{r}$,
$K^1= K^2 =0$ and $Z_3 =1$, and equation (\ref{regcondd})
determines the radius of the supertube in terms of its charges,
and (\ref{regcondb})  fixes the parameter $m_1$ of
(\ref{GHharmonicfn}), and thus determines the angular momentum of
the supertube in terms of its radius and charges.

\subsection{Supertube regularity and spectral flow}

As explained in \cite{Bena:2008wt}, one can obtain a solution with
a supertube inside a general three-charge solution by spectrally
flowing a smooth horizonless bubbling solution. Since spectral
flow is implemented by a coordinate change in six dimensions, it
cannot affect the smoothness or the regularity of the solution.
Equivalently, regularity is determined by placing conditions on
quadratic and quartic $E_7$ invariants, and as shown in
\cite{Bena:2008wt}, these are invariant under spectral flow
transformations.

We therefore expect that the equations that determine the
smoothness of supertubes,  (\ref{regconda}), (\ref{regcondb}) and
(\ref{regcondd}), should be related by spectral flow to the
equations that determine the smoothness of a usual bubbling
solution. Indeed,  consider   the spectral flow transformation
(see \cite{Bena:2008wt} for more detail):
\begin{eqnarray}
\widetilde V &=& V +  \gamma \, K^3\,, \qquad \widetilde K^1 = K^1
- \gamma  \,L_2 \,, \qquad \widetilde K^2  = K^2 - \gamma \,L_1
\,, \qquad \widetilde K^3 = K^3  \\ \widetilde L_1 &=& L_1 \,,
\qquad \widetilde L_2 =   L_2    \,, \qquad \widetilde L_3 = L_3 -
2\,\gamma\,M \,, \qquad \widetilde M = M  \,, \label{GHspecflow}
\end{eqnarray}
with
\begin{equation}
\gamma ~=~ - \ds\frac{q_1}{k^3_1} \,. \label{flowparam}
\end{equation}
This transformation maps a GH bubbled solution to a GH bubbled
solution with a supertube at $r_1 =0$.  Under this spectral flow
one also has:
\begin{eqnarray}
\widetilde Z_1 &~=~& \left(\ds\frac{V}{\widetilde V} \right) \,
Z_1 \,, \qquad \widetilde Z_2  ~=~ \left(\ds\frac{V}{\widetilde V}
\right) \, Z_2 \,, \qquad \tilde \mu
~=~\left(\ds\frac{V}{\widetilde V} \right) \,  \left( \mu  ~-~
\gamma \ds\frac{Z_1 Z_2}{\widetilde  V} \right) \,,   \\
\widetilde Z_3  &~=~& \left(\ds\frac{\widetilde V}{V} \right) \,
Z_3 ~+~ \gamma^2 \left(\ds\frac{Z_1 Z_2}{\widetilde  V} \right)
~-~ 2 \, \gamma \mu \,.\label{Fnflow}
\end{eqnarray}

In the usual bubbling solution, regularity requires that the $Z_I$
are finite and $\mu \to 0$ as $r_1 \to 0$.  In the solution with
the supertube one can use this and (\ref{Fnflow}) to verify that:
\begin{eqnarray}
\lim_{r_1 \to 0}\,  r_1  \left[  \widetilde V  \tilde \mu ~-~
\widetilde Z_3\,   \widetilde K^3 \right]  &~=~&  - \gamma\,
\lim_{r_1 \to 0}\,  r_1\, \left(\ds\frac{V Z_1 Z_2}{\widetilde V }
\right) \,  \left(1 +  \gamma \, \ds\frac{K^3}{V} \right) \,, \\
 \lim_{r_1 \to 0}\,  r_1^2  \left[ \widetilde V \widetilde Z_1
\widetilde Z_2  ~-~ \widetilde Z_3  \, (\widetilde K^3)^2 \right]
& ~=~&
 \lim_{r_1 \to 0}\,  r_1^2 \,  \left(\ds\frac{Z_1 Z_2}{\widetilde V } \right) \,
 \left(V^2   - \gamma^2  (K^3)^2 \right)  \,.
\label{stconds}
\end{eqnarray}

Both  of these vanish by virtue of (\ref{flowparam}) and the
finiteness of the $Z_I$ and $\widetilde V$ as $r_1 \to 0$. Hence,
the equations determining the smoothness and regularity of
two-charge supertubes are related by spectral flow to those
determining the smoothness and regularity of usual three charge
bubbling solution.

\section{Supertube probes and mergers in BPS solutions}

We now turn to the description of supertubes in terms of the
Dirac-Born-Infeld (DBI) action. Our purpose is four-fold: to show
that the supertubes that are solutions of the DBI action
back-react into smooth horizonless geometries; to identify the
Born-Infeld charges of supertube with those of the corresponding
solution with a Gibbons-Hawking base;  to facilitate the analysis
of chronology protection in Section 5, and to set the stage for
the entropy enhancement calculation in Section 6.

We begin with a review of supertubes in the background of a three-charge rotating BPS (BMPV) black hole \cite{Bena:2004wt,Marolf:2005cx}, and then extend this
to a black ring, and to  more general three-charge backgrounds.
The first goal is to show that the Born-Infeld calculation
captures the same essential data that is given by the regularity
conditions of the fully back-reacted supergravity solution.  We
will also show that the Born-Infeld analysis and exact supergravity
analysis give the same merger conditions for supertubes with black
rings.

\subsection{Supertubes in a three-charge black hole background}
\label{BMPVprobe}

As a warm up exercise, we first consider a probe supertube with
two charges and one dipole charge in the background of a 
three-charge (BMPV) black hole. This example was considered in \cite{Bena:2004wt,
Marolf:2005cx} and was generalized to a probe supertube with three
charges and two dipole charges in \cite{Finch:2006sb}. The full
supergravity solution describing a BMPV black hole on the symmetry
axis of a black ring with three charges and three dipole charges
was found in \cite{Bena:2004de,Gauntlett:2004qy}, and a more
general solution in which the black hole is not at the center of
the ring was found in  \cite{Bena:2005zy}

First, we need the BMPV black hole solution in the D0-D4-F1
duality frame. The metric (in the string frame) is:
\begin{eqnarray}
ds^2_{10} =  - \ds\frac{1}{\sqrt{Z_1Z_2}Z_3} \, (dt + k)^2 &+&
\sqrt{Z_1Z_2}\, ( d\rho^2 + \rho^2(d\vartheta^2 + \sin^2\vartheta
d\varphi_1^2 + \cos^2\vartheta d\varphi_2^2) ) \notag\\ &+&
\ds\frac{\sqrt{Z_1Z_2}}{Z_3}\, dz^2 ~+~
\ds\sqrt{\ds\frac{Z_1}{Z_2}}\, ds^2_{T^4}
\end{eqnarray}
and the dilaton and the Kalb-Ramond field are  given by:
\begin{equation}
\Phi = \ds\frac{1}{4}\log \left( \ds\frac{Z_1^3}{Z_2Z_3^2}
\right)~, \qquad B = (Z_3^{-1} - 1) dt \wedge dz + Z_3^{-1} k
\wedge dz  \,.
\end{equation}
The non-trivial RR potentials are:
\begin{equation}
C^{(1)} = (Z_1^{-1} - 1) dt + Z_1^{-1}k~, \qquad C^{(3)} = -(Z_2 -
1) \rho^2 \cos^2\vartheta d\varphi_1 \wedge d\varphi_2 \wedge dz +
Z_3^{-1} dt \wedge k \wedge dz \,.
\end{equation}
The one-form $k$ and the functions $Z_I$ are given by
\begin{equation}
k = k_1d\varphi_1 + k_2d\varphi_2 = \ds\frac{J}{\rho^2}
(\sin^2\vartheta d\varphi_1 - \cos^2\vartheta d\varphi_2)~,
\qquad\qquad Z_I = 1 + \ds\frac{Q_I}{\rho^2} \,,
\end{equation}
where  $J$ is the angular momentum of the black hole.  The
charges, $Q_1$, $Q_2$ and $Q_3$ correspond to the respective D0
brane,  D4 brane and F1 string charges of the black hole.

This solution is indeed a BPS, five-dimensional, rotating black
hole \cite{Breckenridge:1996is} with an event horizon at $r=0$,
whose area is proportional to $\sqrt{Q_1Q_2Q_3-J^2}$. For
$J^2>Q_1Q_2Q_3$ the solution has closed time-like curves and is
unphysical.

We  will denote the world-volume coordinates on the supertube by
$\xi^0$,  $\xi^1 $ and $\xi^2\equiv \theta$.  To make the
supertube wrap $z$ we take $\xi^1=z$ and we will fix a gauge in
which  $ \xi^0=t$.  Note that $z\in (0,2\pi L_z)$. The profile of
the tube, parameterized by $\theta$, lies in the four-dimensional
non-compact $\IR^4$ parameterized by $(\rho, \vartheta, \varphi_1,
\varphi_2 )$ and for a generic profile  all four of these
coordinates will depend on $\theta$.

It is convenient to use polar coordinate $(u, \varphi_1)$ and $(v,
\varphi_2)$ in $\IR^4 = \IR^2 \times \IR^2$, where the $\IR^4$
metric takes the form:
\begin{equation}
ds^2_4 = d\rho^2 + \rho^2(d\vartheta^2 + \sin^2\vartheta
d\varphi_1^2 + \cos^2\vartheta d\varphi_2^2) = du^2 + u^2
d\varphi_1^2 + dv^2 + v^2 d\varphi_2^2 \,. \label{Flatmets}
\end{equation}

There is also a gauge field, $ \mathcal{F}$,  on the world-volume
of the supertube. Supersymmetry requires that $\mathcal{F}$
essentially has constant components and we can then boost the
frames so that $\mathcal{F}_{t \theta} =0$.

In this frame supersymmetry also requires $\mathcal{F} _{tz} =1$
\cite{Mateos:2001qs}.  For the present  we take
\begin{equation}
2\pi\alpha' F \equiv \mathcal{F} = \mathcal{F}_{tz} dt\wedge dz +
\mathcal{F}_{z\theta } dz\wedge d\theta \,,
\end{equation}
where the components are constant. Keeping $\mathcal{F}_{tz} $ as
a variable will enable us to extract the charges below.

The supertube action is a sum of the DBI and Wess-Zumino (WZ)actions:
\begin{equation}
S =  - T_{D2} \int d^3\xi e^{-\Phi}\sqrt{ - \text{det}
\left(\widetilde{G}_{ab} + \widetilde{B}_{ab} + \mathcal{F}_{ab}
\right)} + T_{D2} \int d^3\xi [\widetilde{C}^{(3)} +
\widetilde{C}^{(1)}\wedge (\mathcal{F} + \widetilde{B})] \,,
\label{DBIWZ}
\end{equation}
where, as usual, $\widetilde{G}_{ab}$ and $\widetilde{B}_{ab}$ are
the induced metric and Kalb-Ramond field.  We have also chosen the
orientation such that $\epsilon_{tz\theta}=1$. It is also
convenient to define the following induced quantities on the
world-volume:
\begin{equation}
\Delta_{\mu\nu} = \partial_{\mu}u \partial_{\nu}u + u^2
\partial_{\mu}\varphi_1 \partial_{\nu}\varphi_1 + \partial_{\mu}v
\partial_{\nu}v
+ v^2 \partial_{\mu}\varphi_2 \partial_{\nu}\varphi_2~,\qquad
\gamma_{\mu} = k_1 \partial_{\mu} \varphi_1 + k_2 \partial_{\mu}
\varphi_2 \,,
\end{equation}
where $\partial_\mu \equiv {\partial \over \partial \xi^\mu}$.

After some algebra, the  DBI part of the action simplifies to:
\begin{multline}
S_{DBI} = -T_{D2}\ds\int dt dz d\theta \left\{ \ds\frac{1}{Z_1^2}
\left( \mathcal{F}_{z\theta} -\gamma_{\theta} ( \mathcal{F}_{tz}
-1) \right)^2 + \ds\frac{Z_2}{Z_1} \Delta_{\theta\theta} [ 2 (1 -
\mathcal{F}_{t z}) - Z_3 (\mathcal{F}_{t z} - 1)^2 ]\right\}^{1/2}
\,,
\end{multline}
while the WZ piece of the action takes the form
\begin{equation}
S_{WZ} = T_{D2} \ds\int dt dz d\theta  \left[
(1-\mathcal{F}_{tz})\ds\frac{\gamma_{\theta}}{Z_1} +
\mathcal{F}_{z\theta} \left(\ds\frac{1}{Z_1} - 1\right) \right]
\,.
\end{equation}
For a supersymmetric configuration ($\mathcal{F}_{tz} = 1$) we
have
\begin{equation}
S_{\mathcal{F}_{tz}=1} = S_{DBI} + S_{WZ} = -T_{D2} \int dt dz
d\theta \, \mathcal{F}_{z\theta} \label{SimpAction}
\end{equation}

The foregoing supertube carries D0 and F1 ``electric" charges,
given by
\begin{equation}
N_{1}^{ST} = \ds\frac{T_{D2}}{T_{D0}} \ds\int dz d\theta ~
\mathcal{F}_{z\theta}, \qquad\qquad N_{3}^{ST} =
\ds\frac{1}{T_{F1}} \ds\int d\theta  ~ \ds\frac{\partial
\mathcal{L}}{\partial
\mathcal{F}_{tz}}\bigg|_{\mathcal{F}_{tz}=1}  \,.
\label{STcharges}
\end{equation}
The Hamiltonian density is:
\begin{equation}
\mathcal{H}|_{\mathcal{F}_{tz}=1} =
\left[\ds\frac{\partial{\mathcal{L}}}{\partial\mathcal{F}_{tz}}
\mathcal{F}_{tz}-\mathcal{L}\right]_{\mathcal{F}_{tz}=1} =~~
T_{D2}\mathcal{F}_{z\theta} + \ds\frac{\partial
\mathcal{L}}{\partial \mathcal{F}_{tz}}\bigg|_{\mathcal{F}_{tz}=1}
\,. \label{HamDensity}
\end{equation}
One can easily integrate this to get the total Hamiltonian of the
supertube\footnote{See Appendix D for details about our units and
conventions.} (we assume constant charge density
$\mathcal{F}_{z\theta}$)
\begin{equation}
\int d z d \theta ~ \mathcal{H}|_{\mathcal{F}_{tz}=1} =
N_1^{ST} + N_3^{ST}\,.
\end{equation}
Thus the energy of the supertube is the sum of its conserved
charges which shows that the supertube is indeed a BPS object.

Now choose a static \textit{round supertube} profile
$u'=v'=\varphi_2'=0$, $\varphi_1 = \theta$. One then has:
\begin{equation}
\gamma_{\theta} = k_1 = J\,  \ds\frac{u^2}{(u^2+v^2)^2}~,
\qquad\qquad \Delta_{\theta\theta} = u^2
\end{equation}
and the supertube ``electric" charges are:
\begin{equation}
N_{1}^{ST} = n_2^{ST}\mathcal{F}_{z\theta}~, \qquad\qquad
N_{3}^{ST} = n_2^{ST}\ds\frac{Z_2 u^2}{\mathcal{F}_{z\theta }} \,.
\end{equation}
So we find
\begin{equation}
N_{1}^{ST} N_{3}^{ST} = (n_2^{ST})^2u^2\,Z_2  \,.
\label{STBHradius}
\end{equation}
This is an important relation in that it fixes the location of the
supertube in terms of its intrinsic charges.

This computation was used in  \cite{Bena:2004wt} to study the
merger of a supertube and a  black hole.  In particular, a
supertube can merge with a black hole if and  only if $N_{1}^{ST}
N_{3}^{ST}\leq (n_2^{ST})^2 N_2$,  where $N_2$ is the number of D4
branes in the black hole.  Moreover,  the supertube will ``crown''
the black hole at ``latitude'', $\vartheta= \alpha$, given by:
\begin{equation}
\sin\alpha = \ds\sqrt{\frac{N_{1}^{ST}
N_{3}^{ST}}{(n_2^{ST})^2N_2}} \,. \label{mergeangleBH}
\end{equation}
One can also show that one cannot violate chronology protection by
throwing a supertube into the black hole, that is,  one cannot
over-spin the black hole and that the bound $J^2\leq N_1N_2N_3$ is
preserved after the merger.

\subsection{Supertubes in a black-ring background}
\label{BRprobe}

We now repeat the foregoing analysis in the background of a
supersymmetric black ring where there will be new physical effects
due to the interaction between the dipole charges of the black
ring and the dipole charge of the supertube.   We will also
examine the symmetric merger of the supertube with the black ring
and show that chronology protection is not violated. In Section
\ref{threechgST} we will perform a more general analysis by
considering a probe supertube that has three charges and two
dipole charges.

\subsubsection{The black-ring solution}

The three-charge, three-dipole charge black ring solution
\cite{Bena:2004de,Elvang:2004ds,Elvang:2004rt,Bena:2004wv,Gauntlett:2004qy}
in a IIA duality frame where the ring has D0, D4 and F1 electric
charges and D6, D2 and NS5 dipole charges is given by:
\begin{eqnarray}
ds^2 &=& -(Z_2Z_1)^{-1/2}Z_3^{-1}(dt+k)^2+
(Z_2Z_1)^{1/2}ds^2_{\mathbb{R}^4}+(Z_2Z_1)^{1/2}Z_3^{-1}dz^2+
Z_2^{-1/2}Z_1^{1/2}ds^2_{T^4}, \notag\\
e^{2\Phi} &=& Z_2^{-1/2}Z_1^{3/2}Z_3^{-1}, \\
B &=& (Z_3^{-1}-1)dt \wedge dz+Z_3^{-1}k \wedge dz-B^{(3)} \wedge
dz~,\notag
\end{eqnarray}
for the NS-NS fields, and
\begin{eqnarray}
C^{(1)} &=& (Z_1^{-1}-1)dt +Z_1^{-1}k - B^{(1)}, \\
C^{(3)} &=& Z_3^{-1}dt \wedge k\wedge dz -Z_3^{-1}(dt+k) \wedge
B^{(1)} \wedge dz + B^{(3)} \wedge dt \wedge dz - \gamma_1\wedge
dz ~,
\end{eqnarray}
for the R-R fields.   The one-forms, $B^{(I)}$, are the potentials
defined in section \ref{MtheoryBPS}  with $dB^{(I)} =
\Theta^{(I)}$.  These fields are the magnetic sources of the ring.
The two-form,  $\gamma_1$,  must satisfy:
\begin{equation}
d\gamma_1=\star_4dZ_2 - B^{(1)}\wedge \Theta^{(3)}\,.
\label{gammaeqn}
\end{equation}

We use the canonical coordinates that are adapted to the symmetries
of the black ring in the flat metric of the $\mathbb{R}^4$ base
\cite{Elvang:2004rt}:
\begin{eqnarray}
ds^2_{\mathbb{R}^4}=
g_{\mu\nu}dy^{\mu}dy^{\nu}=\frac{R^2}{(x-y)^2} \left(
\frac{dy^2}{y^2-1} + (y^2-1) d\varphi_1^2 + \frac{dx^2}{1-x^2} +
(1-x^2)d\varphi_2^2 \right).
\end{eqnarray}
We will also use the orientation:
$\epsilon_{yx\varphi_1\varphi_2}=1$. In these coordinates, the
black ring horizon is located at  $y\rightarrow -\infty$.  It is
useful to recall that the change of coordinates:
\begin{equation}
x =  -{ u^2+v^2 - R^2 \over  \sqrt{((u-R)^2 + v^2)( (u+R)^2 + v^2
)}} \,, \qquad y =  -{ u^2+v^2 + R^2 \over  \sqrt{((u-R)^2 + v^2)(
(u+R)^2 + v^2 )}} \label{uvcoords}
\end{equation}
takes one back to the standard flat metric on $\IR^2 \times \IR^2$
(\ref{Flatmets}) parameterized by  $(u, \varphi_1)$ and $(v,
\varphi_2)$ with the ring horizon at  $u=R, v=0$.

The warp factors $Z_I$ are
\begin{equation}
Z_I=1+\frac{\ov{Q}_I}{2R^2}(x-y)-\frac{C_{IJK}}{2}\frac{q^Jq^K}{4R^2}(x^2-y^2),
\label{constituent}
\end{equation}
where $\ov{Q}_I$ are what we refer to as ``constituent charges''
of the black ring, and differ from the charges measured at
infinity. The angular momentum vector is given by
\begin{eqnarray}
k&=&k_1d\varphi_1+k_2d\varphi_2 \nonumber \\
&=&-\left((y^2-1)\left(C(x+y)+D\right) -A(y+1)\right)d\varphi_1 -
\left((x^2-1)\left(C(x+y)+D\right)\right)d\varphi_2
\end{eqnarray}
with $A=(q^1+q^2+q^3)/2$,
$D=(q^1\overline{Q}_1+q^2\ov{Q}_2+q^3\ov{Q}_3)/8R^2$ and
$C=-q^1q^2q^3/8R^2$.
The vector fields, $B^{(I)}$, are given by
\begin{equation}\label{Bi}
B^{(I)}=\frac{q^I}{2}\left((y+d)d\varphi_1 - (x+c)d\varphi_2
\right) \,.
\end{equation}
The constants $c$ and $d$ are locally pure gauge and are not fixed
by the equations of motion.  Indeed, because the ring carries a
magnetic current there will Dirac strings in any attempt at a
global definition of $B^{(I)}$.   In the $(u,v, \varphi_1,
\varphi_2)$ coordinate patch, defined by (\ref{uvcoords}), the
vector fields, $B^{(I)}$, are potentially singular at either
$u=0$, or $v=0$. To remove these singularities we must have $(y+d)
= 0$ at $u=0$ and $(x+c) = 0$ at $v=0$.  From  (\ref{uvcoords}) we
see that this unambiguously requires $d=+1$ but that one has $x =
+ 1$ for $v=0, u<R$ and $x = -1$ for $v=0, u> R$ and so to remove
the Dirac strings we must take:
\begin{equation}
d = +1\,, \ c= -1 \quad {\rm inside \ the\ ring}\,;  \qquad d
=+1\,,  \ c= + 1 \quad{\rm outside \ the\ ring}\,. \label{cdres}
\end{equation}
The coordinates $(x, \varphi_2)$ in fact define a Gaussian
two-sphere around the ring and the choices (\ref{cdres}) represent
the familiar gauge field patches surrounding a magnetic monopole.
In the following we will set $d=1$ and retain $c$ with the
understanding that it is to be chosen as in (\ref{cdres}).

The two-form $\gamma_1$ in $C^{(3)}$ has the form $\gamma_1
=f(x,y)d\varphi_1 \wedge d\varphi_2$ where
\begin{equation}\label{fBR}
f(x,y)=-\frac{\ov{Q}_2}{2}\frac{1-xy}{x-y}
+\frac{q_1q_3}{4}\left[\frac{(1-xy)(x+y)}{x-y}+cy-dx\right]+f_0.
\end{equation}
where $f_0$ is another integration constant.  It is  shown in
Appendix A that $\gamma_1$ satisfies (\ref{gammaeqn}).

We want to stress that our conventions are such that
\begin{equation}
\ov{Q}_{I}~=~\ov{N}_I  \qquad \text{and} \qquad q_I=n_I
\label{QvsN}
\end{equation}
where $\ov{N}_I$ and $n_I$ are integers and specify the number of
``electric" and ``dipole" D-branes comprising the black ring. It
is also useful to note that the angular momentum of the black ring
is related to its dipole charges by
\begin{equation}
J = 4 (q_1+q_2+q_3) R  \,.
\end{equation}
%

\subsubsection{The black ring as a solution with a Gibbons-Hawking base.}
\label{GHBRsolution}

Since Gibbons-Hawking (GH) geometries play an important role in
bubbled solutions, and in our discussion here, it is useful to
re-write the foregoing solution in terms of these geometries. The
change of variables between the ordinary flat $\mathbb{R}^4$
coordinates ($u,\varphi_1,v,\varphi_2$) and the GH coordinates
$(\psi,r,\chi,\phi)$:
\begin{equation}\label{BR-GHcoord}
r~=~\frac{1}{4}(u^2+v^2)~, \qquad \chi~=~2\arctan{\frac{u}{v}}~,
\qquad  \psi~=~2\, \varphi_1~, \qquad    \phi~=~-(\varphi_2 +
\varphi_1) \, ,
\end{equation}
and recall that $u$ and $v$ are related to $x$ and $y$ by
(\ref{uvcoords}). The metric in the new coordinates is:
\begin{equation}
ds^2_{\mathbb{R}^4} = r(d\psi+(\cos\chi+1)d\phi)^2 +
\ds\frac{1}{r} (dr^2+r^2d\chi^2+r^2\sin^2\chi d\phi^2)
\end{equation}
The black ring solution is written in terms of eight harmonic
functions $V$, $L_I$, $K^I$ and $M$ \cite{
Gauntlett:2004wh,Gauntlett:2004qy,Elvang:2005sa,Gaiotto:2005xt,Bena:2005ni}. However, as we noted in the
last subsection, the black ring has a monopolar magnetic field and
so we need two patches that are related by a gauge transformation.
Remembering that the vector potentials in solutions with a GH base
are given by
\be B^{(I)}=V^{-1}K^I(d\psi +A) + \xi^I\,, \ee
one can easily identify the $K^I$ that give these fields, and
observe that changing the patch from $c\!=\!-1$ to $c\!=\!+1$
corresponds, in the GH solution, to the gauge transformation:
\bea \label{gaugetsfo} K^I &\rightarrow& K^I+ c^I V~, \qquad\qquad
L_I\rightarrow
L_I-C_{IJK}c^JK^K-\frac{1}{2}C_{IJK}c^Jc^KV, \\
M &\rightarrow&  M - {1 \over 2} \,c^I \, L_I +{1 \over 12}\,
C_{IJK}\left( V \, c^I \, c^J \, c^K +3 \,  c^I\, c^J\, K^K\right)
\,, \nonumber \eea
with $c^I= q^I/2$. Thus, we can now completely specify the eight
harmonic functions, once we choose a patch. For $c=-1$, we have
\bea V &=& {1 \over r} ~,~~~
K^I = - {q_I \over 2|\vec r - \vec r_{BR}|} ~, \nonumber\\
L_I &=& 1 + {\ov{Q}_I \over 4|\vec r - \vec r_{BR}|} ~,~~~  M = -
{J \over 16|\vec r - \vec r_{BR}|} + {J \over 16 R}\,,
{\label{BRGHsol}} \eea
and for $c=+1$ they become
\bea V &=& {1 \over r} ~, \qquad K^I = - {q_I \over 2|\vec r - \vec r_{BR}|} + {q_I \over 2r}~, \notag\\
L_I &=& 1 + {\ov{Q}_I + C_{IJK}q^Jq^K \over 4|\vec r - \vec
r_{BR}|} - {C_{IJK} q^J q^K \over 8r}~, \qquad M = - {J +
q^I\ov{Q}_I + 3 q^1q^2q^3 \over 16|\vec r - \vec r_{BR}|} -
{q^1q^2q^3 \over 16r} ~.\notag {\label{BRGHsol2}} \eea
As
noted earlier, these formulae  {\it define} the GH charges of the
black ring and these, in turn, define the electric charges of the
four-dimensional black hole corresponding to the ring.  The
electric GH charges $Q_I^{GH}$ are four times the coefficients of
the pole at the location of the ring in the $L_I$ functions, the
GH dipole charges $q_I^{GH}$ are minus two times the coefficients
of the pole in the $K^I$ functions, and the GH angular momentum
$J^{GH}$ is minus sixteen times the coefficient of the pole in $M$
(we use the conventions of \cite{Bena:2005ni}). Thus, we have:
\bea Q_I^{GH}&=&\ov{Q}_I \,, \qquad\qquad q_I^{GH}=q_I\,,
\qquad\qquad J^{GH}=J {\label{BRGHcharges}} \eea
for $c=-1$ and
\bea Q_I^{GH}&=&\ov{Q}_I +C_{IJK}q^Jq^K\,, \qquad q_I^{GH}=q_I\,
\qquad  J^{GH}=J + q^I\ov{Q}_I + 3q^1q^2q^3 {\label{BRGHcharges2}}
\eea
for $c=+1$.

The dipole charges are patch-independent, but the GH electric
charges and the GH angular momentum are {\it gauge dependent}
notions, and are different in different patches. This will be
important in the following discussion.

\subsubsection{Probing the black ring with two-charge supertubes}
\label{probeBR}

We now probe the black ring background with a two-charge supertube
\cite{Mateos:2001qs,Emparan:2001ux}. The calculation proceeds in
much the same way as for the supertube in a black hole background.
As before, we parameterize  the tube by ($t,z,\theta$), and  define
an {\it a priori} arbitrary supertube profile  in $\IR^4$ by
$\vec{y}(\theta)$.    Since we are ultimately going to consider a
supertube that winds multiple times around the ring direction it
will be convenient to take  $\theta \in (0,2\pi n_2^{ST})$  where
$n_2^{ST}$ will become this winding number. Thus the supertube will
have a dipole charge proportional to $n_2^{ST}$, and two net charges
proportional to $N_{1}^{ST}$ and $N_{3}^{ST}$. Its action is a sum
of a DBI and a  WZ  term
\begin{multline}
S=S_{DBI}+S_{WZ} = -T_{D2}\int dt dz d\theta
e^{-\Phi}\sqrt{-\det(\widetilde{G}_{ab}+\widetilde{B}_{ab}+\mathcal{F}_{ab})}
\\+ T_{D2}\int dt dz d\theta \left( \widetilde{C}^{(3)}_{tz\theta}
+ \widetilde{C}^{(1)}_{t} (\widetilde{B}_{z\theta} +
\mathcal{F}_{z\theta}) + \widetilde{C}^{(1)}_{\theta}
(\widetilde{B}_{tz} + \mathcal{F}_{tz}) \right)
\label{DBIWZaction}
\end{multline}

For the supersymmetric configuration one once again finds that
$\mathcal{F}_{tz}=1$ and if one imposes this {\it ab initio} then
one again obtains (\ref{SimpAction}),  (\ref{STcharges}) and
(\ref{HamDensity}) and hence the BPS relation for the supertube.
The expression for the derivative of the action  with respect to
$\mathcal{F}_{tz}$ evaluated at  $\mathcal{F}_{tz}=1$ can be most
convenient expressed as:
\begin{equation}
\left(\frac{\partial\mathcal{L}}{\partial
\mathcal{F}_{tz}}\bigg|_{\mathcal{F}_{tz}=1}+
T_{D2}(B^{(1)}_{\varphi_1} \varphi_1'+B^{(1)}_{\varphi_2}
\varphi_2' )\right) \left(\mathcal{F}_{z\theta}+
(B^{(3)}_{\varphi_1} \varphi_1'+B^{(3)}_{\varphi_2} \varphi_2'
)\right) =T_{D2}Z_2g_{\mu\nu}y'^{\mu}y'^{\nu} \,,
 \label{radiuscondBR}
\end{equation}
where $'$ denotes the derivative with respect to $\theta$. As for
the  black hole \cite{Bena:2004wt, Marolf:2005cx}, one can
reinterpret this in terms of charge densities and arrive at a
generalization of the constraint  (\ref{STBHradius}) that relates
the charges to the radius of the supertube. Note that the
condition (\ref{radiuscondBR}) is local and to get a relation
similar to (\ref{STBHradius}) on has to integrate over the profile
of the supertube. There is an important new feature here in that
there is a contribution from the interactions of the dipole
charges of the supertube and background.  This appears through the
pull-back of the $B^{(I)}$ to the world-volume of the supertube
and it gives an added contribution to the basic supertube charges
to yield what we will refer to as the local {\it effective
charges} of the supertube. We will show in
 section \ref{threechgST} that this also happens when supertubes are placed in three-charge
 solutions with a GH base.

It is also important to remember that the Wess-Zumino action of
the supertube is only invariant under local small gauge
transformations, but is not necessarily invariant under {\it large
gauge transformations.}  Indeed, the black ring is a magnetic
object, and as such the gauge fields, $B^{(I)}$ are not defined
globally  but on patches. Their values, and the value of the
supertube action, differ from patch to patch by what can be
thought of as the effect of a large gauge transformation.

More explicitly, the action depends on the  Wilson lines of these
gauge fields taken around latitudes of the two-sphere that
surrounds the black ring (which is the equivalent of the sphere
that contains a monopole charge). The value of these Wilson loops
may then be defined
 using Stokes theorem as the integral of the magnetic flux coming from the black ring through the
 section of the sphere surrounded by the Wilson line. There is,
 however, an obvious ambiguity: does one integrate the flux over the
 upper or the lower cap of the sphere?  The
 difference is, of course, the monopole charge inside the sphere
 multiplied by the number of times the Wilson loop winds around the
 latitude circle.  These ambiguities will manifest themselves in the
definitions of the constituent charges of the supertube.

To analyze the physics of the merger, we consider a supertube
embedded in spacetime along the curve $\vec{y}(\theta)$ given by:
\begin{equation}
\varphi_1=-\theta~, \qquad\qquad   \varphi_2=-\nu \,\theta\,
\label{winding}
\end{equation}
$x$ and $y$ being at fixed values. The projections of the
supertube in the $(y,\varphi_1)$ and $(x,\varphi_2)$ planes are
circular, with winding numbers $n_2^{ST}$ and $\nu n_2^{ST}$
respectively.    For $\nu=0$, the supertube is circular and simply
winds around the plane of the ring $n_2^{ST}$ times. For $\nu
\!\neq\!0$, the  details of the winding depend upon the
equilibrium position of the supertube. We also assume, for
simplicity, that the charge densities of the tube are independent
of $\theta$. Under these assumptions the condition
(\ref{radiuscondBR}) becomes:
\begin{multline}
\left[ N_1^{ST} -\ds\frac{1}{2} n_2^{ST} n_3 (y+    1 - \nu (x
+c))
\right] \left[ N_3^{ST} -\ds\frac{1}{2} n_2^{ST} n_1 (y+ 1 -\nu (x +c)) \right] = \\
(n_2^{ST})^2 Z_2 \ds\frac{R^2}{(x-y)^2} ( (y^2-1) + \nu^2 (1-x^2)
) \,. \label{radiuscondBRsimp}
\end{multline}
We will call this equation the radius relation. Note that this
equation is invariant under the exchange of $N_1, n_1$ with
$N_3,n_3$, as expected by U-duality. Comparing this constraint to
the one for a black hole background (\ref{STBHradius}), we see
that the charges of the supertube are enhanced to their {\it
effective charges} via the interactions of the dipole charges.
This is an important result that we will discuss further in the
subsequent sections.

To get a better idea of  the supertube configuration in the
black-ring geometry it is instructive to examine the supertube as
it approaches the  horizon ($y \to - \infty$). In this limit, the
physical metric along the horizon becomes:
\begin{equation}
ds_3^2 ~=~  \big(C^2  R^4  \big)^{1 /3} \, \bigg[\, \big(64\,C^2
R^4  \big)^{-1}  \,{\cal M} \, d \varphi_1^2 ~+~ \,( d \alpha ^2
~+~\sin^2 \alpha \, (d \varphi_1 + d \varphi_2)^2) \,\bigg]  \,,
\label{horizonmet}
\end{equation}
where we have set $x= \cos\alpha$, and the parameter, ${\cal M}$,
is proportional to the square of the black-ring entropy \be
S=\pi\sqrt{\mathcal{M}}, \ee
and is given by
\begin{equation}
\mathcal{M} = 2 n_1n_2\ov{N}_1\ov{N}_2 + 2 n_1n_3\ov{N}_1\ov{N}_3
+ 2 n_2n_3\ov{N}_2\ov{N}_3 - (n_1\ov{N}_1)^2 - (n_2\ov{N}_2)^2
-(n_3\ov{N}_3)^2 - 4 n_1n_2n_3 J \,, \label{BRent}
\end{equation}
where $J$ is the ``intrinsic'' angular momentum of the ring, and
is given by the difference between the two angular momenta of the
five-dimensional solution:
\begin{equation}
J = J_1-J_2 = 4(n_1+n_2+n_3)\, R \,. \label{ringangmom}
\end{equation}

The topology of the horizon is $S^2 \times S^1$, but observe that
for a supertube that winds according to (\ref{winding}), the
winding around the horizon is determined by
\begin{equation}
\varphi_1~=~ -\theta~, \qquad\qquad  \varphi_1+  \varphi_2 ~=~
-(\nu+1)  \,\theta\,. \label{horizonwinding}
\end{equation}
The supertube thus enters the horizon by winding around the $S^1$
but enters at a point on the $S^2$ if and only if $\nu =-1$.
Otherwise it winds around the $S^1$ and ``crowns'' the $S^2$ by
winding $(\nu+1)$ times around a latitude determined by $x$.

If we now examine the constraint (\ref{radiuscondBRsimp}) and send
$y\rightarrow -\infty$ the supertube will merge with the black
ring and the constraint (\ref{radiuscondBRsimp}) will become the
merger condition:
\begin{equation}
N_1^{ST} n_{1} + N_3^{ST}n_3 - \ov{N}_2 n_2^{ST} ~=~  n_2^{ST}
n_1n_3 ((1+ c)  - (\nu +1)(x +c))  \,.
 \label{crudemerger}
\end{equation}
More explicitly, this condition be written as:
\begin{eqnarray}
N_1^{ST} n_{1} + N_3^{ST}n_3 - \ov{N}_2 n_2^{ST}
&=& n_2^{ST} n_1n_3\, (\nu +1)  (1- x)  \quad {\rm for} \quad c=-1 \,. \label{merg1} \\
N_1^{ST} n_{1} + N_3^{ST}n_3 - \ov{N}_2 n_2^{ST} &=& n_2^{ST}
n_1n_3\, (2-(\nu +1)  (1+ x))  \quad {\rm for} \quad c=+1 \,.
\label{merg2}
\end{eqnarray}

\begin{figure}[t]
 \centering
    \includegraphics[width=5cm]{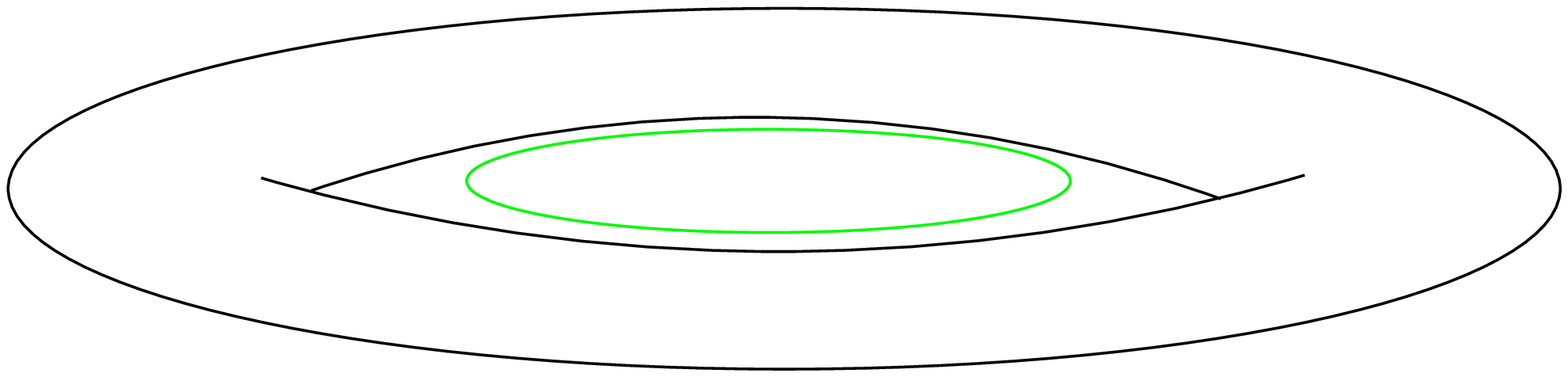}
  \hfill
    \includegraphics[width=5cm]{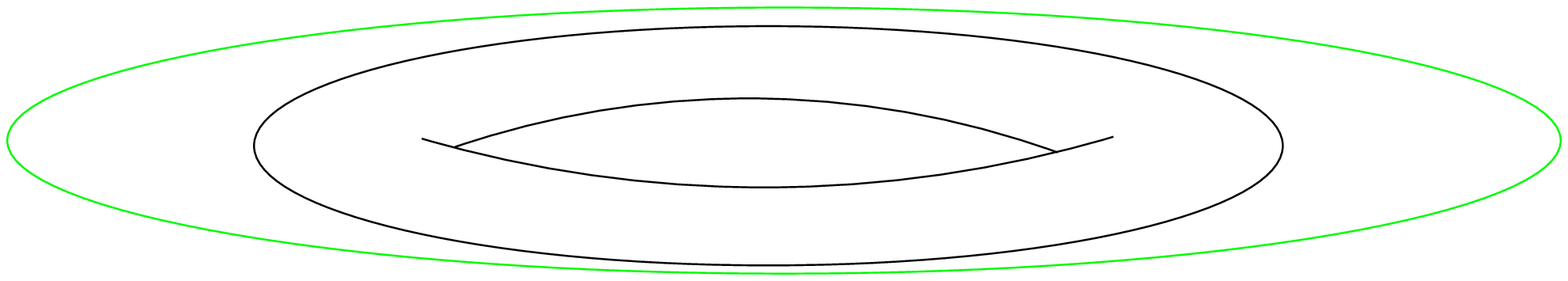}
    \hfill
    \includegraphics[width=7cm]{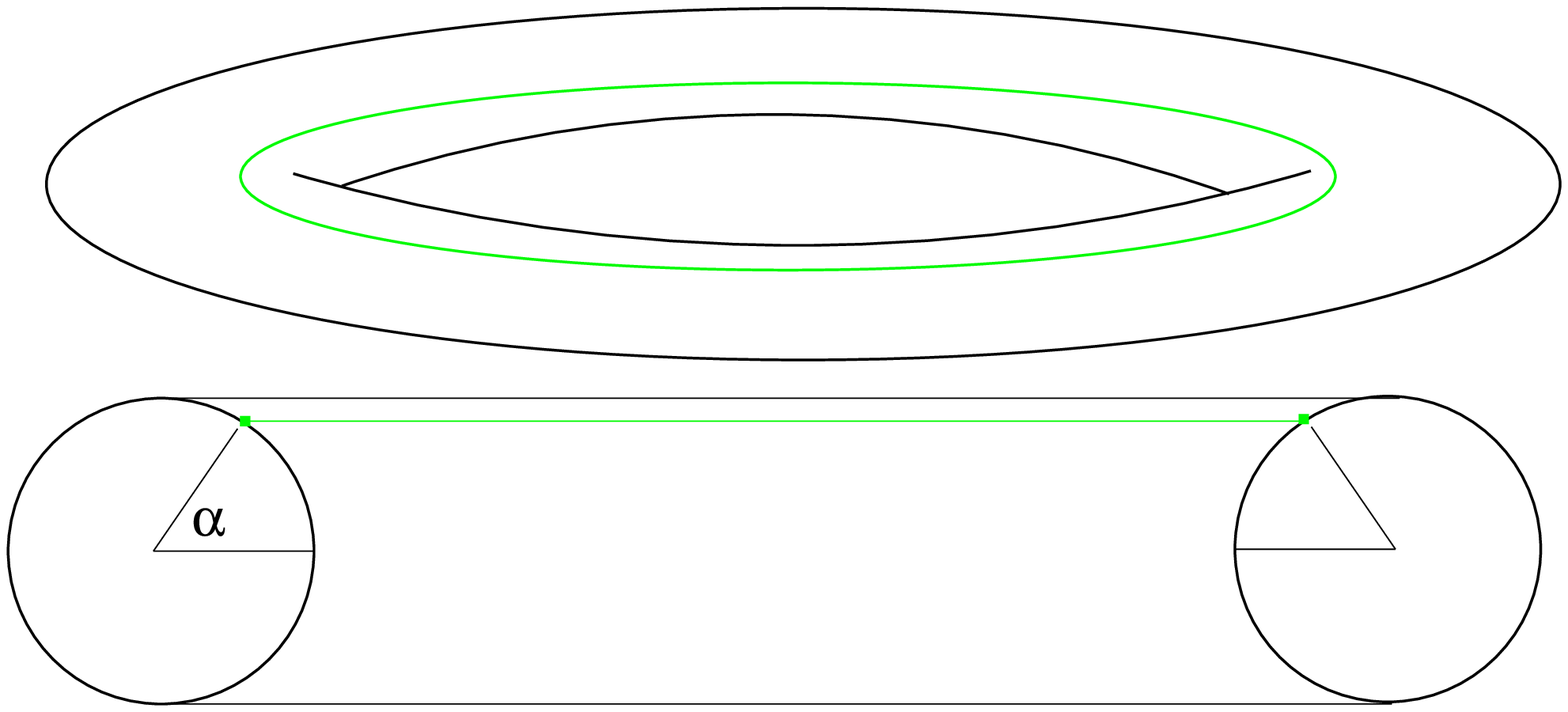}
    \caption{{\it Different black ring and supertube configurations for different values of the supertube charges. In the first picture, the charges of the tube are too small, and hence the tube it is too small, and passes inside the ring. In the second one, the tube is too large and passes on the outside of the ring. In the third picture, the size of the tube is in the correct range for the merger to be possible. The angle $\alpha$ of the merger depends on the tube charges according to (\ref{crudemerger}).}}
\label{fig2}
\end{figure}

The  relation (\ref{crudemerger}) is simply the analogue of the
equation giving the merging angle for the supertube in a
black-hole background (\ref{mergeangleBH}).  In particular, as
depicted in Figure \ref{fig2}, it determines the value of $x$
(which corresponds to an angular variable on the horizon) at which
a supertube with a given set of charges enters the black ring
horizon. Since $-1 \le x \le +1$, this restricts the permissible
charges of supertubes that merge with a given black ring.

\begin{figure}[t]
 \centering
    \includegraphics[width=8.5cm]{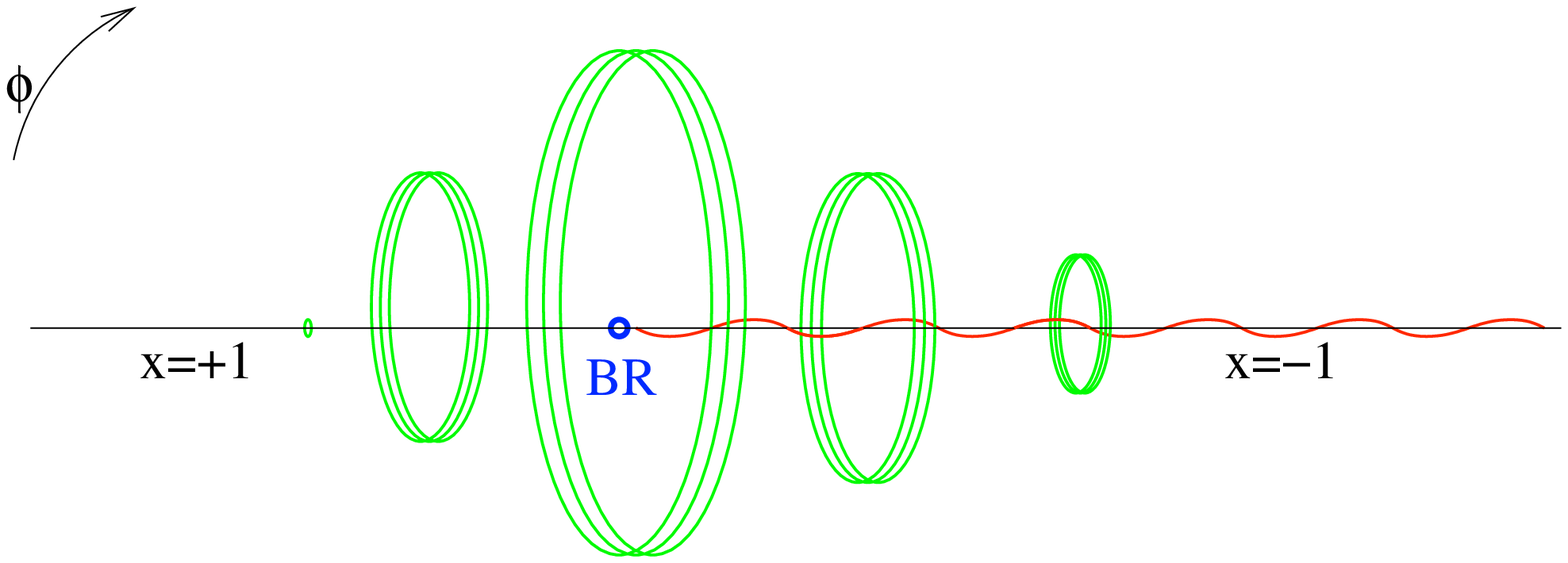}
    \hfill
    \includegraphics[width=8.5cm]{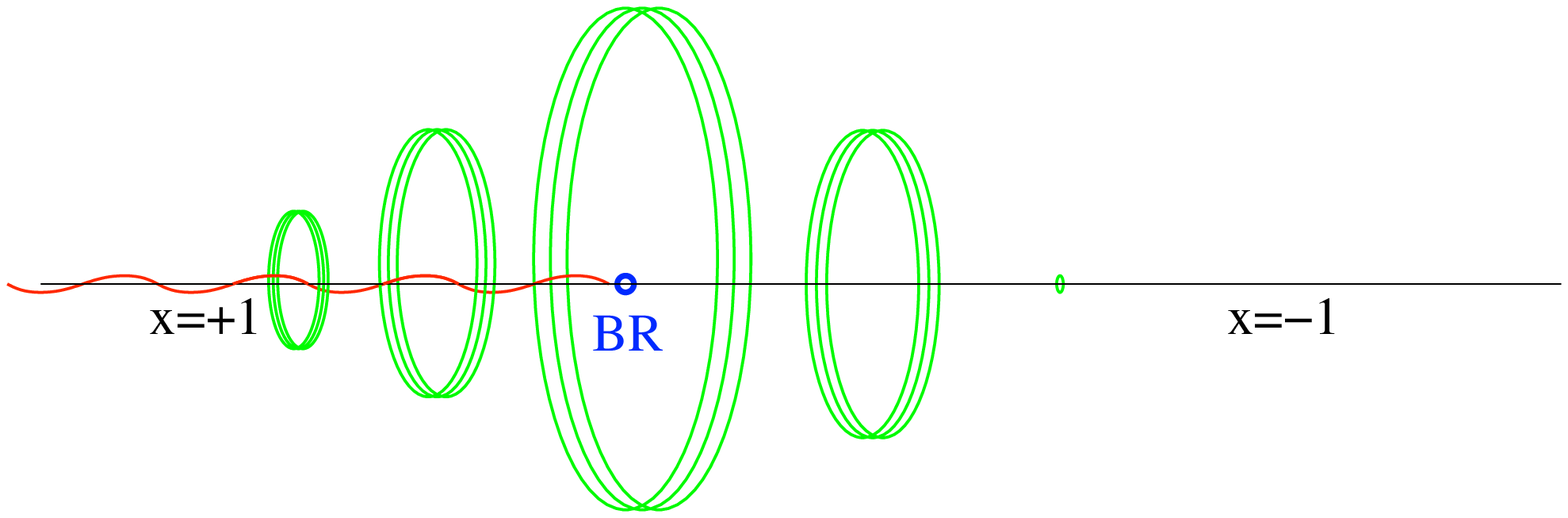}
    \caption{{\it The black ring (in blue) with supertubes (in green) at various positions in the $\mathbb{R}^3$ base of the Gibbons-Hawking space. The black ring is point-like but the tube is point-like only if it lies on the axis $x=\pm1$. Otherwise, it winds $\nu+1$ times the $\phi$ circle. On the left, the Dirac string starts from the ring and extends to infinity. On the right, the Dirac string extends between the center of the space and the ring location.}}
\label{fig1}
\end{figure}

We can see that the radius relation (\ref{radiuscondBRsimp}) and
 the merger condition (\ref{crudemerger}) depend both on the
gauge choice (by an $x$-independent factor) and also on $\nu + 1$.
We can understand this gauge dependance in a physical way: the
gauge choice corresponds to a choice for the location of the Dirac
string. In other words, the gauge dependance comes from the fact
that the tube feels the presence of the Dirac string of the
background. Increasing $x$ then corresponds to the supertube
wrapping, for $c=-1$, or not wrapping, for $c=+1$ the Dirac
string, as can be seen in figure \ref{fig1}.

More precisely, if we choose $c=-1$, that is if we choose the
Dirac string to extend from the ring location to infinity, then we
can put the tube everywhere except on the Dirac string. If we put
it at $x=1$, the $\phi$ circle becomes degenerate and indeed in
(\ref{merg1}) the $\nu$ dependance disappears. This is
expected, because $\nu + 1$ is the winding number of the tube
around a contractible circle. When the size of this circle is
zero, the winding should be irrelevant, which is indeed what
happens.

If we now change the location of the ring to approach $x=-1$
without changing the gauge, the tube winds $\nu+1$ times around
the Dirac string; this winding is physically-relevant, and hence,
as expected, equation (\ref{merg1}) depends on  $\nu$  when $x
\rightarrow 1$. However, if we change the gauge to move the Dirac
string to the inside of the ring, we can see that when the tube is
at $x=1$, where the $\phi$ circle is degenerate, the winding
number is again  irrelevant; as expected the merger formula is
again independent of $\nu$. We should also note that for the
particular value $\nu=-1$, the supertube never wraps the Dirac
string, and hence the merger condition does not depend upon $x$.

In Section \ref{chrono} we will examine the details of such a
merger and discuss chronology protection and black hole
thermodynamics during mergers.

\subsection{The black ring background: comparing the DBI analysis with supergravity.}
\label{comparison1}

We now turn to the main purpose in this section: the relation
between the merger conditions obtained from supergravity and from
the DBI analysis, and the relation between the GH and the DBI
charges of the supertube.

Let begin with the supergravity side. The supergravity solution corresponding to one black ring and one supertube is given as usual the eight harmonic functions $V$, $L_I$, $K^I$ and $M$. The poles of this functions at the location of the ring and of the tube are

\bea
K_1 &=& - {q_1 \over 2|\vec r - \vec r_{BR}|} ~,~~~
K_2 = - {q_2 \over 2|\vec r - \vec r_{BR}|} - {q_2^{ST} \over 2|\vec r - \vec r_{ST}|} ~,~~~
K_3 = - {q_3 \over 2|\vec r - \vec r_{BR}|}~, \nonumber\\
L_1 &=& {Q_1^{GH} \over 4|\vec r - \vec r_{BR}|} + {Q_1^{GH,ST} \over 4|\vec r - \vec r_{ST}|} ~,~~~
L_2 = {Q_2^{GH} \over 4|\vec r - \vec r_{BR}|} ~,~~~
L_3 = {Q_3^{GH} \over 4|\vec r - \vec r_{BR}|} + {Q_3^{GH,ST} \over 4|\vec r - \vec r_{ST}|}  ~,~~~ \nonumber\\
2 M &=& - {J^{GH} \over 8|\vec r - \vec r_{BR}|} - {J^{GH,ST} \over 8 |\vec r - \vec r_{ST}|}
{\label{BRSTGHsol}}
\eea
where $Q^{GH}$ are the GH charges of the black ring defined in Section
\ref{GHBRsolution}, and $Q^{GH,ST}$ are the GH charges of the supertube defined in the same way. Recall once again that the GH charges
depend upon the choice of patch, as in  (\ref{BRGHcharges}) and
(\ref{BRGHcharges2}),  and the GH charges of both the ring and
the tube  transform consistently between the patches.

To obtain the merger condition from supergravity observe that the
bubble (or integrability) equations (\ref{GHbubbleeqns}) contain a
term in which the $E_{7(7)}$ symplectic product of the supertube
and black ring GH charge vectors is divided by their separation.
Hence, these objects only merge if this symplectic product is
zero\footnote{One could also imagine in principle the existence of
a scaling solution, where the distances in $\IR^3$ between the
ring, supertube and the center of Taub-NUT go together to zero. In
such a solution the ring and the supertube would be spinning very
rapidly in opposite directions, which is likely to introduce
closed timelike curves. We leave its exploration for future
work.}. Explicitly, this gives\footnote{As noted in (\ref{QvsN}),
we have adopted a set of conventions in which  the supergravity
charges, $Q^{ST}$,  are the same as the integer charges.}

\bea \label{GHmerger} N_1^{GH,ST} n_{1} + N_3^{GH,ST}n_3 -
N^{GH}_2 n_2^{ST}=0. \eea
Note that the GH charges of the ring and of the tube are gauge
dependent, but the symplectic product is invariant.

To compare the GH merger conditions  (\ref{GHmerger}) to the
merger conditions obtained in the previous section using the DBI
action, one should recall that this condition describes only those
supertubes that correspond to point sources on the $\IR^3$ of the
GH base.  That is,  the supertubes are embedded into $\IR^4$ so as
to wind around the GH fiber, and thus preserve the same
triholomorphic $U(1)$ isometry as the black ring. From
(\ref{winding}) and (\ref{BR-GHcoord}) we see that the winding
numbers of the supertube in the  GH patch are given by $(1,
\nu+1)$. (Remember that $\psi$ has period $4 \pi$.)  Thus a
supertube is point-like in the $\IR^3$ if and only if it has
either $\nu = -1$ or it lies on the polar axis with $x = \pm 1$.
We therefore restrict ourselves to   mergers with $x=\pm1$ for any
value of $\nu$, or mergers with  $\nu=-1$.

For $x=1$, we need to be on the patch $c=-1$, and
(\ref{crudemerger}) gives:
\bea \label{mergx1} N_1^{ST} n_{1} + N_3^{ST}n_3 - \ov{N}_2
n_2^{ST}~=~ 0~. \eea
For $x=-1$, we need to be on the patch $c=+1$, and thus have:
\bea \label{mergx-1} N_1^{ST} n_{1} + N_3^{ST}n_3 - \ov{N}_2
n_2^{ST}~=~ 2n_2^{ST}n_1n_3~. \eea
But using the relation (\ref{BRGHcharges2}), we can rewrite it as
\bea \label{merger2} N_1^{ST} n_{1} + N_3^{ST}n_3 - N^{GH}_2
n_2^{ST}~=~ 0 \eea
on both patches. The extra term in (\ref{mergx-1}) is simply the
shift in $N^{GH}$ induced by changing patches. Thus, if we
identify the DBI charge of the supertube with the GH charge of the
corresponding supergravity solution,
\bea \label{identification} N_I^{ST}=N_I^{GH,ST}, \eea
we have a perfect agreement between the supergravity approach
(\ref{GHmerger}) and the DBI approach (\ref{merger2}).

The supertubes with $\nu=-1$  do not wrap the $\phi$ circle of the
$\mathbb{R}^3$ base of the GH space, and thus are point-like in
this base for any value of $x$, and they source a supergravity
solution with a GH base for any location. Moreover, since these
tubes do not wrap the Dirac string, the merger relations become
$x$ independent. Equations (\ref{merg1}) and (\ref{merg2}) then
become
\bea
N_1^{ST} n_{1} + N_3^{ST}n_3 - \ov{N}_2 n_2^{ST}~=~ 0 ~~~ \text{for} ~~~c=-1\,, \\
N_1^{ST} n_{1} + N_3^{ST}n_3 - \ov{N}_2 n_2^{ST}~=~
2n_2^{ST}n_1n_3 ~~~ \text{for} ~~~c=+1\,, \eea
which once again can be re-written as
\be N_1^{ST} n_{1} + N_3^{ST}n_3 - N^{GH}_2 n_2^{ST}~=~ 0\,.
 \ee
Hence we arrive at the same conclusion as for supertubes at
$x=\pm1$: the DBI charges of the supertube give the GH charges of
the corresponding supergravity solution:
\be N_I^{ST}~\equiv~ N_I^{GH,ST}\,. \ee
%

\subsection{Black rings and three-charge two-dipole-charge supertubes}
 \label{threechgST}

One can generalize the foregoing discussion of mergers to examine
a three-charge, two dipole charge supertube \cite{Bena:2004wt}
merging with a generic black ring.  This can be done both in the
probe approximation, using the DBI action,  and in the exact
supergravity solution.  This supertube is more general than the
two-charge supertube, and although it does not source a smooth
supergravity solution in any duality frame, it can be used to
study rather more general classes of mergers.

The best duality frame to study this merger is that in which the
three-charge supertube is a dipolar D6-brane carrying electric D4,
D0 and F1 charges.  We take our tube to be along the
$(x_1,x_2,x_3,x_4,z,\vec{y}(\theta)))$, where  $\vec{y}(\theta)$
describes a closed curve in the non-compact space. As before, we
take $\theta \in$ (0,$2\pi n_1^{ST}$) with $n_1^{ST}$ being the
winding number of the supertube which is also its D6 dipole
charge. We introduce world-volume electric fields:
$\cF_{z\theta}$, $\cF_{tz}$, $\cF_{12}$ and $\cF_{34}$. where
$\cF_{tz}$ and $\cF_{z\theta} $ generate the F1 and D4 charges
respectively and $\cF_{12}$ and $\cF_{34}$ are needed for the D0
charge. The integer charges are given by
\begin{eqnarray}
    N_{1}^{ST}&=&N_{D0}~=~ \frac{1}{2\pi} \int d\theta\, \cF_{z\theta}\cF_{12} \cF_{34} \,,\\
    N_{2}^{ST}&=&N_{D4}~=~ \frac{1}{2\pi} \int d\theta\,\cF_{z\theta}\,,  \\
    N_{3}^{ST}&=&N_{F1}~=~ \frac{1}{2\pi} \int d\theta \, \frac{\partial\mathcal{L}}{\partial \cF_{tz}}  \bigg|_{\cF_{tz}=1} \,,  \\
    n_2^{ST} &=& n_{D2}~=~ n_1^{ST} \cF_{12}\cF_{34}\,.
\end{eqnarray}
Note that we can take the D4 dipole moments and D2 charges of the
tube to be zero by taking $\cF_{12}$ and $\cF_{34}$ to be
traceless. Supersymmetry requires  that $\cF_{tz}=1$ and $\cF_{12}
= \cF_{34}$ \cite{Bena:2004wt}, and then one can show that
\begin{equation}
\mathcal{H}|_{\cF_{tz}=1, \cF_{12}=\cF_{34}} ~=~ T_{D6}
\cF_{z\theta}\cF_{12} \cF_{34} ~+~  T_{D6} \cF_{z\theta} ~+~
\frac{\partial\mathcal{L}}{\partial \cF_{tz}}  \bigg|_{\cF_{tz}=1,
\cF_{12}=\cF_{34}} \,,
\end{equation}
or equivalently

\begin{equation}
\int d^4{x}d zd\theta\,\mathcal{H}|_{\cF_{tz}=1, \cF_{12}=
\cF_{34}} ~=~ N_1^{ST}  + N_2^{ST} + N_3^{ST}
\,,
\end{equation}
where $\mathcal{H}$ is the energy per unit five-dimensional
volume.

As before, we will assume constant charge densities on the
supertube worldvolume and the interesting physical condition that
generalizes (\ref{radiuscondBRsimp}) comes from the variation that
define the F1-charge, $N_3^{ST}$:
\begin{multline}
\left[ N_3^{ST} -\ds\frac{1}{2} (n_1^{ST} n_2 +n_2^{ST} n_1 )( y+1
- \nu (x+c))
\right] \left[ N_2^{ST} -\ds\frac{1}{2} n_1^{ST} n_3 (y+1 -\nu (x+c)) \right] = \\
n_1^{ST} \left(n_1^{ST} Z_1 +  n_2^{ST} Z_2 \right)
\ds\frac{R^2}{(x-y)^2} ( (y^2-1) + \nu^2 (1-x^2) ) \,.
\end{multline}

Note that, using $n_1^{ST}N_1^{ST}=n_2^{ST}N_2^{ST}$, there is a
symmetry between  (D0,D6) and (D4,D2) charges and dipole moments,
as expected from U-duality. However since the tube has no NS5
dipole moment, there is no exchange symmetry between the F1 charge
and other charges.

One can extract the merger condition from this as before and one
finds that, for a merger with a black ring, (\ref{crudemerger})
generalizes to:
\begin{multline}
n_1N_1^{ST} +n_2N_2^{ST} +n_3N_3^{ST}  - n_1^{ST}\ov{N}_1 -
n_2^{ST}\ov{N}_2 = n_3(n_1n_2^{ST} +n_2n_1^{ST})((1+c) - (\nu
+1)(x+c) ). \label{3chgmerger}
\end{multline}

When the three-charge supertube respects the GH isometry ($x=\pm1$
for any $\nu$ or $\nu=-1$ for any $x$), one can also describe this
merger in supergravity. The solution is given by the same harmonic
functions as in (\ref{BRSTGHsol}), except that now $K_1$ and $L_2$
also have poles at the supertube location: \be K_1 \rightarrow -
{q_1^{GH} \over 2|\vec r - \vec r_{BR}|} - {q_1^{GH,ST} \over 2
|\vec r - \vec r_{ST}|} ~,~~~ L_2 \rightarrow {Q_2^{GH} \over 4
|\vec r - \vec r_{BR}|} + {Q_2^{GH,ST} \over 4 |\vec r - \vec
r_{ST}|}\,. \label{3CHsol} \ee One can see that equation
(\ref{3chgmerger}) is equivalent to the vanishing of the
$E_{7(7)}$ symplectic product of the GH charges of the black ring
and those of the three-charge supertube, and hence the merger
conditions obtained from supergravity and from the Born-Infeld
analysis of the three-charge supertube are the same. The
subtleties associated to the dependence of the charges upon the
patch are identical to those for the two-charge supertube, and we
will not discuss them again.

\subsection{Supertubes in a general solution with a Gibbons-Hawking base}

We now consider two-charge supertubes probing a general
three-charge BPS solution with a Gibbons-Hawking base and we will
again work in the D0-D4-F1 duality frame. The general BPS solution
with three charges and three dipole charges and a GH base is given
in Sections \ref{MtheoryBPS} and \ref{D0D4F1BPS} and we  proceed
as we did for the black-hole and black-ring backgrounds in
Sections \ref{BMPVprobe} and \ref{BRprobe}.  We denote the
supertube coordinates as $\xi^0$, $\xi^1 $ and $\xi^2\equiv
\theta$ and consider the simplified case of a circular supertube
along the $U(1)$ fiber of the GH base:
\begin{equation}
\xi^{0}~=~ t \,,  \qquad \xi^1 ~=~ z  \,, \qquad \theta = \psi~.
\end{equation}
The supertube action (\ref{DBIWZaction}) takes the explicit form
\begin{multline}
S = T_{D2} \int d^3\xi  \,\bigg\{ \left[ \left(\ds\frac{1}{Z_1} -
1\right) \mathcal{F}_{z\theta} + \ds\frac{K^3}{Z_1V} + \left(
\ds\frac{\mu}{Z_1} - \ds\frac{K^1}{V} \right)(\mathcal{F}_{tz}-1)
\right] \\- \bigg[ \ds\frac{1}{V^2Z_1^2} \big[(K^3  -  V
(\mu(1-\mathcal{F}_{tz})- \mathcal{F}_{z\theta}) )^2 + V Z_1Z_2
(1-\mathcal{F}_{tz})(2-Z_3(1-\mathcal{F}_{tz})) \big]
\bigg]^{1/2}\bigg\}  \,.
\end{multline}
For $\mathcal{F}_{tz}=1$ the tube is supersymmetric and, as
before, the Hamiltonian density is the sum of the charge densities
(\ref{HamDensity}). Due to the supersymmetry there is a constraint
similar to (\ref{radiuscondBRsimp}), which determines the location
of the supertube in terms of its charges
\begin{equation}
\left[ N_1^{ST} +  n_2^{ST} \ds\frac{K^3}{V} \right] \left[
N_3^{ST} + \ds\frac{K^1}{V} \right] = (n_2^{ST})^2
\ds\frac{Z_2}{V} \,, \label{radiuscondGH}
\end{equation}
where the charges are still defined by (\ref{STcharges}).

\subsection{Gibbons-Hawking backgrounds: comparing the DBI analysis with supergravity.}

Equation (\ref{radiuscondGH}) determines the position of a
supertube in an arbitrary three-charge background with a
triholomorphic $U(1)$ isometry. Since both the supertube and the
background preserve this isometry, their fully back-reacted
supergravity solution will have a Gibbons-Hawking base, and its
form is well-known. Hence, one can compare (\ref{radiuscondGH}) to
the corresponding condition coming from the supergravity analysis
of the supertube, and confirm that supertubes that are solutions
of the Born-Infeld action always give rise to smooth supergravity
solutions.

To do this,  it is useful to remember that in any Gibbons-Hawking
solution the singularities in the harmonic functions $K_2, L_1,
L_3$ and $M$ at the supertube location are given by
(\ref{BRSTGHsol}). If one now takes equation (\ref{regcondd}) for
a supertube with charges $Q_1^{GH,ST}, Q^{GH,ST}_3$ and $q_2^{ST}$
and uses the asymptotic behavior of these harmonic functions near
the supertube one obtains:
\be \left[Q_1^{GH,ST} -2 q_2^{ST} \ds\frac{K^3}{V} \right]
\left[Q_3^{GH,ST} -2 q_2^{ST} \ds\frac{K^1}{V} \right] =
(q_2^{ST})^2 \ds\frac{Z_2}{V}\,. \ee
Since the supergravity GH charges, $Q_1^{GH,ST}, Q_3^{GH,ST},
q_2^{ST}$, are the same as the integer charges $N_1^{GH,ST},
N_3^{GH,ST}, n_2^{ST}$, one sees that this agrees exactly with the
DBI calculation.

It is interesting to observe that the DBI action only gives one
equation of motion for the supertube, (\ref{radiuscondGH}), while
the supergravity analysis of the supertube gives two independent
equations, that can be chosen to be any two of (\ref{regconda}),
(\ref{regcondb}) and (\ref{regcondd}). This is because in the
Born-Infeld analysis the inputs are the supertube charges and
dipole charge, which one first uses to find the embedding, and
then one derives the angular momentum of the supertube, $J^{ST}$,
from that solution.

By contrast, in the supergravity analysis the angular momentum of
the supertube along the Gibbons-Hawking fiber appears as the
coefficient of the singular part in the harmonic function $M$, and
is one of the {\it inputs} of the calculation. Indeed, in
supergravity one can build ``supertube'' solutions for any value
of $J_T$. However most of these solutions will be singular: if
$J_T$ is too large the solutions will have closed timelike curves,
and if $J_T$ is too small the solutions will have a naked
singularity\footnote{Such a singularity might be cloaked by a
Planck-sized horizon \cite{shigemori}.}. Only one specific value
of $J_T$ gives a supergravity solution that is smooth and
horizonless in the duality frame in which the supertube charges
correspond to D1 and D5 branes.

To find this value it is most convenient to use equation
(\ref{regconda}), and the expansion of the harmonic functions
(\ref{BRSTGHsol}) near the supertube location to find the
supertube angular momentum as a function of the supertube charges
$Q_1^{GH,ST}, Q_3^{GH,ST}$ and dipole charge $q_2^{ST}$:
\be J^{GH,ST} = {N_1^{GH,ST} N_3^{GH,ST} \over n_2^{ST}} \ee

To obtain this equation from the DBI analysis one needs to
calculate the angular momentum of the supertube along the
Gibbons-Hawking fiber. This calculation is partially shown in Appendix C\footnote{For supertubes in $\IR^4$ in the presence of arbitrary charges and dipole charges} and gives
\be J^{ST}={N_1^{ST}N_3^{ST} \over n_2^{ST}}. \ee

This indicates that when supertubes are embedded in a
solution with a Gibbons-Hawking base, respecting the
triholomorphic $U(1)$ isometry of this solution, their Born-Infeld
analysis gives the equations needed for the fully back-reacted
supergravity solution of these supertubes to be smooth and free of
closed timelike curves.

\subsection{A comment on black rings in Taub-NUT and their four-dimensional charges.}
\label{FourDcharges}

An interesting by-product of our results in Section  \ref{GHBRsolution} is that a given five-dimensional supersymmetric black ring can be embedded in Taub-NUT \cite{Elvang:2005sa,Gaiotto:2005xt,Bena:2005ni} in many ways depending upon the choice of the gauge field for the in the magnetic flux\footnote{For the embedding of nonsupersymmetric black rings in Taub-NUT see \cite{nonbps}}.   We considered patches and gauge choices that preserve the $U(1)$ of the GH base and this still left a free parameter, $c$, in (\ref{Bi}).  The two natural patches, with $c=+1$ and $c=-1$ have a single Dirac string, and together they provide a complete cover of the solution.  Other choices of $c$ split the Dirac strings into two parts, one at each pole of the $S^2$.  If one  compactifies the black ring down to a four-dimensional black hole  then we saw that  the electric charges of the black hole are given by the GH electric charges at the ring location.  We also saw that the GH charges depended upon the choice of patch and if one uses the $c=+1$ or $c=-1$ then the black-hole charges are not the same as the electric charges, measured at infinity, of the  five-dimensional black ring.

Hence, from a four-dimensional perspective the black ring can correspond to an infinite family of black holes, whose D2 and D0 charges are related via the gauge transformation (\ref{gaugetsfo}). The effect of this transformation is to introduce Wilson lines for the gauge fields along the Taub-NUT circle at infinity, and to create or remove Dirac strings at the north or south pole of the black hole.  Nevertheless, even if the four-dimensional charges depend upon the choice of gauge, the warp factors $Z_I$ and the symplectic products that determine the metric, the field strengths, and the location of the black ring, are invariant under (\ref{gaugetsfo}).

One can also take a peculiar gauge with $c=0$ for which the solution has two Dirac strings but for this choice the four-dimensional electric charges are the same as the asymptotic charges in the five-dimensional solution \cite{Hanaki:2007mb}. On the other hand, in this gauge the D0 charge is given neither by the five-dimensional ``ring angular momentum'' (which was the difference between the two angular momenta in five dimensions), nor by the total angular momentum in the plane of the ring, $J_1$ (as assumed in \cite{Cyrier:2004hj}), but rather it is given by a combination of the five-dimensional charges and angular momenta that has no obvious interpretation in five dimensions:
\be
J_{c=0} = J_1-J_2 + {1 \over 2} q^I \overline Q_I + {3 \over 4} q^1 q^2 q^3 \,.
\ee

It is not hard to see that all the shifts of charges brought about by gauge transformations leave the $E_{7(7)}$ quartic invariant unchanged. The entropy of the ring is still determined by this invariant \cite{Bena:2004tk}, but now as a function of the shifted electric charges, and the shifted angular momentum. Therefore, the entropy of all the four-dimensional black holes related to the ring can be understood microscopically by an MSW analysis \cite{msw} that is done without the shift of $L_0$. Hence, the observation of \cite{Hanaki:2007mb} that the  five-dimensional asymptotic electric charges of the black ring can be related to those of a four-dimensional black hole does not solve the discrepancy between the two microscopic descriptions of black rings\footnote{One might get this impression from \cite{Hanaki:2007mb}.} \cite{Bena:2004tk,Cyrier:2004hj}.

Our analysis thus establishes that the four-dimensional charges that one uses in the $E_{7(7)}$ quartic invariant to obtain the black ring entropy, depend on the choice of patch, and one can switch between various charges (like the asymptotic charges of the ring and the intrinsic charges) by gauge
transformations.  Nevertheless, this transformation generically also changes the angular momentum parameter (or the D0 charge). Therefore, in trying to find the microscopic description of extremal non-BPS black rings (as was done recently in \cite{Emparan}) one should not focus on the fact that a certain charge appears in the quartic invariant, but rather on a gauge-independent concept like why, for a given choice of charges, does a certain angular momentum parameter appear in the quartic  invariant.

\section{Chronology protection }
\label{chrono}

Having obtained the condition under which a supertube and a black
ring can merge, both using the Born-Infeld description of
supertubes, and (where appropriate) also using the supergravity
solution corresponding to the merger, we now turn to verifying
that supertube mergers preserve the physical properties of the
black ring. For simplicity, and because it is sufficient for
capturing all the relevant physics of the merger, we will
primarily focus on circular embeddings for the tube
(\ref{winding}).

\subsection{Mergers of black rings with two-charge supertubes}

We begin by considering the merger of a black ring with a
two-charge supertube of arbitrary shape. To do this one must first
establish what shape can the supertube have when it crosses the
black ring horizon.  Based on our intuition from supertubes
merging with black holes \cite{Bena:2004wt} we expect that the
supertube will be parallel to the horizon, and that it should not
be possible to have a part of the supertube inside the black ring
horizon and a part of it is outside.

To see this we can analyze equation (\ref{radiuscondBR}) and
change variables to $w=\frac{1}{y}$; the merger then happens at
$w\rightarrow 0$. After some algebra one can see that for $w
\rightarrow 0$ the leading divergent term in (\ref{radiuscondBR})
imposes the constraint $\frac{\partial w}{\partial\theta}=0$,
which implies that the supertube is always tangent to the horizon
when it merges to a black ring.

It is particularly important to examine the thermodynamics of
mergers and see whether by ``throwing in'' supertubes one could
decrease the entropy of a black ring, or overspin it and introduce
closed timelike curves (violating chronology protection). To do
this one must determine what are the charges that a supertube
brings into a ring. As we saw in the section \ref{comparison1},
there are some subtleties in this determination and we cannot
always add the DBI charges of the supertube to the constituent
charges, the $\ov{N}$'s, of the ring. We have   learned that the
DBI charges  have to be identified with the GH charges of the
supertube, which are patch-dependent, and are not the same as the
constituent ones.   We have seen this explicitly from the
supergravity solution for concentric mergers (when $x=\pm1$) or
alternatively when we take $\nu = -1$ so that the supertube does
not wind around latitude circles and crosses the ring horizon at a
point on the $S^2$ of the horizon. We will first focus on mergers
where the supertube merges at a point on the $S^2$, and discuss
the other ones at the end of this subsection.

The entropy of the black ring is given by
$S=\pi\sqrt{\mathcal{M}}$ where ${\cal M}$ is defined in
(\ref{BRent})
\begin{equation}
\mathcal{M} = 2 n_1n_2\ov{N}_1\ov{N}_2 + 2 n_1n_3\ov{N}_1\ov{N}_3
+ 2 n_2n_3\ov{N}_2\ov{N}_3 - (n_1\ov{N}_1)^2 - (n_2\ov{N}_2)^2
-(n_3\ov{N}_3)^2 - 4 n_1n_2n_3 J \,.
\end{equation}
Note that $\mathcal{M}$ is in fact the $E_{7(7)}$ quartic
invariant and is therefore invariant under a gauge transformation
(\ref{gaugetsfo}). In  terms of GH charges of the ring, we have
\bea
\mathcal{M} &=& 2 n_1n_2N^{GH}_1N^{GH}_2 + 2 n_1n_3N^{GH}_1N^{GH}_3 + 2 n_2n_3N^{GH}_2N^{GH}_3 \nonumber \\
& & - (n_1N^{GH}_1)^2 - (n_2N^{GH}_2)^2 -(n_3N^{GH}_3)^2 - 4
n_1n_2n_3 J^{GH} \,. \eea

From the analysis in  the previous section, we know that the
supertube DBI charges correspond to GH charges, and thus should be
directly added to the GH charges of the ring.

To keep the expressions simple we will take the three electric and
the three dipole charges of the black ring charges to be equal, we
will also assume that the two electric charges of the supertube
are equal, namely:
\begin{equation}
N^{GH}_1 = N^{GH}_2 = N^{GH}_3~\equiv~ N~, \qquad   n_1 = n_2 =
n_3~\equiv~ n ~, \qquad  N_1^{ST} = N_3^{ST}~\equiv~\Delta N~.
\end{equation}
Then we have
\begin{equation}
\mathcal{M} ~=~  n^2 ( 3N^2 - 4 n J )
\end{equation}
and the charges of physical black rings satisfy: $3N^2\geq  4nJ $.

Let   $\Delta n$ denote the dipole charge of the tube and $\Delta
J$ its angular momentum. The new horizon area parameter,
$\widetilde{\mathcal{M}}$, after the merger  is then
\begin{eqnarray}
\widetilde{\mathcal{M}} &=&  4\, n\,N\, (n+\Delta n )(N+ {\Delta
N})
+2\, n^2(N+\Delta N)^2 -(n+\Delta n )^2 N^2 \nonumber \\
& & \qquad \qquad  \qquad \qquad  -2\, n^2(N+\Delta N)^2
- 4\, n^2 (n+\Delta n )(J+\Delta J)\nonumber  \\
&=& \mathcal{M} ~+~ n\, \Delta n  (3N^2-4nJ) \\
&& \qquad \qquad -~ \frac{(n+\Delta n )}{\Delta n  } \, \Big[ \,
(2 \, n\, \Delta N  - N \Delta n )^2~+~
 4\, n^2 \Delta n  \Big(\Delta J - \frac{(\Delta N)^2}{\Delta n } \Big )\Big] \notag\,.
\end{eqnarray}
We now need to remember that the angular momentum of the tube is
given by (\ref{JSTBR})
\begin{equation}
\Delta J= {(\Delta N)^2 \over \Delta n}\,, \label{STangmom}
\end{equation}
and also that that for the charges we consider the merger
condition (\ref{merger2}) becomes
\begin{equation}
2 \, n \, \Delta N = \Delta n \, N \, .
\end{equation}
Using these two equations, we finally have
\begin{equation}
\Delta \mathcal{M} ~\equiv~ \widetilde{\mathcal{M}}   -\mathcal{M}
~=~ n\, \Delta n ( 3N^2 - 4 n J ) ~\ge~ 0\,,
  \label{posentcond}
\end{equation}
with equality if and only if the original black ring has vanishing
horizon area. Hence, for mergers with $\nu = -1$ or $x=\pm1$, we
have proved that chronology is protected, and that the second law
of black hole thermodynamics holds. This conclusion is similar to
that of \cite{Bena:2004wt, Marolf:2005cx, Bena:2005zy} for
supertube-black hole mergers.

However for $\nu \ne -1$ the situation is rather more subtle.
First, the complete supergravity solution is not known for mergers
in which the supertube winds around an $S^1$ in the $S^2$ of the
horizon. As a result we cannot identify the supertube DBI charges
with   simple supergravity charges. In addition it is not clear
how to identify directly the charges carried across the horizon
during the merger.  If one simply chooses one of the patches
discussed above and assumes that the supertube carries its
constituent DBI or GH charges across the horizon then the
$x$-dependence in the merger condition (\ref{merg1}) can lead to
mergers in which the
 horizon area of the black ring decreases, thus
contradicting black hole thermodynamics.

The most likely solution to this conundrum is that the charges
carried by the supertube across the horizon are not the same as
the constituent supertube charges $\ov{N}^{ST}, \ov{J}^{ST}$, but
are modified {\it in an $x$-dependent way}, so as not to decrease
the horizon area. This would imply that in $\nu \ne -1$,
$x\ne\pm1$ mergers the supertube brings in not only its intrinsic
charges, but also some of the charge and angular momentum
dissolved in supergravity fluxes. Since it is unclear how the
dynamics of this charge can be captured via a Born-Infeld
analysis, we believe that the understanding of this phenomenon and
a resolution of this puzzle will probably come from finding the
fully back-reacted supergravity solution corresponding to the $\nu
\ne -1$ mergers\footnote{Such mergers do not have a
tri-holomorphic $U(1)$ invariance and hence the supergravity
solution will be more complicated than the solutions with a
Gibbons-Hawking base presented here.}.

\subsection{Mergers of black rings with three-charge two-dipole-charge supertubes}

Another interesting example for illustrating chronology protection
is the merger of a three-charge two-dipole charge supertube with
another supertube of the same kind, that can also be thought of as
a singular black ring that has one zero dipole charge $n_3^{BR}
=0$.  Such a singular black ring must have vanishing horizon area,
and to avoid closed timelike curves it must satisfy the charge
condition \cite{hairs}:
\begin{equation}
n_1^{BR} N_1^{BR} ~=~ n_2^{BR}  N_2^{BR}  \,. \label{2ChgNoCTC}
\end{equation}
Similarly, the three-charge supertube considered above has no NS5
dipole charge ($n_3 =0$) and also satisfies
\begin{equation}
n_1^{ST} N_1^{ST} ~=~ n_2^{ST}  N_2^{ST}  \,. \label{Stubecond}
\end{equation}
Since the merger produces another two-dipole three-charge tube, it
must also satisfy the regularity condition:
\begin{equation}
(n_1^{BR} + n_1^{ST} ) (N_1^{BR}+N_1^{ST} )  ~-~ (n_2^{BR} +
n_2^{ST} ) (N_2^{BR}+N_2^{ST} )  ~=~0 \,, \label{necessconda}
\end{equation}
which is equivalent to
\begin{equation}
n_1^{BR}  N_1^{ST}  + n_1^{ST}   N_1^{BR}   ~-~ (n_2^{BR} N_2^{ST}
+ n_2^{ST} N_2^{BR} )     ~=~0 \,. \label{necesscondb}
\end{equation}

On the other hand, the merger condition (\ref{3chgmerger}) for
$n^{BR}_3 =0$ yields:
\begin{equation}
(n_1^{BR}N_1^{ST} +n_2^{BR}N_2^{ST} ) ~-~
(n_1^{ST}N_1^{BR}+n_2^{ST}N_2^{BR}) ~=~0 \,. \label{3chgmergerred}
\end{equation}

To establish chronology protection one must show that
(\ref{3chgmergerred}) implies (\ref{necesscondb}).

However, one also knows that the two merging objects obey
(\ref{2ChgNoCTC}) and (\ref{Stubecond}).   Multiplying
(\ref{3chgmergerred}) by $n_2^{BR} n_2^{ST}$ and using
(\ref{2ChgNoCTC})  and (\ref{Stubecond}) one obtains:
\begin{equation}
 (n_2^{BR} N_1^{ST} -  n_2^{ST} N_1^{BR} )  \,(n_1^{ST}   \, n_2^{BR} + n_2^{ST}  \, n_1^{BR} )      ~=~0 \,.
\label{mergerb}
\end{equation}
Similarly, one finds that (\ref{necesscondb}) is equivalent to
\begin{equation}
(n_2^{BR} N_1^{ST} -  n_2^{ST} N_1^{BR} )  \, (n_1^{ST}   \,
n_2^{BR} -  n_2^{ST}  \,  n_1^{BR} )      ~=~0 \,.
\label{necesscondc}
\end{equation}
Since all the $n$'s are positive, we see that (\ref{mergerb})
implies (\ref{necesscondc}) and so the merger condition
(\ref{3chgmergerred}) implies
 that the regularity condition (\ref{necesscondb})  is
satisfied. Hence, the merger of two three-charge two-dipole charge
supertubes always respects chronology protection.

We can also consider a merger of a three-charge two-dipole charge
supertube with a fully fledged black ring, we take for simplicity
equal charges and dipoles: $n_1^{BR}=n_2^{BR}=n_3^{BR}=n$,
$N_1^{BR}=N_2^{BR}=N_3^{BR}=N$, $ N_1^{ST} = N_2^{ST} = N_3^{ST}
=\Delta N$ and $ n_1^{ST} = n_2^{ST} =\Delta n$.  The
non-negativity of the initial black ring entropy implies that
$3N^2\geq 4nJ $ and the merger condition\footnote{We consider $\nu=-1$ tubes in the $c=-1$ patch; all the subtleties having to do with changing patches are the same as for two-charge supertubes.} becomes $3n \Delta N =
2\Delta n N$. Also remembering
that angular momentum of the three-charge supertube is given by
\begin{equation}
J^{ST} ~=~ \frac{N_1^{ST} N_3^{ST}} {n_2^{ST}}~=~\frac{N_2^{ST}
N_3^{ST}} {n_1^{ST}} \,
\end{equation}
and hence $\Delta J=\Delta N^2/\Delta n$, we obtain
\begin{equation}
\Delta \mathcal{M} ~\equiv~ \widetilde{\mathcal{M}}   -
\mathcal{M}~=~ \frac{4}{9}\,(7 N^2- 9 nJ) \,(2 n\Delta n +  \Delta
n^2 )\,  \,.
\end{equation}
Since $ N^2\geq \frac{4}{3} nJ $ this merger is always
irreversible, and does not violate chronology protection.

\section{Fluctuating supertubes and entropy enhancement}
\label{FluctuatingStubes}

This section is devoted to an in-depth review of the Born-Infeld
calculation of the entropy coming from the shape modes of
supertubes, as well as to an extension of this calculation to a
supertube in a black-ring background. This calculation
demonstrates that one can equally obtain an enhanced entropy from
fluctuations along the compact internal directions of the solution
and fluctuations in the non-compact directions of the solution.
Furthermore, as we have shown in the previous sections of this
paper, we expect the latter supertube fluctuations to give rise to
smooth horizonless solutions. Hence, our analysis strongly
supports the  existence of smooth horizonless three-charge
solutions that depend on arbitrary continuous functions, and whose
entropy is much larger than their typical charge, and might even
be as large as the square root of the cube of their charge.  That
is, it might be black-hole-like.

Our goal is to quantize the small oscillations about round
two-charge supertubes in flat space, black-hole, black-ring, and
generic three-charge backgrounds, and to examine the entropy coming
from these fluctuations. We find it convenient to work in the
D0-D4-F1 duality frame, and our approach follows that of
\cite{Palmer:2004gu,enhance} (see also \cite{Bak}).

We begin by reviewing the Marolf-Palmer entropy calculation for a
supertube in flat space, and in the following subsections extend
this calculation for a supertube in a 3-charge black hole
background and in a black ring background. In the last subsection
we also include, for completeness, the entropy calculation in the
background of a general solution with a Gibbons-Hawking base space
\cite{enhance}.

As first reported in \cite{enhance}, in the latter two backgrounds
we find a non-trivial enhancement of the entropy of a supertube
when the dipole magnetic fields are large. This enhancement arises
because the entropy that can be stored in a supertube is governed
not by the electric charges of the supertube (as in flat space or
in a black hole background) but by its locally-defined {\it
effective charges}, that can get large contributions from the
interactions of the dipole moment of the supertube with the
magnetic fluxes of the background.

\subsection{Flat space}

In the absence of background fluxes, the WZ action of the
supertube is zero, and the DBI action (\ref{DBIWZ}) reduces to
\begin{equation}
S  = - T_{D2} \int dtdzd\theta
\sqrt{R^{2}(1-\mathcal{F}_{tz}^2)+\mathcal{F}_{z\theta}^2} \,,
\end{equation}
where $R$ is the radius of the supertube and its embedding is
\begin{equation}
t = \xi^0~, \qquad z=\xi^1~, \qquad \varphi_1=\theta ~.
\label{STembedding}
\end{equation}
The charges of the tube are given by (\ref{STcharges}):
\begin{equation}
N_{1}^{ST} ~=~  n_2^{ST}\mathcal{F}_{z\theta} \,,  \qquad
N_{3}^{ST} ~=~ n_2^{ST}\ds\frac{R^2}{\mathcal{F}_{z\theta}} \,,
\end{equation}
where the factors of $n_2^{ST}$ come from multiple windings in
$\theta$. Similarly  the radius relation  (\ref{STBHradius})
reduces to:
\begin{equation}
N_1^{ST}N_3^{ST} ~=~ \big(n_2^{ST} \big)^2 R^2\,.
\end{equation}
The angular momentum of the supertube is  (\ref{JST}):
\begin{equation}
J ~=~ {N_1^{ST}N_3^{ST} \over n_2^{ST}}  ~=~ n_2^{ST} R^2   \,.
\end{equation}
The foregoing results apply to round (maximally spinning)
supertubes. Supertubes of arbitrary shape will have more
complicated expressions for their conserved quantities and will
generically have smaller angular momentum.

In this subsection we will perform a simplified version of the
analysis in \cite{Palmer:2004gu}, which will be enough to give us
the correct dependence of the entropy on the supertube charges. We
consider small fluctuations of the supertube in the six directions
transverse to its world-volume:
\begin{equation}
x_i \rightarrow x_i + \eta_i(t,\theta)~, \qquad\qquad i=1, \dots,
6\,,
\end{equation}
where four of these fluctuations take place on the compact $T^4$
and the other two are radial coordinates in the non-compact space.
In general there are eight independent fluctuation modes for the
supertube, consisting of seven transverse coordinate motions and a
charge density fluctuation (which also affects the shape). To keep
the computations simple here, we have restricted to a
representative sample of oscillations in both the compactification
space and in the space-time. Since we are only interested in BPS
fluctuations we will also restrict $\eta_i$ to depend only upon
$t$ and $\theta$ \cite{Palmer:2004gu}\footnote{The  time dependent
modes will break supersymmetry. Hence, we will retain  the time
dependence of $\eta_i$ to compute momenta and quantize the system
but then we will set $\partial_t \eta_i \equiv \dot{\eta}_i =0$.}.

The effective Lagrangian for the fluctuations is obtained by
expanding the DBI Lagrangian of the supertube
\begin{equation}
L_{\eta} = - T_{D2} \left[
(1-\mathcal{F}^2_{tz}-\dot\eta_i\dot\eta_i)(R^2 +\eta_i'\eta_i') -
2 \mathcal{F}_{tz}\mathcal{F}_{z\theta}\dot{\eta}_i\eta'_i +
\mathcal{F}_{z\theta}^2 (1- \dot\eta_i\dot\eta_i) +
(\dot{\eta}_i\eta'_i)^2\right]^{1/2} \,,
\end{equation}
where the repeated index $i$ is summed over. The canonical momenta
conjugate to $\eta_i$ are:
\begin{equation}
\Pi_i = \int_{0}^{2\pi L_z} dz~ \ds\frac{\partial
L_{\eta}}{\partial \dot{\eta}_i}\bigg|_{\dot{\eta}_i=0\,, \,
\mathcal{F}_{tz}=1} ~=~ \ds\frac{1}{2\pi} \eta'_i \,,
\end{equation}
and the canonical  commutation relations are:
\begin{equation}
[ \eta_j(t,\theta), \Pi_k (t,\theta') ] = i\delta_{jk}
\delta(\theta-\theta') \,.
\end{equation}
The BPS modes $\eta_i$ then can be expanded as:
\begin{equation}
\eta_{i} = \ds\frac{1}{\sqrt{2}} \bigg[ \ds\sum_{k>0}
e^{ik\theta/n_2^{ST}} \ds\frac{(a_k^i)^{\dagger}}{\sqrt{|k|}}
~~+~~ \text{h.c.} \bigg] \label{etaexp}
\end{equation}
where $(a_k^i)^{\dagger}$ and $a_k^i$ are creation and
annihilation operators for the $k^{\text{th}}$ harmonic.   The
normalization has been chosen such that\footnote{Technically, to
get this normalization correct we need to include the mode
expansion of the non-BPS modes  in (\ref{etaexp}). Ignoring the
non-BPS modes gives an incorrect factor of $\sqrt{2}$ in the
normalization of the $\eta_i$.  Here we have given the correctly
normalized expressions that one would obtain
 if one included the non-BPS modes.}:
\begin{equation}
[(a_k^i)^{\dagger},  a_{k'}^j ] = \delta^{ij}\delta_{k,k'}
\end{equation}
It is not hard to see that the fluctuations do not change
$N_1^{ST}$ and the angular momentum $J$. The charge $N_3^{ST}$
becomes:
\begin{equation}
N_{3}^{ST} ~=~  \frac{1}{T_{F1}} \int_{0}^{2\pi n_2^{ST}}d\theta~
\frac{\partial \mathcal{L}}{\partial
\mathcal{F}_{tz}}\bigg|_{\mathcal{F}_{tz}=1} \,  ~~=~~
\ds\frac{T_{D2}}{T_{F1}}\int_{0}^{2\pi n_2^{ST}} d\theta~
\ds\frac{(R^2 + \eta'_i\eta'_i)}{\mathcal{F}_{z\theta}}\, \,,
\end{equation}
from which one finds
\begin{eqnarray}
\sum_{i=1}^{6}\sum_{k>0} k (a_k^i)^{\dagger} a_k^i &=& L_z T_{D2}
\int_{0}^{2\pi n_2^{ST}} d\theta \int_{0}^{2\pi n_2^{ST}} d\theta'
~
\sum_{i=1}^{6} \eta'_i\eta'_i  \\
&=&  N_1^{ST} N_3^{ST} - (n_2^{ST})^2 R^2 ~=~ N_1^{ST} N_3^{ST} -
n_2^{ST} J~.
\end{eqnarray}

The left hand side of this expression can be thought of as the
energy of a system of six massless bosons in $(1\!+\! 1)$
dimensions. Due to supersymmetry there will also be six
corresponding fermionic degrees of freedom. The total central
charge of the system is thus $c=9$, and so the entropy of this
system  is given by the Cardy formula:
\begin{equation}
S = 2\pi \sqrt{\ds\frac{c}{6}} \sqrt{  N_1^{ST} N_3^{ST} -
n_2^{ST} J  } = 2\pi \sqrt{\ds\frac{3}{2}} \sqrt{ N_1^{ST}
N_3^{ST} - n_2^{ST} J } \,.
\end{equation}
If we had included all eight bosonic fluctuation modes then we
would have had eight bosons and eight fermions and hence a theory
with  $c=12$ and with the entropy:
\begin{equation}
S_{ST}  = 2\pi\sqrt{2} \sqrt{N_1^{ST} N_3^{ST} - n_2^{ST} J}~.
\end{equation}
This is the correct central charge and it yields   the correct
supertube entropy \cite{Palmer:2004gu}.  By restricting  our
analysis to six of the shape modes and ignoring the other
supersymmetric modes we have obtained a finite, but well
understood, fraction of the supertube entropy.  Since our purpose
here is to analyze when entropy enhancement happens, and when it
does not, we will only be interested on the dependence of the
supertube entropy on the macroscopic charges, and not pay
particular attention to numerical coefficients. Restricting our
analysis in more general backgrounds to transverse BPS
fluctuations and counting the entropy coming from these modes will
therefore be enough to illustrate the physics of entropy
enhancement.

\subsection{The three-charge black hole}

A two-charge round supertube in the background of a three-charge BPS rotating (BMPV) black
hole was discussed in section \ref{BMPVprobe}. Here we will use
the metric and background fields presented in section
\ref{BMPVprobe} and consider small shape fluctuations in the
directions transverse to the world-volume of the supertube. We are
again interested only in BPS excitations, which have the following
form
\begin{equation}
x_i \to x_i + \eta_i (t,\theta)\,, \ \  i=1,2,3,4 \,,   \qquad u
\to u + \eta_5(t,\theta)\,, \qquad v \to v + \eta_6 (t,\theta) \,,
\label{flucts}
\end{equation}
where we have defined the metric on the four-torus to be
\begin{equation}
ds^2_{T^4} = dx_1^2 + dx_2^2 + dx_3^2 + dx_4^2 \,.
\label{metric_torus}
\end{equation}
and the supertube embedding is the same as (\ref{STembedding}).
One can use the sum of the DBI and WZ actions, find an effective
action for the supertube fluctuations and compute the momenta
conjugate to $\eta_5$, $\eta_6$ and $\eta_i$:
\begin{eqnarray}
\Pi_{\eta_5} &=& \int
dz\bigg(\ds\frac{\partial\mathcal{L}}{\partial
\dot\eta_5} \bigg)\bigg|_{BPS}~=~ \ds\frac{Z_2}{2\pi} \eta_5' \,,\\
\Pi_{\eta_6} &=& \int dz
\bigg(\ds\frac{\partial\mathcal{L}}{\partial
\dot\eta_6} \bigg)\bigg|_{BPS}~=~  \ds\frac{Z_2}{2\pi} \eta_6' \,, \\
\Pi_{\eta_i} &=& \int dz
\bigg(\ds\frac{\partial\mathcal{L}}{\partial \dot\eta_i}
\bigg)\bigg|_{BPS}~=~ \ds\frac{1}{2\pi} \eta_i' \,,
\end{eqnarray}
where the subscript ``BPS'' means that we have evaluated
everything ``on shell,'' which means we have imposed the BPS
conditions of no time dependence and $\mathcal{F}_{tz} = 1$.

The BPS modes $\eta_i$, $\eta_5$ and $\eta_6$ then can be expanded
as
\begin{equation}
\begin{array}{l}
\eta_{i} = \ds\frac{1}{\sqrt{2}} \bigg[ \ \ds\sum_{k>0}
e^{ik\theta/n_2^{ST}} \ds\frac{(a_k^i)^{\dagger}}{\sqrt{|k|}}
~~+~~
\text{h.c.} \ \bigg] \,, \\\\
\eta_{5} = \ds\frac{1}{\sqrt{2}} \bigg[\ \ds\sum_{k>0}
e^{ik\theta/n_2^{ST}} \ds\frac{(a_k^5)^{\dagger}}{\sqrt{|k|}}
~~+~~
\text{h.c.}\ \bigg] \,, \\\\
\eta_{6} = \ds\frac{1}{\sqrt{2}} \bigg[ \ \ds\sum_{k>0}
e^{ik\theta/n_2^{ST}} \ds\frac{(a_k^6)^{\dagger}}{\sqrt{|k|}}
~~+~~ \text{h.c.}\ \bigg] \,.
\end{array}
\label{modexBMPV}
\end{equation}
At first glance, the physics of the $\eta_i$ fluctuations along
the torus appears very different from that of the fluctuations in
the spacetime direction, $\eta_{5}$ and $\eta_{6}$; indeed the
latter have a factor of $Z_2$ in the denominator, and this factor
becomes arbitrarily large when the supertube is near the horizon
of a black hole.

The charge $N_1^{ST}$ is the same as that of the round supertube,
but the charge $N_3^{ST}$ is modified to:
\begin{equation}
N_{3}^{ST} = \frac{1}{T_{F1}}
 \int d\theta~ \frac{\partial \mathcal{L}}{\partial
\mathcal{F}_{tz}}\bigg|_{BPS} =
\ds\frac{T_{D2}}{T_{F1}\mathcal{F}_{z\theta}}\int d\theta
\left(Z_2u^2 + Z_2 [(\eta'_5)^2+(\eta'_6)^2 ] +
\ds\sum_{i=1}^{4}(\eta'_i)^2 \right) \,.
\end{equation}
Using similar arguments to those given for the flat space
background one finds the entropy of the BPS shape modes to be:
\begin{equation}
S =  2\pi \sqrt{\ds\frac{3}{2}}\sqrt{ N_1^{ST} N_3^{ST} -
(n_2^{ST})^2 Z_2u^2 } \,.
\end{equation}
Hence, despite the presence of the warp factor $Z_2$ in the radius
relation and in the mode expansions (\ref{modexBMPV}), the entropy
of the supertube depends on its charges in exactly the same way as
in flat space, and hence there is no entropy enhancement.

\subsection{The three-charge black ring background}

We now consider small shape fluctuations around the  round
supertube in a black ring background presented in section
\ref{BRprobe}. The important new element is that this background
has non-zero magnetic dipole charges and these will enter the
calculation in some very non-trivial ways.

Again we consider the fluctuations (\ref{flucts}) and use the DBI
and WZ actions to find an effective action for the fluctuations.
After straightforward calculations on can compute the momenta
conjugate to $\eta_5$, $\eta_6$ and $\eta_i$:
\begin{eqnarray}
\Pi_{\eta_5} &=& \int
dz\bigg(\ds\frac{\partial\mathcal{L}}{\partial \dot\eta_5}
\bigg)\bigg|_{BPS}~=~ \ds\frac{Z_2}{2\pi}
\ds\frac{R^2}{(y^2-1)(x-y)^2}
\eta'_5 \,,\\
\Pi_{\eta_6} &=& \int dz
\bigg(\ds\frac{\partial\mathcal{L}}{\partial \dot\eta_6}
\bigg)\bigg|_{BPS}~=~ \ds\frac{Z_2}{2\pi}
\ds\frac{R^2}{(1-x^2)(x-y)^2}
\eta'_6 \,, \\
\Pi_{\eta_i} &=& \int dz
\bigg(\ds\frac{\partial\mathcal{L}}{\partial \dot\eta_i}
\bigg)\bigg|_{BPS}~=~\ds\frac{1}{2\pi} \eta_i' \,,
\end{eqnarray}

The BPS modes $\eta_i$, $\eta_5$ and $\eta_6$ can be expanded as:
\begin{equation}
\begin{array}{l}
\eta_{i} ~=~ \ds\frac{1}{\sqrt{2}} \, \bigg[ \ \ds\sum_{k>0}
e^{ik\theta/n_2^{ST}} \ds\frac{(a_k^i)^{\dagger}}{\sqrt{|k|}}
~~+~~
\text{h.c.}\ \bigg] \,, \\\\
\eta_{5} ~=~ \ds\sqrt{\ds\frac{(y^2-1)(x-y)^2}{2Z_2R^2}} \,
\bigg[\  \ds\sum_{k>0} e^{ik\theta/n_2^{ST}}
\ds\frac{(a_k^5)^{\dagger}}{\sqrt{|k|}} ~~+~~
\text{h.c.}\ \bigg]\,, \\\\
\eta_{6} ~=~  \ds\sqrt{\ds\frac{(1-x^2)(x-y)^2}{2Z_2R^2}}\, \bigg[
\ \ds\sum_{k>0} e^{ik\theta/n_2^{ST}}
\ds\frac{(a_k^6)^{\dagger}}{\sqrt{|k|}} ~~+~~ \text{h.c.}\ \bigg]
\,.
\end{array}
\label{etamodes}
\end{equation}
Suppose that we have a round supertube parallel to the ring ($t =
\xi^0,~ z=\xi^1,~\varphi_1=-\theta$), then for the F1 charge of
the supertube one finds
\begin{eqnarray}
N_3^{ST} &=&  \ds\frac{1}{T_{F1}}\int_{0}^{2\pi n_2^{ST}}
\bigg(\ds\frac{\partial\mathcal{L}}{\partial
\mathcal{F}_{tz}} \bigg) \bigg|_{BPS} \\
 &=&    \ds\frac{T_{D2}}{T_{F1}} n_2^{ST}n_1 (1+y) ~+~
  \ds\frac{ T_{D2}}{T_{F1}(\mathcal{F}_{z\theta} -
\frac{n_3}{2}(1+y))} \bigg[ \ds\frac{Z_2 R^2 (y^2-1)}{(x-y)^2}  \\
&& + ~Z_2 \ds\frac{R^2}{(y^2-1)(x-y)^2} (\eta'_5)^2 + Z_4
\ds\frac{R^2}{(1-x^2)(x-y)^2} (\eta'_6)^2 + (\eta'_i\eta'_i)
\bigg] \,.
\end{eqnarray}
The expression for the entropy coming from the shape oscillations
now becomes:
\begin{multline}
S ~=~ 2\pi \sqrt{\ds\frac{3}{2}} \bigg\{
\big[N_1^{ST}-\ds\frac{1}{2}n_2^{ST}n_3 (1+y) \big] \, \big[
N_3^{ST}- \ds\frac{1}{2}n_2^{ST}n_1 (1+y) \big]~-~ (n_2^{ST})^2
\ds\frac{Z_2R^2(y^2-1)}{(x-y)^2}\bigg\}^{1 \over 2}
\end{multline}
Note that for a supertube located near the black ring
($y\rightarrow-\infty$) one has a huge entropy enhancement due to
the dipole-dipole interaction.

For completeness, it is equally easy to consider  a round
supertube orthogonal to the black ring ($t = \xi^0,~
z=\xi^1,~\varphi_2=-\theta$).  One then finds that the entropy of
the shape modes is:
\begin{multline}
S = 2\pi \sqrt{\ds\frac{3}{2}} \bigg\{
\big[N_1^{ST}+\ds\frac{1}{2}n_2^{ST}n_3 (x+c) \big] \, \big[
N_3^{ST}+ \ds\frac{1}{2}n_2^{ST}n_1 (x+c) \big]~-~(n_2^{ST})^2
\ds\frac{Z_2R^2(1-x^2)}{(x-y)^2}\bigg\}^{1 \over 2}
\label{BRResult}
\end{multline}
While there is still a dipole-dipole interaction, the entropy
enhancement does not grow arbitrarily large because the coordinate
$x$ has a finite range ($x\in (-1,1)$).

\subsection{Solution with a general Gibbons-Hawking base}

For the sake of completeness, it is worth reviewing also the
entropy enhancement for a supertube in a three-charge background
with a Gibbons-Hawking base. For this background, one can only
calculate easily the entropy coming from the internal fluctuations
of the supertube. The entropy coming from fluctuations of the
supertube in the spacetime directions is more complicated than for
the black ring background.

For this background the supertube action becomes:
\begin{multline}
S = T_{D2} \int d^3\xi  \,\bigg\{ \left[ \left(\ds\frac{1}{Z_1} -
1\right) \mathcal{F}_{z\theta} + \ds\frac{K^3}{Z_1V} + \left(
\ds\frac{\mu}{Z_1} - \ds\frac{K^1}{V} \right)(\mathcal{F}_{tz}-1)
\right] \\\\
- \bigg[ \ds\frac{1}{V^2Z_1^2} \big[(K^3  -  V
(\mu(1-\mathcal{F}_{tz})- \mathcal{F}_{z\theta}) )^2    + V Z_1Z_2
(1-\mathcal{F}_{tz})(2-Z_3(1-\mathcal{F}_{tz})) \big]
\bigg]^{1/2}\bigg\}  \,.
\end{multline}
Because of the complexity of this background, we consider small
shape oscillations in the compactification manifold, $T^4$, around
a round supertube along the GH fiber :
\begin{equation}
t= \xi^0 ~, \qquad z=\xi^1~,\qquad \psi=\theta~,\qquad\qquad
x_i\rightarrow x_i + \eta_{i} (t,\theta) \qquad i=1,2,3,4 \,.
\end{equation}

The quantization proceeds exactly as before and the conserved
electric charges are now:
\begin{equation}
N_{1}^{ST} =  \frac{T_{D2}}{T_{D0}}  \ds\int_{0}^{2\pi L_z}dz
\ds\int_{0}^{2\pi n_2^{ST}}d\theta \, \mathcal{F}_{z \theta} =
n_2^{ST}\mathcal{F}_{z\theta} \,, \label{None}
\end{equation}
\begin{equation}
N_{3}^{ST} =  \frac{T_{D2}}{T_{F1}} \ds\int_{0}^{2\pi
n_2^{ST}}d\theta \left[ -\ds\frac{K^1}{V} +
\ds\frac{1}{\mathcal{F}_{z\theta} + V^{-1}K^3} \left(
\ds\frac{Z_2}{V} + \ds\sum_{i}^{4}(\eta'_i)^2 \right) \right]  \,.
 \label{Nthree}
\end{equation}
Substituting (\ref{etamodes}) into (\ref{Nthree})  and rearranging
using (\ref{None})  leads to:
\begin{multline}
\ds\sum_{i=1}^{4}\ds\sum_{k>0} k (a^i_k)^{\dagger} a^i_k = L_z
T_{D2} \int_{0}^{2\pi n_2^{ST}} d\theta \int_{0}^{2\pi n_2^{ST}}
d\theta \ds\sum_{i=1}^{4} \eta'_i\eta'_i \\ = \left[ N_{1}^{ST} +
n_2^{ST} \ds\frac{K^3}{V}\right]\left[ N_{3}^{ST} + n_2^{ST}
\ds\frac{K^1}{V}\right]  - (n_2^{ST})^2 \ds\frac{Z_2}{V}
\,.\label{EnResult}
\end{multline}
and this leads to the following expression for the entropy:
\begin{equation}
S = 2\pi \sqrt{\left[ N_{1}^{ST} +  n_2^{ST}
\ds\frac{K^3}{V}\right]\left[ N_{3}^{ST} +  n_2^{ST}
\ds\frac{K^1}{V}\right]  - (n_2^{ST})^2 \ds\frac{Z_2}{V} } \,.
\label{GHResult}
\end{equation}
%

\subsection{Comments on the supertube effective charges}

As we have seen, in flat space and in a BMPV black hole
background, the entropy of the two-charge supertube, when
expressed in terms of its charges, is simply
\begin{equation}
S  ~\sim~ \sqrt{ Q_{1} \, Q_{3}  ~-~ J } \,. \label{Stubeentsimp}
\end{equation}
However, if the background has non-trivial dipole magnetic fields
the entropy is given by equations (\ref{BRResult}) and
(\ref{GHResult}), and can be written as:
\begin{equation}
S  ~\sim~ \sqrt{ Q_{1}^{eff} \, Q_{3}^{eff}  ~-~ J^{eff}} \,.
\label{Stubeent}
\end{equation}
Here the {\it effective charges}, $Q_{I}^{eff} $ and $J^{eff}$,
involve a non-trivial interaction between the dipoles of the
supertube and the dipoles of the background.  These effective
charges can become arbitrarily large if the supertube moves
suitably close to the background dipole sources.

From the perspective of the supertube DBI--WZ action, these
effective charges are:
\begin{equation}
Q_{1}^{eff} ~\equiv~  Q_{1}^{ST}  ~+~ n_2^{ST} \, \tilde \xi^{(1)}
\,, \qquad Q_{3}^{eff} ~\equiv~  Q_{3}^{ST} ~+~ n_2^{ST} \,
\tilde\xi^{(2)} \,, \label{DBIQeff}
\end{equation}
where the $\xi^{(I)}$ are defined in (\ref{vecpotdefns}) and
$\tilde {\xi}^{(I)}$ denotes the pull-back onto the supertube.
There is another way to think about these effective charges when
considering the fully back-reacted solution found for a round
supertube in GH backgrounds \cite{enhance} -- they give the
leading divergence of the warp factors $ Z_I$ near the supertube:
\begin{equation}
Q_I^{eff} ~\equiv~ 4 \, \lim_{r_N \to 0} \,  r_N \, Z_I
\,, \qquad I =1,3\,, \label{QIeffdefn}
\end{equation}
where the supertube is located at $r_N = 0$. Nicely enough, even
if the DBI--WZ action of the supertube is perturbative, it does
capture these effective charges via the pull back in
(\ref{DBIQeff}).

As discussed in \cite{enhance}, the crucial insight coming from
this analysis is that the entropy of the supertube is not
determined in terms of its asymptotic charges (measured at
infinity) but in terms of its local effective charges, which
depend on the location of the supertube. Hence, the entropy can
become very large when the magnetic fields are very strong -- this
happens for example when the supertube is near the horizon of a
black ring, or when it is in a deep scaling horizonless solution 
\cite{Bena:2006kb,Bena:2007qc,denef1,Denef:2007vg}.

Our analysis also demonstrates that entropy enhancement affects
{\it both} the fluctuations of the supertube in the internal
(torus) directions, as well as the fluctuations of the supertube
in the non-compact transverse space. In a general three-charge
background the latter are very hard to analyze, as the non-trivial
magnetic field mixes the fluctuation modes. However, in a black
ring background this mixing is not present for the supertube
fluctuations in the plane transverse to the ring. Our calculation
shows that these fluctuations exhibit {\it the same} amount of
entropy enhancement as the torus fluctuations, and hence indicate
that entropy enhancement is a feature of all the supertube modes,
and not just some. It would be interesting to calculate whether,
in a general background, some modes are more enhanced than others,
as this would indicate whether the typical microstates of
``enhanced'' fluctuating supertubes are smooth in supergravity or
not.

\section{Conclusions}

Our purpose in this paper has been four-fold:

First, we proved that if one takes supertubes that are solutions
of the Born-Infeld action to a regime of parameters where their
back-reaction is important, the fully back-reacted supergravity
solution is smooth in the duality frame where the supertubes have
D1 and D5 electric charges. The two conditions necessary for the
supergravity solution to be free of closed timelike curves and to
be smooth are reproduced exactly by the Born-Infeld analysis.

Our analysis strengthens the case for the existence of families of
supergravity solutions that have the same charges as black holes,
and that depend on arbitrary continuous functions (and hence have
a moduli space of infinite dimension). Furthermore, these
solutions are smooth and horizonless in the regime of parameters
in which the corresponding black hole has a macroscopic horizon.

The second purpose of the paper has been to identify the relation
between the charges of supertubes and black rings that appear in
the exact supergravity description, and those that appear in the
microscopic (Born Infeld) description.

We have seen in Section \ref{FourDcharges} that a given
five-dimensional black ring can be embedded in Taub-NUT in two
ways, that differ from each other by the choice of the location of
the Dirac string in the gauge potentials. One can furthermore find
black ring embeddings with multiple Dirac strings, that depend on
several parameters, and these can be related to each other by
gauge transformations. The Gibbons-Hawking charges of the black
ring, which give the electrical charges of the corresponding
four-dimensional black hole, are different in different patches
(\ref{BRGHcharges},\ref{BRGHcharges2}).  Nevertheless, the $E_{7(7)}$
quartic that gives the microscopic entropy of the black ring, is
independent of the choice of patch.

It is interesting to note that the entropy of extremal non-BPS
black rings has been recently expressed in terms of the $E_{7(7)}$
quartic invariant as a function of the asymptotic charges, and a
certain angular momentum parameter $J$  \cite{Emparan}. Our
analysis establishes that the apparent four-dimensional charges
(that appear in this invariant) depend on the location of the
Dirac string, and that one can switch between the asymptotic
charges of the ring and the intrinsic charges by a gauge
transformation. This transformation nevertheless also changes the
angular momentum parameter, and thus the question that should be
asked in trying to find the microscopic description of extremal
non-BPS black rings is not ``Why does a certain charge appear in
the quartic invariant?'' but rather ``Why, for a given choice of
charges, does a certain angular momentum parameter appear in the
quartic  invariant?''

We have also found the relation between the charges of supertubes
that appear in their Born-Infeld description, and those that
appear in their supergravity description. We have established that
if a supertube that gives rise to a solution with a
Gibbons-Hawking base is put at a smooth location\footnote{More
precisely, not exactly on top of a Dirac string.}, its Born-Infeld
electric charges are equal to the Gibbons-Hawking charges of the supergravity
solution. Since the Gibbons-Hawking charges are the ones that
contributes to the asymptotic charge of a solution, and since these
charges are much smaller than the enhanced charges (that give the
supertube entropy in a three-charge background) our analysis
definitively establishes the phenomenon of entropy enhancement: a
given two-charge supertube in a three-charge two-dipole charge
background has an entropy much larger than one would expect from
the amount of charge visible from infinity.

The third aim of our paper has been to analyze issues related to
black-hole thermodynamics and chronology protection when a
supertube is merged with a black ring. If supertubes respect the
triholomrphic $U(1)$ isometry of the ring, and are able to merge
with a black ring, then this neither  decreases the ring entropy
nor creates closed timelike curves. The supertubes that might do
this, and hence are ``dangerous'' for chronology protection and
thermodynamics, are unable to merge with the ring.

The situation is a bit more subtle with supertubes that {\it do not}
respect the triholomorphic $U(1)$ isometry of the ring, and wind
around $S^1$ latitude circles in the $S^2$ of the black ring
horizon. We have found that if the charge these supertubes carry
into a black ring is given by their Born-Infeld charge, then
chronology protection and black-hole thermodynamics can be violated!
The only way these are not violated is if the charge brought into
the black ring depends continuously on the angle at which the
supertube merges with the ring (which is the angle of the $S^1$
latitude circle it wraps). It would be interesting to understand the
origin of this very puzzling fact, by constructed the fully
back-reacted solution corresponding to this merger. This solution
will have a $U(1)$ isometry, but not a triholomorphic one, and will
hence not be a Gibbons-Hawking solution, but a more general one of
the type constructed in \cite{Bena:2007ju,Bena:2005zy,mann}.

The fourth aim of the paper was to extend the entropy enhancement
calculation of \cite{enhance} to supertubes that oscillate both in
the internal compact directions and in spacetime non-compact
directions. Such a calculation is generically quite complicated:
if a solution depends on these directions, this mixes the
corresponding oscillator modes of the supertube, which makes the
counting much more involved. Nevertheless, we have found a class
of examples in which this mixing is not present, and the
calculation of the entropy coming from the spacetime modes of the
supertube is as simple as that coming from the internal modes.

Our results show that the two kind of modes contribute to the
enhanced entropy equally, despite the presence of different
(large) factors in the mode expansions. If, as we expect, the
entropy coming from these fluctuations will be black-hole-like,
and therefore the fluctuating supertubes will give the typical
microstates of the corresponding black hole, these microstates
will have a non-trivial transverse size, and the smooth
horizonless microstates will act as representatives for all the
black hole microstates \cite{enhance,Warner:2008ma}.

The obvious question left unanswered by our analysis is what is
the enhanced entropy coming from the modes that mix. This question
requires a more tedious analysis than we have done, but its answer
could have dramatic consequences.  If this enhanced entropy is
equal or less than that coming from the internal modes, then most
likely the typical black-hole microstate  geometries will be given
by a combination of internal and transverse space oscillations,
which in general will not be smooth (but may have smooth
representatives). However, if the entropy coming from the
transverse modes that mix is greater than the one coming from the
internal directions, then the typical microstates might all be
given by smooth horiozonless supergravity solutions.

To recapitulate, we have proven that the supergravity and the Born
Infeld descriptions of supertube agree, found the four-dimensional
charges of five-dimensional black rings and supertubes, analyzed
chronology protection and black hole thermodynamics during
black-ring supertube mergers, and established that the entropies
of supertube modes in the internal directions in the spacetime
directions are enhanced equally, and hence these modes contribute
equally to the entropy of the supertube.

We have also filled in a  few details in the analysis of
supertubes and black rings solutions: we have dualized the black
ring and the more general multi-center solutions with a
Gibbons-Hawking base to various duality frames (in Appendix A), and have found (to
our knowledge for the first time) the exact form of the magnetic
potentials in these solutions. We have also calculated (in Appendix C) the angular
momenta of a supertube of arbitrary shape in a general solution
with an $\IR^4$ base, and shown that the contribution of a piece
of an arbitrarily-shaped supertube to the angular momentum along
the direction of this piece is the same as for a piece of a
circular supertube, and is in fact a universal quantity, as
suggested also by the supergravity analysis. 

Last, but not least, we have shown (in Appendix B) that all the three-charge,
three-dipole charge solutions with a Gibbons-Hawking base
constructed so far can be dualized to the duality frame where they
have D1, D5 and momentum charges, and can be scaled\footnote{This
has been done before for black rings \cite{Bena:2004tk}, but not
for general multi-center solutions.} in such a way as to become
asymptotically $AdS_3 \times S^3$. Hence all these smooth
horizonless solutions are dual via the $AdS$/CFT correspondence to
microstates of the D1-D5 CFT. It would be very interesting to
extend the holographic methods of \cite{Skenderis:2006ah} (that
were successfully used in \cite{Kanitscheider:2007wq} for
two-charge microstates) to the analysis of these three-charge
geometries. This would enable one to establish whether the
geometries constructed so far are dual to typical CFT microstates,
whether the geometries dual to these microstates have Planck-scale
curvature or are well-described in supergravity, and whether the
smooth microstate geometries constructed so far can act as
representatives of the typical microstates.

\newpage
\bigskip
\leftline{\bf Acknowledgements}
\smallskip
We would like to thank Roberto Emparan, Stefano Giusto, and Ashish Saxena for
interesting discussions. NB and NPW are grateful to the
IPhT(SPhT), CEA-Saclay for kind hospitality while this work was
completed. IB is grateful to the CERN Theory Institute for
hospitality during the Black Hole workshop. The work of NB and NPW
was supported in part by DOE grant DE-FG03-84ER-40168. The work of
IB and CR was supported in part by the DSM CEA-Saclay, by the ANR
grants BLAN 06-3-137168 and JCJC ERCS07-12, and by the Marie Curie IRG
046430. The work of NB was also supported by the John Stauffer
Fellowship, the Dean Joan M. Schaefer Research Scholarship and the
USC International Summer Field Research Award.

\section*{Appendix A. Three charge solutions and T-duality}
\appendix
\renewcommand{\theequation}{A.\arabic{equation}}
\setcounter{equation}{0} \addcontentsline{toc}{section}{Appendix
A. Three charge solutions and T-duality}

\subsection*{Appendix A1. T-duality transformations}
\appendix
\renewcommand{\theequation}{A1.\arabic{equation}}
\setcounter{equation}{0}
\addcontentsline{toc}{subsection}{Appendix A1. T-duality
transformations}

In this Appendix we summarize the T-duality transformation rules
for type II theories with non-zero RR fields. These rules are
derived in \cite{Bergshoeff:1995as} and can be considered a
generalization of the Buscher rules \cite{Buscher:1987sk}. In the
expressions below we will adopt the conventions and notation of
\cite{Myers:1999ps}, the different RR forms are denoted with
$C^{(n)}$ and the fields obtained after the T-duality
transformations are denoted with a tilde, $w=x_{9}$ is the
M-theory compactification direction and $x$ is the T-duality
direction.

The set of fields in the low energy limit of M-theory, {\it i.e.}
eleven-dimensional supergravity, are:
\begin{equation}
G_{\mu\nu} \qquad \text{and} \qquad A_{\mu\nu\rho} \,.
\end{equation}
After the compactification along $w=x_{9}$ we are left with type
IIA supergravity with the fields
\begin{equation}
g_{\mu\nu} , \qquad C^{(3)}_{\mu\nu\rho}, \qquad B_{\mu\nu},
\qquad C^{(1)}_{\mu}, \qquad \Phi \,,
\end{equation}
which are related to the eleven-dimensional fields as follows
(note that we are working in string frame):
\begin{equation}
\begin{array}{l}
g_{\mu\nu} = \sqrt{G_{ww}} \left( G_{\mu\nu}  + \ds\frac{ G_{\mu
w} G_{\nu w} }{ G_{ww} } \right)\,,  \qquad\qquad  C^{(1)}_{\mu} =
\ds\frac{ G_{\mu w}}{G_{ww}}\,, \\\\
C^{(3)}_{\mu\nu\rho} = A_{\mu\nu\rho} \,, \qquad\qquad B_{\mu\nu}
= A_{\mu\nu w} \,, \qquad\qquad \Phi = \ds\frac{3}{4}\log(G_{ww})
\,.
\end{array}
\end{equation}
The type IIB fields are:
\begin{equation}
g_{\mu\nu}, \qquad B_{\mu\nu}, \qquad \Phi, \qquad C^{(0)}, \qquad
C^{(2)}_{\mu\nu}, \qquad C^{(4)}_{\mu\nu\rho\sigma} \,.
\end{equation}
The T-duality rules for the metric and the NS-NS fields are:
\begin{equation}
\begin{array}{l}
\tilde{g}_{xx} = \ds\frac{1}{g_{xx}}, \qquad \tilde{g}_{\mu x} =
\ds\frac{B_{\mu x}}{g_{xx}}, \qquad \tilde{g}_{\mu\nu} =
g_{\mu\nu}
- \ds\frac{g_{\mu x} g_{\nu x} - B_{\mu x}B_{\nu x}}{g_{xx}}\,,  \\\\
\tilde{B}_{\mu x} = \ds\frac{g_{\mu x}}{g_{xx}} ,\qquad
\tilde{B}_{\mu\nu} = B_{\mu\nu} - \ds\frac{B_{\mu x} g_{\nu x} -
g_{\mu x}B_{\nu x}}{g_{xx}}\,, \qquad \tilde{\Phi} = \Phi -
\ds\frac{1}{2}\log g_{xx}  \,.
\end{array}
\end{equation}
The RR forms transform under T-duality as:
\begin{equation}
\begin{array}{l}
\tilde{C}^{(n)}_{\mu...\nu\alpha x} = C^{(n-1)}_{\mu...\nu\alpha}
- (n-1)\ds\frac{C^{(n-1)}_{[\mu...\nu | x} g_{|\alpha]x}}{g_{xx}}\,, \\\\
\tilde{C}^{(n)}_{\mu...\nu\alpha\beta} =
C^{(n+1)}_{\mu...\nu\alpha\beta x} + n
C^{(n-1)}_{[\mu...\nu\alpha}B_{\beta]x} +
n(n-1)\ds\frac{C^{(n-1)}_{[\mu...\nu | x}B_{|\alpha|x}
g_{|\beta]x}}{g_{xx}} \,.
\end{array}
\end{equation}
Alternatively one can transform the RR field strengths as follows
(for a detailed derivation of these rules see Appendix A of
\cite{Kanitscheider:2007wq})
\begin{equation}
\begin{array}{l}
\tilde{F}^{(n)}_{\mu_1...\mu_{n-1} x} =
F^{(n-1)}_{\mu_1...\mu_{n-1}} +
(n-1)(-1)^n\ds\frac{ g_{x[\mu_1} F^{(n-1)}_{\mu_2...\mu_{n-1}] x} }{g_{xx}}\,, \\\\
\tilde{F}^{(n)}_{\mu_1...\mu_n} = F^{(n+1)}_{\mu_1...\mu_n x} -
n(-1)^n B_{x[\mu_1} F^{(n-1)}_{\mu_2...\mu_n]} -
n(n-1)\ds\frac{B_{x[\mu_1}g_{\mu_2|x|}F^{(n-1)}_{\mu_3..\mu_n]x}}{g_{xx}}
\,.
\end{array}
\end{equation}

We now give the explicit transformations that take us from the
M-theory duality frame in Section 2, to solutions in other useful
duality frames. In Table 1 we specify the directions along which
the M2 branes and the M5 branes are wrapped or smeared.

\begin{table}[h]
 \centering
\begin{tabular}{|c|c|c|c|c|c|c|c|c|c|c|c|}
\hline
 & 0 & 1 & 2 & 3 & 4 & 5 & 6 & 7 & 8 & 9 & 10\\
\hline
 (D1)\, M2& $\updownarrow$ & $\bullet$ & $\bullet$ & $\bullet$ & $\bullet$ & $\updownarrow$ & $\updownarrow$ & $\leftrightarrow$ & $\leftrightarrow$ & $\leftrightarrow$ & $\leftrightarrow$\\
\hline
 (D5)\, M2& $\updownarrow$ & $\bullet$ & $\bullet$ & $\bullet$ & $\bullet$ & $\leftrightarrow$ & $\leftrightarrow$ & $\updownarrow$ & $\updownarrow$ & $\leftrightarrow$ & $\leftrightarrow$\\
\hline
 (P)\, M2& $\updownarrow$ & $\bullet$ & $\bullet$ & $\bullet$ & $\bullet$ & $\leftrightarrow$ & $\leftrightarrow$ & $\leftrightarrow$ & $\leftrightarrow$ & $\updownarrow$ & $\updownarrow$\\
\hline
 (d5)\, M5& $\updownarrow$ & $y^{\mu}(\phi)$ & $y^{\mu}(\phi)$ & $y^{\mu}(\phi)$ & $y^{\mu}(\phi)$ & $\leftrightarrow$ & $\leftrightarrow$ & $\updownarrow$ & $\updownarrow$ & $\updownarrow$ & $\updownarrow$\\
\hline
 (d1)\, M5& $\updownarrow$ & $y^{\mu}(\phi)$ & $y^{\mu}(\phi)$ & $y^{\mu}(\phi)$ & $y^{\mu}(\phi)$ & $\updownarrow$ & $\updownarrow$ & $\leftrightarrow$ & $\leftrightarrow$ & $\updownarrow$ & $\updownarrow$\\
\hline
 (kkm)\, M5& $\updownarrow$ & $y^{\mu}(\phi)$ & $y^{\mu}(\phi)$ & $y^{\mu}(\phi)$ & $y^{\mu}(\phi)$ & $\updownarrow$ & $\updownarrow$ & $\updownarrow$ & $\updownarrow$ & $\leftrightarrow$ & $\leftrightarrow$\\
\hline
\end{tabular}
\caption{\textit{The configuration of branes in M-theory that
preserves the four supersymmetries of the M2-M2-M2 three-charge
black hole \cite{Bena:2004de}. The vertical arrows represent the
directions along which the branes are extended and the horizontal
arrows represent smearing directions. The functions
$y^{\mu}(\phi)$ describe a closed curve which is wrapped by the M5
branes. In the first column we have indicated also the brane
identification in the D1-D5-P duality frame.}}
\end{table}
%

\subsection*{Appendix A2. Three charge solutions in different
duality frames}
\renewcommand{\theequation}{A2.\arabic{equation}}
\setcounter{equation}{0}
\addcontentsline{toc}{subsection}{Appendix A2. Three charge
solutions in different duality frames}

\bigskip

\noindent{\bf Compactification along $x_{9}$ }

\bigskip

The first step is to compactify the eleven-dimensional solution,
presented in Section 2, along $x_{9}$, in this way we obtain the
following combination of ``electric"\footnote{We are choosing
$x_9$ to be the M-theory circle in order to match the conventions
in the literature for the global signs of the B-field and the RR
potentials for the BMPV black hole \cite{Breckenridge:1996is} and
the supersymmetric black ring solutions \cite{Elvang:2004ds}.}
\begin{equation}
N_1: \, D2 \, (56) \qquad  N_2: \, D2 \, (78) \qquad N_3: \, F1
\,(z)
 \label{D2D2F1}
\end{equation}
and ``dipole" branes
\begin{equation}
n_1: \, D4 \, (y78z) \qquad  n_2: \, D4 \, (y56z) \qquad n_3: \,
NS5 \, (y5678) \label{d4d4ns5}
\end{equation}
in Type IIA. From now on we will denote $x_{10}=z$. The
ten-dimensional \textit{string frame} metric is
\begin{equation}
ds^2_{10} = - \ds\frac{1}{Z_3\sqrt{Z_1Z_2}}(dt+k)^2 +
\sqrt{Z_1Z_2}ds^2_4 +\ds\frac{\sqrt{Z_1Z_2}}{Z_3}dz^2 +
 \ds\sqrt{\frac{Z_2}{Z_1}}(dx_5^2+dx_6^2) +
 \ds\sqrt{\frac{Z_1}{Z_2}}(dx_7^2+dx_8^2) \label{D2D2F1metric}
\end{equation}
The dilaton and the Kalb-Ramond field are
\begin{equation}
\Phi = \ds\frac{1}{4} \log \left( \ds\frac{Z_1Z_2}{Z_3^2}\right),
\qquad\qquad B = -A^{(3)}\wedge dz \,. \label{D2D2F1NSNS}
\end{equation}
The RR (``electric") forms are
\begin{equation}
C^{(1)} = 0\,, \qquad\qquad C^{(3)} = A^{(1)}\wedge dx_5 \wedge
dx_6 + A^{(2)}\wedge dx_7 \wedge dx_8 \,,
\label{D2D2F1RRpotentials}
\end{equation}
and the four-form field strength is\footnote{Note that we are
using the notation of \cite{Polchinski:1998rr}
$\widetilde{F}^{(4)} = dC^{(3)} + dB \wedge C^{(1)}$.}
\begin{eqnarray}
\widetilde{F}^{(4)} &=& dC^{(3)} + dB \wedge  C^{(1)} ~=~
A^{(1)}\wedge dx_5 \wedge dx_6  + dA^{(2)}\wedge dx_7 \wedge dx_8 \\
&=& d\mathcal{F}^{(1)}\wedge dx_5 \wedge dx_6  +
\mathcal{F}^{(2)}\wedge dx_7 \wedge dx_8 \,,
 \label{D2D2F1RR}
\end{eqnarray}
where we have used the notation $\mathcal{F}^{(I)} = dA^{(I)}$.
Now we will perform a chain of T-dualities in order to arrive at
the desired frame.

\bigskip
\noindent{\bf T-duality along $x_5$}
\bigskip

A T-duality along the $x_5$ direction brings us to Type IIB with
the following sets of ``electric"
\begin{equation}
N_1: \, D1 \, (6) \qquad  N_2: \, D3 \, (578) \qquad N_3: \, F1 \,
(z) \label{D1D3F1}
\end{equation}
and ``dipole" branes
\begin{equation}
n_1: \, D5 \, (y578z) \qquad  n_2: \, D3 \, (y6z) \qquad n_3: \,
NS5 \, (y5678) \,. \label{d5d3ns5}
\end{equation}
The metric is
\begin{equation}
ds^2_{10} = - \ds\frac{1}{Z_3\sqrt{Z_1Z_2}}(dt+k)^2 +
\sqrt{Z_1Z_2}ds^2_4 +\ds\frac{\sqrt{Z_1Z_2}}{Z_3}dz^2 +
 \ds\sqrt{\frac{Z_2}{Z_1}}dx_6^2 +
 \ds\sqrt{\frac{Z_1}{Z_2}}(dx_5^2+dx_7^2+dx_8^2)\,.
 \label{D1D3F1metric}
\end{equation}
The other NS-NS fields are
\begin{equation}
\Phi = \ds\frac{1}{4} \log \left( \ds\frac{Z_1^2}{Z_3^2}\right),
\qquad\qquad B = -A^{(3)}\wedge dz  \,. \label{D1D3F1NSNS}
\end{equation}
The RR field strengths are
\begin{equation}
\begin{array}{c}
F^{(3)} = - \mathcal{F}^{(1)}\wedge dx_6\,, \\\\
\widetilde{F}^{(5)} =  \mathcal{F}^{(2)} \wedge dx_5 \wedge dx_7
\wedge dx_8 + \star_{10} ( \mathcal{F}^{(2)} \wedge dx_5 \wedge
dx_7 \wedge dx_8 ) \,,
\end{array} \label{D1D3F1RR}
\end{equation}
where in the expression for $\widetilde{F}^{(5)}$  we have added
the Hodge dual piece by hand to ensure self-duality
\cite{Schwarz:1983qr}. Note that if one is working in the
``democratic formalism" ({\it i.e.} with both electric and
magnetic field strengths) $\widetilde{F}^{(5)}$ will be
automatically self-dual, however since we have chosen to T-dualize
explicitly only the electric field strengths we have to add the
self-dual piece by hand whenever we encounter a five-form field
strength after T-dualizing a four-form field strength.

Using the form of the ten-dimensional metric (\ref{D1D3F1metric})
one can show that
\begin{equation}
\star_{10} ( dA^{(2)} \wedge dx_5 \wedge dx_7 \wedge dx_8 ) = -
\left(\ds\frac{Z_2^5}{Z_1^3Z_3^2}\right)^{1/4} \star_5 (dA^{(2)}
\wedge dz \wedge dx_6) \,, \label{10to5hodge}
\end{equation}
where $\star_5$ is the Hodge dual on the five-dimensional subspace
given by the metric
\begin{equation}
ds^2_{5} = - \ds\frac{1}{Z_3\sqrt{Z_1Z_2}}(dt+k)^2 +
\sqrt{Z_1Z_2}ds^2_4~. \label{App5dmetric}
\end{equation}

\bigskip

\noindent{\bf T-duality along $x_6$}
\bigskip

Now perform T-duality along $x_6$ to get
\begin{equation}
N_1: \, D0   \qquad  N_2: \, D4 \, (5678) \qquad N_3: \, F1 \, (z)
\label{D0D4F1-app}
\end{equation}
``electric"
\begin{equation}
n_1: \, D6 \, (y5678z) \qquad  n_2: \, D2 \, (yz) \qquad n_3: \,
NS5 \, (y5678) \label{d6d2ns5-app}
\end{equation}
and ``dipole" branes in Type IIA. The metric is
\begin{equation}
ds^2_{10} = - \ds\frac{1}{Z_3\sqrt{Z_1Z_2}}(dt+k)^2 +
\sqrt{Z_1Z_2}ds^2_4 +\ds\frac{\sqrt{Z_1Z_2}}{Z_3}dz^2 +
\ds\sqrt{\frac{Z_1}{Z_2}}(dx_5^2+dx_6^2+dx_7^2+dx_8^2) \,.
\label{D0D4F1metric-app}
\end{equation}
The dilaton and the Kalb-Ramond fields are
\begin{equation}
\Phi = \ds\frac{1}{4} \log \left(
\ds\frac{Z_1^3}{Z_2Z_3^2}\right), \qquad\qquad B = -A^{(3)}\wedge
dz \,. \label{D0D341NSNS-app}
\end{equation}
The RR field strengths are
\begin{equation}
 F^{(2)} = - \mathcal{F}^{(1)}\,, \qquad \qquad
\widetilde{F}^{(4)} =  -
\left(\ds\frac{Z_2^5}{Z_1^3Z_3^2}\right)^{1/4} \star_5
(\mathcal{F}^{(2)}) \wedge dz \,.
 \label{D0D4F1RR-app}
\end{equation}
Since we are interested in studying probe two charge supertubes in
this background, we will also need the RR potentials since they
enter the Wess-Zumino action of the supertube.

\bigskip

\noindent{\bf Finding the RR and NS-NS potentials in the D0-D4-F1
frame}

\bigskip

If everything is consistent, then the Bianchi identities for the
field strengths should be satisfied. For the solution given by
\eqref{D0D4F1metric}--\eqref{D0D4F1RR}, the non-trivial Bianchi
identity is:\footnote{See \cite{Polchinski:1998rr} p. 86.}
\begin{equation}
d\widetilde{F}^{(4)} = -F^{(2)} \wedge dB \,.
\end{equation}
Indeed we can use the BPS equations to show that
\begin{eqnarray}
d\widetilde{F}^{(4)} &=& - d
\bigg(\bigg(\ds\frac{Z_2^5}{Z_1^3Z_3^2} \bigg)^{1/4} \star_5
(\mathcal{F}^{(2)}) \bigg) \wedge dz \notag\\\\ &=& - \bigg[ d
\bigg( \ds\frac{1}{Z_1Z_3} \bigg) \wedge dk \wedge (dt+k) - d
\bigg( \ds\frac{(dt+k)}{Z_1} \bigg) \wedge \Theta^3 \\
&& \qquad  \qquad- d \bigg( \ds\frac{(dt+k)}{Z_3} \bigg) \wedge
\Theta^1 + \Theta^3\wedge\Theta^1 \bigg]\wedge dz \,. \notag
\end{eqnarray}
On the other hand
\begin{eqnarray}
F^{(2)} \wedge dB &=& dA^{(1)} \wedge dA^{(3)} \wedge  dz
\notag\\\\ &=& \bigg[d \bigg( \ds\frac{1}{Z_1Z_3} \bigg) \wedge dk
\wedge (dt+k) - d \bigg( \ds\frac{(dt+k)}{Z_1} \bigg) \wedge
\Theta^3 \\
& & \qquad  \qquad- d \bigg( \ds\frac{(dt+k)}{Z_3} \bigg) \wedge
\Theta^1 + \Theta^3\wedge\Theta^1 \bigg]\wedge dz \,. \notag
\end{eqnarray}
So the Bianchi identity is obeyed and it can be checked in a
similar manner that the equations of motion of type IIA
supergravity are obeyed.  Thus confirms the consistency of our
calculations.

We will now find the RR three-form potential $C^{(3)}$ in the same
duality frame. It satisfies the following differential equation
\begin{equation}  d C^{(3)} ~\equiv~  \tilde{F}^{(4)} ~+~  C^{(1)}
\wedge H^{(3)}  \,.  \end{equation}
Note that this depends upon a gauge choice for $C^{(1)}$, we
choose a gauge in which $C^{(1)}$ is vanishing at asymptotic
infinity, namely\footnote{We have fixed $Z_I \sim 1 +
\mathcal{O}(r^{-1})$.}
\begin{equation}
C^{(1)} ~=~ - A^{1}~-~dt \,.
\end{equation}
Computing explicitly one finds
\begin{equation} d C^{(3)} ~=~ \big[  \big( - \star_4 d Z_2
~+~B^{(1)} \wedge \Theta^{(3)} \big) ~-~d \big(Z_3^{-1} ( dt +k)
\wedge B^{(1)}  ~+~ dt \wedge A^{(3)} \big)  \big] \wedge dx_5 \,,
\end{equation}
and hence
\begin{equation}  C^{(3)} ~=~ - \big(\gamma ~+~ Z_3^{-1} ( dt +k)
\wedge B^{(1)} ~+~ dt \wedge A^{(3)}  \big) \wedge dx_5  \,,
\end{equation}
where
\begin{equation} d\gamma  ~=~  \big(  \star_4 d Z_2 ~-~B^{(1)}
\wedge \Theta^{(3)} \big)  \,.
\end{equation}
So the calculation boils down to integrating for the 2-form
$\gamma$. Up to this stage we have not assumed any particular form
of the four-dimensional base space. If this space is
Gibbons-Hawking then the equation for $\gamma$ can be integrated
explicitly. Using the BPS supergravity solutions presented in
Section 2 it is not hard to show that
\begin{align}
\star_4 d Z_2 ~-~B^{(1)} \wedge \Theta^{(3)}  ~=~&
\left(-\partial_a Z_2 + K^1 \partial_a(V^{-1} K^3) \right)
\frac{1}{2} \epsilon_{abc} (d \psi + A) \wedge dy^b \wedge dy^c
\cr &~-~ \xi^{(1)}_a \,\big(
\partial_b(V^{-1} K^3) \big) (d \psi + A) \wedge dy^a \wedge dy^b
\cr &~+~ V \big(\vec \xi^{(1)} \cdot  \vec \nabla  (V^{-1} K^3)
\big)\, dy^1 \wedge dy^2 \wedge dy^3\,.
\end{align}
Recall that $Z_2 =L_2 + V^{-1} K^1 K^3$ and define $\vec\zeta$ by:
\begin{equation}
 \vec \nabla \times \vec \zeta   ~\equiv~ - \vec
\nabla L_2 \,, \label{zetadefn}
\end{equation}
then using
\begin{equation}
\Omega_{\pm}^{(a)} = \hat{e}^{1}\wedge \hat{e}^{a+1} \pm
\ds\frac{1}{2}\epsilon_{abc} \hat{e}^{b+1}\wedge \hat{e}^{c+1} \,,
\end{equation}
one can show that:
\begin{align}
\star_4 d Z_2 ~-~B^{(1)} \wedge \Theta^{(3)}  ~=~&
d\big[\big(-\zeta_a ~-~  V^{-1} K^3 \xi^{(1)}_a \big)
\Omega_-^{(a)} \big]  \cr & ~-~\big(V \, \vec \nabla \cdot \vec
\zeta  ~+~ K^3 \, \vec \nabla \cdot \vec \xi^{(1)}   \big) \, dy^1
\wedge dy^2 \wedge dy^3  \,.
\end{align}
The last term is a multiple of the volume form on $\IR^3$ and so
is necessarily exact, however, it can be simplified if we chose a
gauge for $\vec \xi^{(1)}$ and $ \vec \zeta$:
\begin{equation}  \vec \nabla \cdot \vec \zeta  ~=~  \vec \nabla
\cdot \vec \xi^{(1)} ~=~ 0\,.
\end{equation}
Then one has:
\begin{equation} \gamma ~=~  - \big[\big(\zeta_a ~+~  V^{-1} K^3
\xi^{(1)}_a  \big) \,
 \Omega_-^{(a)} \big]    \,.
\end{equation}
Finally, let $\vec r_i = (y_1 -a_i, y_2-b_i, y_3-c_i)$ and let $F
\equiv {1 \over r_i}$ and then define $\vec w$ by $\vec \nabla
\times \vec w   ~\equiv~ - \vec  \nabla F$, then the standard
solution for $\vec w$ is:
\begin{equation} w   ~=~  - {y_3 - c_i \over r_i} \,  {(y_1 - a_i ) \, dy_2 -
(y_2 - b_i ) \, dy_1 \over ((y_1 - a_i )^2  + (y_2 - b_i )^2)} \,.
\end{equation}
It is elementary to verify that $\vec \nabla \cdot \vec w =0$ and
so this is the requisite gauge. Finally the explicit form of the
RR three-form potential for a solution with GH base in the
D0-D4-F1 frame is
\begin{equation}
C^{(3)} ~=~ \big(\zeta_a ~+~  V^{-1} K^3 \xi^{(1)}_a  \big) \,
\Omega_-^{(a)}  \wedge dz  - \big(Z_3^{-1} ( dt +k) \wedge B^{(1)}
~+~ dt \wedge A^{(3)}  \big) \wedge dz \,.\label{C3IIAA}
\end{equation}

\bigskip

\noindent{\bf T-duality along $z$}

\bigskip

Another T-duality along $z$ transforms the system into D1-D5-P
frame with
\begin{equation}
N_1: \, D1 \,(z)  \qquad  N_2: \, D5 \, (5678z) \qquad N_3: \, P
\, (z) \label{D1D5PA}
\end{equation}
``electric"
\begin{equation}
n_1: \, D5 \, (y5678) \qquad  n_2: \, D1 \, (y) \qquad n_3: \, kkm
\, (y5678z) \label{d5d1kkmA}
\end{equation}
and ``dipole" branes. The metric is
\begin{equation}
ds^2_{IIB} = - \ds\frac{1}{Z_3\sqrt{Z_1Z_2}}(dt+k)^2  +
\sqrt{Z_1Z_2}ds^2_4 + \ds\frac{Z_3}{\sqrt{Z_1Z_2}}(dz+A^3)^2 +
\ds\sqrt{\frac{Z_1}{Z_2}}(dx_5^2+dx_6^2+dx_7^2+dx_8^2) \,.
\label{D1D5PmetricA}
\end{equation}
The dilaton and the Kalb-Ramond  field are:
\begin{equation}
\Phi = \ds\frac{1}{2} \log \left( \ds\frac{Z_1}{Z_2}\right),
\qquad\qquad B = 0\,. \label{D1D5PNSNSA}
\end{equation}
The RR three-form field strength (it is the only non-zero field
strength) is:
\begin{equation}
F^{(3)} = -\left(\ds\frac{Z_2^5}{Z_1^3Z_3^2}\right)^{1/4} \star_5
(\mathcal{F}^{(2)}) - \mathcal{F}^{(1)}\wedge (dz - A^{(3)}) \,.
\label{D1D5PRRA}
\end{equation}
For the supersymmetric black ring solution in D1-D5-P frame then
our general result agrees (up to conventions) with
\cite{Elvang:2004ds}. We can also easily find the RR 2-form
potential by T-dualizing (\ref{C3IIAA})
\begin{multline}
C^{(2)} ~=~ \left(\zeta_a ~+~  V^{-1} K^3 \xi^{(1)}_a  \right) \,
\Omega_-^{(a)}    - \left(Z_3^{-1} ( dt +k) \wedge B^{(1)} ~+~ dt
\wedge A^{(3)}  \right) \\+ A^{(1)}\wedge(A^{(3)}-dz -dt) ~+~
dt\wedge(A^3-dz) \,.
\end{multline}
%

\section*{Appendix B. BPS solutions in D1-D5-P frame and their decoupling limit}
\appendix
\renewcommand{\theequation}{B.\arabic{equation}}
\setcounter{equation}{0} \addcontentsline{toc}{section}{Appendix
B. BPS solutions in D1-D5-P frame and their decoupling limit}

In this Appendix we consider the decoupling limit of the
three-charge metric in the D1-D5-P duality frame
(\ref{D1D5PmetricA}). As shown in \cite{Bena:2004tk,Elvang:2004ds}, for a
supersymmetric black ring, such a limit takes an
asymptotically-flat solution into a solution that is
asymptotically $AdS_3 \times S^3 \times T^4$, and is thus dual to
a state or an ensemble of states in the D1-D5 CFT.

Like for three-charge black holes and black rings, one can take this limit by sending $\alpha' \to 0$ and scaling
the coordinates and the parameters of the solution in such a way that the
type IIB metric scales as $\alpha'$. At this point it is useful to
give the form of the ``electric" charges $Q_{I}$ in terms of the
parameters of the eleven-dimensional solution:
\begin{equation}
Q_{I} = -2 C_{IJK} \ds\sum_{j=1}^{N}
\ds\frac{\tilde{k}_{j}^{J}\tilde{k}_{j}^{K}}{q_{j}}  \qquad
\text{where} \qquad  \tilde{k}_{j}^{I} = k_{j}^{I} -
q_{j}\ds\sum_{i=1}^{N}k_{i}^{I} \,.
\end{equation}
The angular momenta are obtained by expanding the one-form $k$ at
infinity and one finds:
\begin{equation}
J_R \equiv J_1 + J_2 = C_{IJK}\ds\sum_{j=1}^{N}
\ds\frac{\tilde{k}_{j}^{I}\tilde{k}_{j}^{J}\tilde{k}_{j}^{K}}{q_{j}^2},
\qquad\qquad J_{L} = J_1 - J_2 = 8 \Big
|\ds\sum_{j=1}^{N}\ds\sum_{I=1}^3 \tilde{k}_{j}^{I}\vec{y}^{(j)}
\Big |  \,,
\end{equation}
where the $\vec{y}^{(j)}$ are the positions of the GH centers. The scaling with $\alpha'$ of the coordinates is the same as for the black hole solution 
\begin{equation}
y_1\sim y_2 \sim y_3 \sim (\alpha')^2 \,, \qquad x_a\sim
(\alpha')^{1/2}, \qquad a=5,6,7,8 \,, \qquad t \sim z \sim \psi
\sim (\alpha')^{0}
\end{equation}
where we have written the four-dimensional base  as a GH space \eqref{GHmetric}. 

The electric charges have also the same scaling as for the black hole:  
\begin{equation}
\qquad Q_1 \sim Q_2 \sim
\alpha', \qquad Q_3 \sim (\alpha')^2 \,.
\end{equation}
Hence, to preserve the fact that the charges of bubbling solutions come entirely from magnetic fluxes, the latter need to scale as
\be
k_{j}^{1} \sim k_{j}^{2} \sim \alpha', \qquad k_{j}^{3} \sim
(\alpha')^{0} 
\ee
In particular, we have $r^2 = y_1^2 + y_2^2 +y_3^2 $, so $r \sim
(\alpha')^2$. At infinity in the M-theory solution the functions
$Z_I$ behave like
\begin{equation}
Z_{I} \sim 1 + \ds\frac{Q_{I}}{4 r} + ...
\end{equation}
and so
\begin{equation}
Z_1 \sim \ds\frac{1}{\alpha'}  \qquad  Z_2 \sim
\ds\frac{1}{\alpha'} \qquad  Z_3 \sim \text{const}\,.
\end{equation}
So in the limit $\alpha' \to 0$ we can ignore the constant
 in $Z_1$ and $Z_2$ but we should keep it in $Z_3$.  It can be
shown that $k \sim A^3 \sim (\alpha')^0$ which finally leads to
the desired scaling
\begin{equation}
ds^{2}_{IIB} \sim \alpha' \,.
\end{equation}
After we have taken the $\alpha'  \to 0$ limit we can take the
large $r=\ds\frac{\rho^2}{4}$ limit and switch to four-dimensional
spherical polar coordinates (\ref{Flatmets}), with radial
coordinate $\rho$, in which we have:
\begin{eqnarray}
ds_{IIB}^2 &\sim& \ds\frac{\rho^2}{\sqrt{Q_1Q_2}}(-dt^2 + dz^2) +
\sqrt{Q_1Q_2}\ds\frac{d\rho^2}{\rho^2} \\
&& \qquad \qquad \qquad + ~\sqrt{Q_1Q_2} (d\vartheta^2
+\sin^2\vartheta d\varphi_1^2 + \cos^2\vartheta d\varphi_2^2) +
\sqrt{\ds\frac{Q_1}{Q_2}}ds^2_{T^4}
\end{eqnarray}
where we have used the freedom to change $A^3$ by pure gauge
transformations. This metric is indeed that of the product space
$AdS_3\times S^3 \times T^4$, where the radius of the $AdS_3$ and
the $S^3$ is the same and is equal to $(Q_1Q_2)^{1/4}$. So the
bubbling solutions in the decoupling limit are asymptotic to
$AdS_3\times S^3 \times T^4$ and thus should be described by the
D1-D5 CFT as expected\footnote{See \cite{deBoer:2008fk} for a
discussion of a different decoupling limit in which some of these bubbling solutions become dual to microstates of the MSW CFT \cite{msw}}

Note that the asymptotic metric in the decoupling limit of
\textit{any} of the BPS solutions of section \ref{3ChgSummary} is
the same as the metric of the three-charge BPS black hole in the decoupling
limit. This implies that the geometries we are analyzing have a
field theory description in the same D1-D5 CFT as the three-charge black
hole with identical electric charge.  The same result was found for supersymmetric black rings
\cite{Bena:2004tk,Elvang:2004ds}. 

We should also emphasize that in the
decoupling limit only the three-charge black holes and the two-charge
supertubes have metrics that are everywhere locally $AdS_3\times S^3
\times T^4$. A general BPS solution like a black ring or a
horizonless bubbling solution will have non-trivial geometry and
topology.

\section*{Appendix C. The angular momentum of the supertube}
\appendix
\renewcommand{\theequation}{C.\arabic{equation}}
\setcounter{equation}{0} \addcontentsline{toc}{section}{Appendix
C. The Angular Momentum of the Supertube}


\subsection*{Generalities}

Our goal in this Appendix is to compute the angular momentum of a
supertube in the background of three-charge black holes and black
rings. Once again we will work in the D0-D4-F1 duality frame:
\begin{equation}
ds^2_{IIA} = - \ds\frac{1}{Z_3\sqrt{Z_1Z_2}}(dt+k)^2 +
\sqrt{Z_1Z_2}ds^2_4 +\ds\frac{\sqrt{Z_1Z_2}}{Z_3}dz^2 +
\ds\sqrt{\frac{Z_1}{Z_2}}(dx_5^2+dx_6^2+dx_7^2+dx_8^2)~.
\label{D0D4F1metric-again}
\end{equation}
For the purpose of our calculations we can restrict without loss
of generality to a (non-generic) $U(1)\times U(1)$ invariant base
metric of the form:
\begin{equation}
ds^2_4 = g_1(u,v) du^2 + g_2(u,v) d\varphi_1^2 + h_1(u,v) dv^2 +
h_2(u,v) d\varphi_2^2~,
\end{equation}
in which the angular momentum vector has the form
\begin{equation}
k = k_1(u,v) d\varphi_1 +k_2(u,v) d\varphi_2 \,.
\end{equation}
The solutions we consider also have RR and NS-NS fields, which
have the general form
\begin{equation}
B = (Z_3^{-1} -1) dt \wedge dz + Z_3^{-1} k \wedge dz - B^{(3)}
\wedge dz
\end{equation}
\begin{equation}
C^{(1)} = (Z_1^{-1} -1) dt + Z_1^{-1} k  - B^{(1)}
\end{equation}
\begin{equation}
C^{(3)} = Z_3^{-1} dt \wedge k \wedge dz -Z_3^{-1} (dt+k) \wedge
B^{(1)} \wedge dz + B^{(3)} \wedge dt \wedge dz - f(u,v)
d\varphi_1 \wedge d\varphi_2 \wedge dz
\end{equation}
where the self-dual harmonic two-forms are
$\Theta^{(I)}=dB^{(I)}$, $I=1,2,3$ and
\begin{equation}
B^{(I)} = B^{(I)}_{\varphi_1} d\varphi_1 + B^{(I)}_{\varphi_2}
d\varphi_2 \,.
\end{equation}

Consider a probe supertube with world-volume coordinates $\xi=\{
\xi^0,\xi^1 ,\xi^2\equiv \theta\}$ in the above background and
suppose  that the supertube is embedded as follows:
\begin{equation}
t=\xi^0~, \qquad z=\xi^1 ~, \qquad \varphi_1 = \nu_1 \theta~,
\qquad \varphi_2 = \nu_2\theta
\end{equation}
where $0\leq \theta \leq 2\pi n_2^{ST}$ and $0\leq z\leq2\pi L_z$.
 The supertube   ``electric" charges are:
\begin{equation}
N_1^{ST} = \ds\frac{T_{D2}}{T_{D0}} \int dzd\theta
\mathcal{F}_{z\theta} = n_2^{ST} \mathcal{F}_{z\theta}
\label{N1ST}
\end{equation}
\begin{multline}
N_3^{ST} \!=\! \ds\frac{1}{T_{F1}} \int d\theta
\left(\ds\frac{\partial \mathcal{L}_{tot}}{\partial
\mathcal{F}_{tz}}\right)\bigg|_{BPS} \!\!=\! n_2^{ST} \! \left[
Z_2 \left( \ds\frac{\nu_1^2 g_2(u,v) +\nu_2^2
h_2(u,v)}{\mathcal{F}_{z\theta}+ \nu_1 B^{(3)}_{\varphi_1} +\nu_2
B^{(3)}_{\varphi_2}}\right) - (\nu_1 B^{(1)}_{\varphi_1} +\nu_2
B^{(1)}_{\varphi_2}) \right]
\end{multline}
Since the background is independent of $\varphi_1$ and
$\varphi_2$, the supertube has two conserved angular momenta:
\begin{equation}
J_{\varphi_1}^{ST} ~=~  \int dz d\theta  ~ \ds\frac{\partial
\mathcal{L}_{tot}}{\partial \dot{\varphi_1}}~, \qquad\qquad
J_{\varphi_2}^{ST} ~=~   \int dz d\theta  ~ \ds\frac{\partial
\mathcal{L}_{tot}}{\partial \dot{\varphi_2}}\,.
\end{equation}
One can compute them explicitly and find
\begin{multline}
J_{\varphi_1}^{ST} ~=~ n_2^{ST} \Bigg[\nu_1 Z_2g_2~-~
 \mathcal{F}_{z\theta}B^{(1)}_{\varphi_1} ~-~ Z_2
B^{(3)}_{\varphi_1} \bigg( \ds\frac{\nu_1^2 g_2 + \nu_2^2
h_2}{\mathcal{F}_{z\theta}+ \nu_1 B^{(3)}_{\varphi_1} +\nu_2
B^{(3)}_{\varphi_2}} \bigg) \\
+~ \nu_2(B^{(1)}_{\varphi_2} B^{(3)}_{\varphi_1} -
B^{(1)}_{\varphi_1} B^{(3)}_{\varphi_2})+\nu_2 f(u,v) \Bigg]~,
\label{J1general}
\end{multline}
\begin{multline}
J_{\varphi_2}^{ST} ~=~ n_2^{ST}\Bigg[\nu_2 Z_2 h_2~-~
  \mathcal{F}_{z\theta}B^{(1)}_{\varphi_2} ~-~   Z_2
B^{(3)}_{\varphi_2} \bigg( \ds\frac{\nu_1^2 g_2 + \nu_2^2
h_2}{\mathcal{F}_{z\theta}+ \nu_1 B^{(3)}_{\varphi_1} +\nu_2
B^{(3)}_{\varphi_2}}\bigg) \\
+~ \nu_1(B^{(1)}_{\varphi_1} B^{(3)}_{\varphi_2} -
B^{(1)}_{\varphi_2} B^{(3)}_{\varphi_1})-\nu_1 f(u,v)
\Bigg]~.\label{J2general}
\end{multline}
One can also define a ``total'' angular momentum of the supertube,
as the angular momentum along the direction of the supertube
\begin{equation}
J^{ST}_{TOT}~=~  \nu_1 J_{\varphi_1}^{ST} + \nu_2
J_{\varphi_2}^{ST}
\end{equation}
and one can show that
\begin{equation}
J^{ST}_{TOT} ~=~ \nu_1 J_{\varphi_1}^{ST} + \nu_2
J_{\varphi_2}^{ST} = \ds\frac{N_1^{ST} N_3^{ST}}{n_2^{ST}}~.
\label{JST}
\end{equation}
%

\subsection*{Flat Space}

For flat space we have
\begin{equation}
Z_{I} = 1~, \qquad B^{(I)}_{\varphi_1}=B^{(I)}_{\varphi_2}=0~,
\qquad k_1(u,v) = k_2(u,v)=0~, \qquad f(u,v)=0~,
\end{equation}
and using the change of variables $u=\rho\sin\vartheta$,
$v=\rho\cos\vartheta$ one has:
\begin{equation}
g_1(u,v) = h_1 (u,v)=1~, \qquad g_2=\rho^2\sin^2\vartheta~, \qquad
h_2 = \rho^2\cos^2\vartheta~.
\end{equation}
The conserved ``electric" charges of the supertube are
\begin{equation}
N_1^{ST} = n_2^{ST} \mathcal{F}_{z\theta}
\end{equation}
\begin{equation}
N_3^{ST} = n_2^{ST} \left( \ds\frac{\nu_1^2 \rho^2\sin^2\vartheta
+\nu_2^2 \rho^2\cos^2\vartheta}{\mathcal{F}_{z\theta}}\right)
\end{equation}
From these expressions one recovers the familiar radius relation
of the supertube
\begin{equation}
\ds\frac{N_1^{ST}N_3^{ST}}{(n_2^{ST})^2} = \rho^2 (\nu_1^2
\sin^2\vartheta + \nu_2^2 \cos^2\vartheta)~.
\end{equation}
The components of the supertube angular momentum are
\begin{equation}
J^{ST}_{\varphi_1} =  \nu_1 n_2^{ST} \rho^2\sin^2\vartheta~,
\end{equation}
\begin{equation}
J^{ST}_{\varphi_2} =  \nu_2 n_2^{ST} \rho^2\cos^2\vartheta~.
\end{equation}
Of course we again have
\begin{equation}
J^{ST}_{TOT} = \nu_1 J_{\varphi_1}^{ST} + \nu_2 J_{\varphi_2}^{ST}
= \ds\frac{N_1^{ST} N_3^{ST}}{n_2^{ST}}~.
\end{equation}
%

\subsection*{BMPV Black Hole}

For a BMPV black hole we have
\begin{eqnarray}
Z_{I}  & =& 1 + \ds\frac{Q_{I}}{\rho^2}~, \qquad
B^{(I)}_{\varphi_1}=B^{(I)}_{\varphi_2}=0~, \qquad k_1 =
\ds\frac{J \sin^2\vartheta}{\rho^2}~, \qquad k_2=-\ds\frac{J
\cos^2\vartheta}{\rho^2}~, \\
 f  & =& (Z_2-1) \rho^2 \cos^2\vartheta~,  \qquad
g_1(u,v) = h_1 (u,v)=1~,   \\
 g_2  & =& \rho^2\sin^2\vartheta~, \qquad  h_2 = \rho^2\cos^2\vartheta\,.
\end{eqnarray}
The conserved ``electric" charges of the supertube are
\begin{equation}
N_1^{ST} = n_2^{ST} \mathcal{F}_{z\theta}~,
\end{equation}
\begin{equation}
N_3^{ST} = n_2^{ST} \left( 1 + \ds\frac{Q_2}{\rho^2} \right)\left(
\ds\frac{\nu_1^2 \rho^2\sin^2\vartheta +\nu_2^2
\rho^2\cos^2\vartheta}{\mathcal{F}_{z\theta}}\right)~.
\end{equation}
These again lead to a radius relation for the supertube in the
background of the BMPV black hole
\begin{equation}
\ds\frac{N_1^{ST}N_3^{ST}}{(n_2^{ST})^2} = \left( 1 +
\ds\frac{Q_2}{\rho^2} \right) \rho^2 (\nu_1^2 \sin^2\vartheta +
\nu_2^2 \cos^2\vartheta)~.
\end{equation}
The components of the supertube angular momentum are
\begin{equation}
J^{ST}_{\varphi_1} = n_2^{ST} \left[\nu_1 \left( 1 +
\ds\frac{Q_2}{\rho^2} \right) \rho^2\sin^2\vartheta  + \nu_2 Q_2
\cos^2\vartheta\right]~,\label{J1_BH}
\end{equation}
\begin{equation}
J^{ST}_{\varphi_2} = n_2^{ST} \left[\nu_2 \left( 1 +
\ds\frac{Q_2}{\rho^2} \right) \rho^2\cos^2\vartheta  - \nu_1 Q_2
\cos^2\vartheta\right]~. \label{J2_BH}
\end{equation}
One can compare this result to the one obtained in
\cite{Marolf:2005cx} where the special case $\nu_1=n_2^{ST}=1$,
$\nu_2=0$ was considered. For these special values \eqref{J1_BH}
and \eqref{J2_BH} are identical to (4.4) and (4.5) in
\cite{Marolf:2005cx}.

\subsection*{Three-charge BPS Black Ring}

For a three-charge BPS black ring we have :
\begin{equation}
g_1 = \ds\frac{R^2}{(x-y)^2 (y^2-1)}~, \quad g_2 = \ds\frac{R^2
(y^2-1)}{(x-y)^2 }~, \quad h_1 = \ds\frac{R^2}{(x-y)^2 (1-x^2)}~,
\quad h_2 = \ds\frac{R^2 (1-x^2)}{(x-y)^2 }~.
\end{equation}
The functions, $Z_I$, appearing in the ten-dimensional metric, the
one-forms $B^{(I)}$ and the function $f(x,y)$ are given by
(\ref{constituent}), (\ref{Bi}) and (\ref{fBR}) respectively. The
explicit form of the angular momentum components of the black
ring, $k_1(x,y)$ and $k_2(x,y)$, is not needed here.

The conserved ``electric" charges of the supertube are
\begin{equation}
N_1^{ST} = n_2^{ST} \mathcal{F}_{z\theta}~,
\end{equation}
\begin{multline}
N_3^{ST} ~=~   n_2^{ST} \bigg[
\ds\frac{n_1}{2} ( -\nu_1(d+y) + \nu_2(c+x))  \\
+~ \ds\frac{Z_2}{\mathcal{F}_{z\theta} +\frac{n_3}{2} (-\nu_2(c+x)
+\nu_1(d+y))} \bigg(\nu_1^2 R^2 \ds\frac{(y^2-1)}{(x-y)^2}+\nu_2^2
R^2 \ds\frac{(1-x^2)}{(x-y)^2} \bigg) \bigg]~,
\end{multline}
which leads to the radius relation
\begin{multline}
\bigg[ N_1^{ST}+\ds\frac{1}{2} n_2^{ST} n_3 (\nu_1 (y+d) -\nu_2(x
+ c)) \bigg] \bigg[ N_3^{ST} +\ds\frac{1}{2} n_2^{ST} n_1 (\nu_1
(y+d) -\nu_2(x + c)) \bigg] ~=~ \\
(n_2^{ST})^2 Z_2 \ds\frac{R^2}{(x-y)^2} ( \nu_1^2(y^2-1) +
\nu_2^2(1-x^2) ) \label{radiuscondBRsimp-b}
\end{multline}
The components of the supertube angular momentum are
\begin{multline}
J_{\varphi_1}^{ST} ~=~   n_2^{ST}  \bigg[ -\mathcal{F}_{z\theta}
\ds\frac{n_1}{2}(d+y) + \nu_1 Z_2 R^2
\ds\frac{(y^2-1)}{(x-y)^2} +  \nu_2 f(x,y) \\
-~Z_2 \ds\frac{n_3(d+y)}{2} \bigg(\ds\frac{\nu_1^2 R^2
\frac{(y^2-1)}{(x-y)^2}+\nu_2^2 R^2
\frac{(1-x^2)}{(x-y)^2}}{\mathcal{F}_{z\theta} +\frac{n_3}{2}
(-\nu_2(c+x) +\nu_1(d+y)) } \bigg)  \bigg]
\end{multline}
\begin{multline}
J_{\varphi_2}^{ST} ~=~ n_2^{ST} \bigg[ \mathcal{F}_{z\theta}
\ds\frac{n_1}{2}(c+x) + \nu_2 Z_2 R^2
\ds\frac{(1-x^2)}{(x-y)^2} -\nu_1 f(x,y) \\
+~Z_2 \ds\frac{n_3(c+x)}{2} \bigg(\ds\frac{\nu_1^2 R^2
\frac{(y^2-1)}{(x-y)^2}+\nu_2^2 R^2
\frac{(1-x^2)}{(x-y)^2}}{\mathcal{F}_{z\theta} +\frac{n_3}{2}
(-\nu_2(c+x) +\nu_1(d+y)) } \bigg) \bigg]
\end{multline}
And we again have
\begin{equation} \label{JSTBR}
J^{ST}_{TOT} = \nu_1 J_{\varphi_1}^{ST} + \nu_2 J_{\varphi_2}^{ST}
= \ds\frac{N_1^{ST} N_3^{ST}}{n_2^{ST}}~.
\end{equation}
%

\section*{Appendix D. Units and conventions}
\appendix
\renewcommand{\theequation}{D.\arabic{equation}}
\setcounter{equation}{0} \addcontentsline{toc}{section}{Appendix
D. Units and Conventions}

Here we summarize some of the conventions we use in this paper (see
\cite{Polchinski:1998rr,Peet:2000hn} for more details).

The tensions of the extended objects in string and M-theory are:
\begin{equation}
T_{F1} = \ds\frac{1}{2\pi\alpha'}~, \qquad T_{Dp} =
\ds\frac{1}{g_s (2\pi)^p (l_s)^{p+1}}~, \qquad T_{NS5} =
\ds\frac{1}{g_s^2 (2\pi)^5 (l_s)^6} \,,
\end{equation}
\begin{equation}
T_{M2} = \ds\frac{1}{(2\pi)^2 (l_{11})^3}~, \qquad\qquad T_{M5} =
\ds\frac{1}{(2\pi)^5 (l_{11})^{6}}
\end{equation}
where $\alpha'=l_s^2$, $l_s$ is the string length, $g_s$ is the
string coupling constant (in the particular duality frame in which
one works) and $l_D$ is the $D$-dimensional Planck length. The
Newton's constant in different dimensions is
\begin{equation}
16\pi G_{11} = (2\pi)^8 (l_{11})^9~, \qquad 16\pi G_{10} =
(2\pi)^7 (g_s)^2 (l_{s})^8~, \qquad 16\pi G_D = (2\pi)^{D-3}
(l_{D})^{D-2} \,.
\end{equation}
One can show that
\begin{equation}
l_{11} = g_s^{1/3} l_s = g_s^{1/3} (\alpha')^{1/2} \,.
\end{equation}
T-duality along a circle of radius $R$ changes the coupling
constants to:
\begin{equation}
\widetilde{R} = \ds\frac{\alpha'}{R}~, \qquad \tilde{g}_s =
\ds\frac{l_s}{R} g_s~,\qquad \tilde{l}_s = l_s \,.
\end{equation}
where $\widetilde{R}$ is the radius after T-duality:

When one compactifies M-theory on a circle of radius $L_{9}$, the
coupling constants of the resulting type IIA string theory
satisfy:
\begin{equation}
L_{9} = g_s l_s \,.
\end{equation}

If one compactifies M-theory on a $T^6$ (along the directions
$5,6,7,8,9,10$) and the radius of each circle is $L_i$
($i=\{5,6,7,8,9,10\}$), the five-dimensional Newton's constant is
\begin{equation}
G_5 = \ds\frac{G_{11}}{vol(T^6)} = \ds\frac{G_{11}}{(2\pi)^6 L_5
L_6L_7L_8L_9L_{10}} = \ds\frac{\pi}{4} \ds\frac{(l_{11})^9}{L_5
L_6L_7L_8L_9L_{10}} \,.
\end{equation}
The relations between the number of M2 and M5 branes, $N_I$ and
$n_I$, and the physical charges of the five-dimensional solution
obtained by compactifying M-theory on a $T^6$, $Q_I$ and $q_I$,
are
\begin{equation}
Q_1=\ds\frac{(l_{11})^6}{L_{7}L_{8}L_{9}L_{10}}~N_1~, \qquad
Q_2=\ds\frac{(l_{11})^6}{L_{5}L_{6}L_{9}L_{10}}~N_2~, \qquad
Q_3=\ds\frac{(l_{11})^6}{L_{5}L_{6}L_{7}L_{8}}~N_3~,
\end{equation}
\begin{equation}
q_1=\ds\frac{(l_{11})^3}{L_{5}L_{6}}~n_1~, \qquad
q_2=\ds\frac{(l_{11})^3}{L_{7}L_{8}}~n_2~, \qquad
q_3=\ds\frac{(l_{11})^3}{L_{9}L_{10}}~n_3~.
\end{equation}
We will choose a system of units in which all three $T^2$ are of
equal volume
\begin{equation}
L_5 L_6 = L_7 L_8 = L_9 L_{10} = (l_{11})^3 \equiv g_s l_s^3 ~,
\end{equation}
note that this is a numerical identity and is not dimensionally
correct since $g_s$ is dimensionless. With this choice we will
have
\begin{equation}
G_5 = \ds\frac{\pi}{4}~, \qquad\qquad Q_I = N_I~, \qquad\qquad
q_I=n_I~.
\end{equation}
and these identities hold in every duality frame we use in the
paper. Furthermore we will choose
\begin{equation}
g_s l_s = 1\,.
\end{equation}
Since we are compactifying M-theory on $L_9$ we will have
$L_9=g_sl_s=1$ and $L_{10}=l_s^2$, this implies (note that
throughout the paper we put $L_{10}\equiv L_z$)
\begin{equation}
T_{D0} = 1~, \qquad 2\pi T_{F1}L_{10} = 1~, \qquad \text{and}
\qquad \ds\frac{2\pi T_{D2}}{T_{F1}}=1\,.
\end{equation}
We have fixed $l_s=g_s^{-1}$ so that a lot of the various brane
tension factors, appearing in the probe supertube calculations
throughout the paper, cancel. Note that with our choices $g_s$ is
still a free parameter but we have fixed the volume of the
compactification torii.



\end{document}